

\def\figsize{9.5cm}
\def\figsiz{8.5cm}
\def\smtopskip{-1.2cm}
\def\smbotskip{-1.0cm}

\def\rn{\noindent\parshape 2 0truecm 8.8truecm 0.3truecm 8.5truecm}
\def\nn#1 #2{#1, #2.}				
\def\nnn#1 #2 #3{#1, #2. #3.}			
\def\nnnn#1 #2 #3 #4{#1, #2. #3. #4.}		
\def\nnnnn#1 #2 #3 #4 #5{#1, #2. #3. #4. #5.}	
\def\dualand{, \&\hbox{ }}				
\def\multiand{, \&\hbox{ }}				

\def\rg#1;#2;#3;#4;#5;#6 {\par\rn#1 #2, {\it #3}, {\bf #4}, #5 (``#6'') \par}
\def\rf#1;#2;#3;#4;#5 {\par\rn#1 #2, {\it #3}, {\bf #4}, #5\par}
\def\rfbook#1;#2;#3;#4;#5 {{\frenchspacing\par\rn#1 #2, {\it #3} (#4: #5)\par}}
\def\rfproc#1;#2;#3;#4;#5;#6 {{\frenchspacing\par\rn#1 #2, in {\it #3}, ed. #4 (#5: #6)\par}}
\def\rfprocp#1;#2;#3;#4;#5;#6;#7 {{\frenchspacing\par\rn#1 #2, in {\it #3}, ed. #4 (#5: #6), p#7\par}}
\def\rfprep#1;#2;#3  {{\par\rn#1 #2, #3\par}}
\def\rfprepp#1;#2;#3 {{\par\rn#1 #2, #3\par}}

\def\Mpc{{\rm Mpc}}
\def\hperMpc{\,h/\Mpc}
\def\hMpc{\,h^{-1}\Mpc}

\def\expec#1{\langle#1\rangle}

\def\etal{{\frenchspacing\it et al.}}
\def\ie{{\frenchspacing\it i.e.}}
\def\eg{{\frenchspacing\it e.g.}}
\def\etc{{\frenchspacing\it etc.}}

\def\beq#1{\begin{equation}\label{#1}}
\def\eeq{\end{equation}}
\def\beqa#1{\begin{eqnarray}\label{#1}}
\def\eeqa{\end{eqnarray}}
\def\eq#1{equation~(\ref{#1})}
\def\Eq#1{Equation~(\ref{#1})}
\def\eqn#1{~(\ref{#1})}

\def\fig#1{Figure~\ref{#1}}
\def\Fig#1{Figure~\ref{#1}}

\def\sec#1{Section~\ref{#1}}
\def\Sec#1{Section~\ref{#1}}

\def\nskip{\hskip-2mm}

\def\spose#1{\hbox to 0pt{#1\hss}}
\def\simlt{\mathrel{\spose{\lower 3pt\hbox{$\mathchar"218$}}
     \raise 2.0pt\hbox{$\mathchar"13C$}}}
\def\simgt{\mathrel{\spose{\lower 3pt\hbox{$\mathchar"218$}}
     \raise 2.0pt\hbox{$\mathchar"13E$}}}
\def\simpropto{\mathrel{\spose{\lower 3pt\hbox{$\mathchar"218$}}
     \raise 2.0pt\hbox{$\propto$}}}

\def\ed{\end{document}}

\def\Ob{\Omega_b}

\def\Ol{\Omega_\Lambda}
\def\Om{\Omega_m}

\def\ns{{n_s}}
\def\Angstrom{{\rm AA}}  

\def\rdim{r_{\mathrm{limit}}}

\def\beq#1{\begin{equation}\label{#1}}
\def\eeq{\end{equation}}
\def\beqa#1{\begin{eqnarray}\label{#1}}
\def\eeqa{\end{eqnarray}}
\def\eq#1{equation~(\ref{#1})}
\def\Eq#1{Equation~(\ref{#1})}
\def\eqn#1{~(\ref{#1})}

\def\tr{\hbox{tr}\>}

\def\deltah{\widehat{\delta}}
\def\nbar{{\bar n}}

\def\ith{i^{\rm th}}

\def\e{{\bf e}}

\def\k{{\bf k}}
\def\r{{\bf r}}
\def\khat{\widehat{\bf k}}
\def\rhat{\widehat{\bf r}}
\def\ahat{\widehat{\bf a}}

\def\p{{\bf p}}
\def\phat{\widehat\p}
\def\ph{\widehat p}

\def\r{{\bf r}}
\def\x{{\bf x}}

\def\z{{\bf z}}

\def\C{{\bf C}}

\def\F{{\bf F}}
\def\I{{\bf I}}

\def\SS{{\bf\Sigma}}

\def\M{{\bf M}}
\def\N{{\bf N}}
\def\P{{\bf P}}

\def\W{{\bf W}}

\def\Q{{\bf Q}}

\def\Sb{{\bf S}}
\def\Sgg{{\bf P}_{\rm gg}}
\def\Sgv{{\bf P}_{\rm gv}}
\def\Svv{{\bf P}_{\rm vv}}

\def\zetah{\widehat{\zeta}}

\def\psih{\widehat{\psi}}

\def\M{{\bf M}}

\def\rh{\widehat{\bf r}}

\def\tr{\hbox{tr}\>}
\def\dV{{d^3k\over (2\upi)^3}}

\def\ngal{N_{\rm gal}}

\def\nx{{N_x}}

\def\l{\ell}

\def\lcut{\l_{\rm cut}}

\def\Pgg{P_{\rm gg}}
\def\Pgv{P_{\rm gv}}
\def\Pvv{P_{\rm vv}}
\def\Pggtrue{P_{\rm gg}^{\rm true}}
\def\Pmono{P^{\rm s}_0}
\def\Pquad{P^{\rm s}_2}
\def\Phexa{P^{\rm s}_4}

\def\ignore#1{}
\def\upi{\pi}		

\def\DM{{\rm DM}}

\def\beff{b_{\rm eff}}
\def\bstar{b_*}
\def\deltaobs{\delta_{\rm obs}}
\def\Mdim{{M_{\rm dim}}}
\def\Mbri{{M_{\rm bri}}}
\def\Mmin{{M_{\rm min}}}
\def\Mmax{{M_{\rm max}}}
\def\Phat{{\tilde P}}

\def\tr{\hbox{tr}\,}

\def\ith{i^{th}}

\newcommand{\lsssample}{\texttt{sample11}}

\def\simless{\mathbin{\lower 3pt\hbox
        {$\,\rlap{\raise 5pt\hbox{$\char'074$}}\mathchar"7218\,$}}} 
\def\simgreat{\mathbin{\lower 3pt\hbox
        {$\,\rlap{\raise 5pt\hbox{$\char'076$}}\mathchar"7218\,$}}} 

\newcommand{\band}[2]{{^{#1}\!{#2}}}

\documentstyle[emulateapj,danonecolfloat]{article}
\def\NoApjSectionMarkInTitle#1{#1.\ }
\begin{document}
\twocolumn[

\journalid{337}{15 January 1989}
\articleid{11}{14}

\def\penn{1}
\def\nyu{2}
\def\pton{3}
\def\drexel{4}
\def\osu{5}
\def\fnal{6}
\def\chicago{7}
\def\hopkins{8}
\def\pitt{9}
\def\tucson{10}
\def\colorado{11}
\def\cmu{12}
\def\hawaii{13}
\def\apo{14}
\def\mit{15}
\def\barcelona{16}
\def\sussex{17}
\def\tokyo{18}
\def\flagstaff{19}
\def\michigan{20}
\def\rochester{21}
\def\psu{22}
\def\efi{23}

\submitted{Submitted to the Astrophysical Journal June 18 2003, report received Sept.~21, resubmitted Oct.~27}

\title{The Three-Dimensional Power Spectrum of Galaxies from the Sloan
Digital Sky Survey}

\author{
Max Tegmark\altaffilmark{\penn}, 
Michael R. Blanton\altaffilmark{\nyu},
Michael A. Strauss\altaffilmark{\pton}, 
Fiona Hoyle\altaffilmark{\drexel},
David Schlegel\altaffilmark{\pton}, 
Roman Scoccimarro\altaffilmark{\nyu}, 
Michael S. Vogeley\altaffilmark{\drexel},
David H. Weinberg\altaffilmark{\osu}, 
Idit Zehavi\altaffilmark{\chicago},
Andreas Berlind\altaffilmark{\chicago},
Tam\'as Budavari\altaffilmark{\hopkins}, 
Andrew Connolly\altaffilmark{\pitt},
Daniel J. Eisenstein\altaffilmark{\tucson},
Douglas Finkbeiner\altaffilmark{\pton},
Joshua A. Frieman\altaffilmark{\chicago,\fnal},
James E. Gunn\altaffilmark{\pton}, 
Andrew J. S. Hamilton\altaffilmark{\colorado}, 
Lam Hui\altaffilmark{\fnal}, 
Bhuvnesh Jain\altaffilmark{\penn},
David Johnston\altaffilmark{\chicago,\fnal}, 
Stephen Kent\altaffilmark{\fnal},
Huan Lin\altaffilmark{\fnal},
Reiko Nakajima\altaffilmark{\penn}, 
Robert C. Nichol\altaffilmark{\cmu}, 
Jeremiah P. Ostriker\altaffilmark{\pton},
Adrian Pope\altaffilmark{\hopkins}, 
Ryan Scranton\altaffilmark{\pitt},
Uro\v s Seljak\altaffilmark{\pton},
Ravi K. Sheth\altaffilmark{\pitt}, 
Albert Stebbins\altaffilmark{\fnal},
Alexander S. Szalay\altaffilmark{\hopkins},
Istv\'an Szapudi\altaffilmark{\hawaii}, 
Licia Verde\altaffilmark{\pton},
Yongzhong Xu\altaffilmark{\penn}, 
James Annis\altaffilmark{\fnal}, 
Neta A. Bahcall\altaffilmark{\pton}, 
J. Brinkmann\altaffilmark{\apo},
Scott Burles\altaffilmark{\mit},
Francisco J. Castander\altaffilmark{\barcelona},
Istvan Csabai\altaffilmark{\hopkins},
Jon Loveday\altaffilmark{\sussex}, 
Mamoru Doi\altaffilmark{\tokyo},  
Masataka Fukugita\altaffilmark{\tokyo},
J. Richard Gott III\altaffilmark{\pton},
Greg Hennessy\altaffilmark{\flagstaff},
David W. Hogg\altaffilmark{\nyu},
\v Zeljko Ivezi\'c\altaffilmark{\pton},
Gillian R. Knapp\altaffilmark{\pton},
Don Q. Lamb\altaffilmark{\chicago},
Brian C. Lee\altaffilmark{\fnal},
Robert H. Lupton\altaffilmark{\pton},
Timothy A. McKay\altaffilmark{\michigan},
Peter Kunszt\altaffilmark{\hopkins},
Jeffrey A. Munn\altaffilmark{\flagstaff}, 
Liam O'Connell\altaffilmark{\sussex},
John Peoples\altaffilmark{\fnal}, 
Jeffrey R. Pier\altaffilmark{\flagstaff},
Michael Richmond\altaffilmark{\rochester},
Constance Rockosi\altaffilmark{\chicago}, 
Donald P. Schneider\altaffilmark{\psu}, 
Christopher Stoughton\altaffilmark{\fnal}, 
Douglas L. Tucker\altaffilmark{\fnal},
Daniel E. Vanden Berk\altaffilmark{\pitt},
Brian Yanny\altaffilmark{\fnal}, 
Donald G. York\altaffilmark{\chicago,\efi},
for the SDSS Collaboration
}
\footnote{\penn}{Department of Physics, University of Pennsylvania,
Philadelphia, PA 19101, USA}
\footnote{\nyu}{Center for Cosmology and Particle Physics,
Department of Physics, New York University, 4 Washington
Place, New York, NY 10003}
\footnote{\pton}{Princeton University Observatory, Princeton, NJ 08544,
USA}
\footnote{\drexel}{Department of Physics, Drexel University, Philadelphia,
PA
19104, USA}
\footnote{\osu}{Department of Astronomy, Ohio State University, 
Columbus, OH 43210, USA}
\footnote{\fnal}{Fermi National Accelerator Laboratory, P.O. Box 500, Batavia,
IL 60510, USA}
\footnote{\chicago}{Center for Cosmological Physics and Department of Astronomy \& Astrophysics, University of
Chicago, Chicago, IL 60637, USA}
\footnote{\hopkins}{Department of Physics and Astronomy, The Johns Hopkins
University, 3701 San Martin Drive, Baltimore, MD 21218, USA}
\footnote{\pitt}{University of Pittsburgh, Department of Physics and
Astronomy, 3941 O'Hara Street, Pittsburgh, PA 15260, USA}
\footnote{\tucson}{Department of Astronomy, University of Arizona, 
Tucson, AZ 85721, USA}
\footnote{\colorado}{JILA and Dept.~of Astrophysical and Planetary Sciences, 
U. Colorado, Boulder, CO 80309, USA, Andrew.Hamilton@colorado.edu}
\footnote{\cmu}{Department of Physics, 5000 Forbes Avenue, Carnegie
Mellon
University, Pittsburgh, PA 15213, USA}
\footnote{\hawaii}{Institute for Astronomy, University of Hawaii, 2680
Woodlawn Drive, Honolulu, HI 96822, USA}
\footnote{\apo}{Apache Point Observatory, 2001 Apache Point Rd, 
Sunspot, NM 88349-0059, USA}
\footnote{\mit}{Dept. of Physics, Massachusetts Institute of Technology, 
Cambridge, MA 02139}
\footnote{\barcelona}{Institut d'Estudis Espacials de Catalunya/CSIC, Gran Capita 2-4, 
08034 Barcelona, Spain}
\footnote{\sussex}{Sussex Astronomy Centre, University of Sussex, Falmer,
Brighton BN1 9QJ, UK}
\footnote{\tokyo}{Inst. for Cosmic Ray Research, 
Univ. of Tokyo, Kashiwa 277-8582, Japan}
\footnote{\flagstaff}{U.S. Naval Observatory, 
Flagstaff Station, Flagstaff, AZ 86002-1149, USA}
\footnote{\michigan}{Dept. of Physics, Univ. of Michigan, 
Ann Arbor, MI 48109-1120, USA}
\footnote{\rochester}{Physics Dept., Rochester Inst. of Technology, 
1 Lomb Memorial Dr, Rochester, NY 14623, USA}
\footnote{\psu}{Dept. of Astronomy and Astrophysics, Pennsylvania State University, 
University Park, PA 16802, USA}
\footnote{\efi}{Enrico Fermi Institute, University of
Chicago, Chicago, IL 60637, USA}

\begin{abstract}
We measure the large-scale real-space power spectrum $P(k)$ using a sample of 205{,}443 galaxies 
from the Sloan Digital Sky Survey, covering 2417 effective square degrees with mean redshift $z\approx 0.1$.
We employ a matrix-based method using pseudo-Karhunen-Lo\`eve eigenmodes,
producing 
uncorrelated minimum-variance measurements in 22 $k$-bands of 
both the clustering power and its anisotropy due to redshift-space distortions,
with narrow and well-behaved window functions in the range
$0.02\hperMpc < k < 0.3\hperMpc$.
We pay particular attention to modeling, quantifying and correcting for 
potential systematic errors,
nonlinear redshift distortions and
the artificial red-tilt caused by 
luminosity-dependent bias. Our results are robust to omitting angular and 
radial density fluctuations and are consistent between different
parts of the sky.
Our final result is a measurement of the real-space matter power spectrum $P(k)$
up to an unknown overall multiplicative bias factor. 
Our calculations suggest that this bias factor 
is independent of scale to better than a few percent for $k<0.1\hperMpc$, 
thereby making our results useful for precision measurements of 
cosmological parameters in conjunction with data from other experiments such as 
the WMAP satellite.  The power spectrum is not well-characterized by a
single power law, but unambiguously shows curvature. 
As a simple characterization of the data, 
our measurements are well fit by a flat scale-invariant adiabatic cosmological model with 
$h\Om =0.213\pm 0.023$ 
and $\sigma_8=0.89\pm 0.02$ for $L_*$ galaxies, 
when fixing the baryon fraction $\Omega_b/\Omega_m=0.17$ and the Hubble parameter $h=0.72$;
cosmological interpretation is given in a companion paper.
\end{abstract}

]

\def\v{{\bf v}}

\setcounter{footnote}{0}

\section{Introduction}

The spectacular recent cosmic microwave background (CMB) measurements from the 
WMAP satellite (Bennett {\etal} 2003) and other experiments 
have increased the importance of non-CMB measurements
for the endeavor to constrain cosmological models and their free parameters.
These non-CMB constraints are crucially needed for breaking CMB degeneracies
(Eisenstein {\etal} 1999; Efstathiou \& Bond 1999; Bridle {\etal} 2003); for instance, 
WMAP alone is consistent with a closed universe with
Hubble parameter $h=0.32$ and no cosmological constant (Spergel {\etal} 2003; Verde {\etal} 2003).
Yet they are currently less reliable and precise than the CMB, making them the limiting factor and 
weakest link in the quest for precision cosmology.
Much of the near-term progress in cosmology will therefore be driven by
reductions in statistical and systematic uncertainties of
non-CMB probes such as Lyman $\alpha$ forest and galaxy clustering and motions,
gravitational lensing, cluster studies, and supernovae Ia distance determinations.
Galaxy redshift surveys can play a key role in breaking degeneracies and providing 
cross checks (Tegmark 1997a; Goldberg \& Strauss 1998; Wang {\etal} 1999; 
Eisenstein {\etal} 1999), 
but only if systematics can be controlled to high precision.
The goal of the present paper is to do just this, using 
over 200{,}000 galaxies from the Sloan Digital Sky Survey (SDSS; York {\etal} 2000)
to measure the shape of the real-space matter power spectrum $P(k)$, 
accurately quantifying and correcting for the effects of light-to-mass bias, 
redshift space distortions, survey geometry effects and other complications.

The cosmological constraining power of 
three-dimensional maps of the Universe provided by
galaxy redshift surveys has motivated ever more ambitious
observational efforts such as 
the CfA/UZC (Huchra {\etal} 1990; Falco {\etal} 1999),
LCRS (Shectman {\etal} 1996), and PSCz (Saunders {\etal} 2000) 
surveys, each well in excess of $10^4$ galaxies.
The current state of the art is the 
AAT two degree field galaxy redshift survey (2dFGRS; Colless {\etal} 2001; Hawkins {\etal} 2003; 
Peacock 2003 and references therein).
Analysis of the first 147{,}000 2dFGRS galaxies 
(Peacock {\etal} 2001; Percival {\etal} 2001, 2002; Norberg {\etal} 2001, 2002; Madgwick {\etal} 2002)
have supported a flat dark-energy
dominated cosmology, as have angular clustering analyses of the 
parent catalogs underlying the 2dFGRS (Efstathiou \& Moody 2001)
and SDSS 
(Scranton {\etal} 2002; Connolly {\etal} 2002; Tegmark {\etal} 2002; 
Szalay {\etal} 2003; Dodelson {\etal} 2002).
Tantalizing evidence for baryonic wiggles in the galaxy power spectrum
is presented by
Percival {\etal} (2001) and Miller {\etal} (2001a,b, 2002),
and cosmological models have been further constrained in conjunction with
cosmic microwave background (CMB) data (\eg, Spergel {\etal} 2003;
Verde {\etal} 2003; Lahav {\etal} 2002).

The SDSS is the most ambitious galaxy redshift survey to date, whose
goal, driven by large-scale structure science, is to measure of order 
$10^6$ galaxy  redshifts. Zehavi {\etal} (2002) computed
the correlation function using about 30{,}000 galaxies from early SDSS data 
(Stoughton {\etal} 2002).
In conjunction with the first major SDSS data release in 2003 (hereafter DR1; Abazajian {\etal} 2003), 
a series of papers will address various aspects of the 3D clustering of a much larger data set involving
over 200{,}000 galaxies with redshifts. 
This paper is focused on measuring the power galaxy spectrum $P(k)$ on large scales,
dealing with complications such as luminosity-dependent bias and redshift distortions only to
the extent necessary to recover an undistorted measurement of the real-space matter power spectrum.
Zehavi {\etal} (2003a) measure and model the real space 
correlation function, mainly on smaller scales, focusing on 
departures from power-law behavior, and 
Zehavi {\etal} (2003b) will study how the correlation function depends on galaxy properties.
Pope {\etal} (2003) measure the parameters which characterize the large-scale power spectrum with a complementary
approach involving direct likelihood analysis on Karhunen-Lo\`eve eigenmodes, as opposed to the quadratic
estimator technique employed in the present paper. 

This paper is organized as follows.
In \sec{DataSec}, we describe the SDSS data used and how we model it;
the technical details can be found in Appendix~\ref{DataAppendix}.
In \sec{AnalysisSec}, we describe our methodology and present our basic
measurements of both the power spectrum and its redshift-space
anisotropy.  The details of the formalism for doing this are described
in Appendix~\ref{powerdetails}. 
In \sec{zspaceSec} we focus on this anisotropy to model, quantify and correct for 
the effects of redshift-space distortions, producing an estimate of
the real-space galaxy power spectrum and testing our procedure with Monte-Carlo simulations.
In \sec{BiasSec}, we model, quantify and correct for the effects of
luminosity-dependent biasing, producing an estimate of the 
real-space matter power spectrum.
In \sec{SystematicsSec}, we test for 
a variety of systematic errors.
In \sec{ConcSec}, we discuss our results.
The cosmological interpretation of our measurements is given in a companion paper
(Tegmark {\etal} 2003, hereafter ``Paper II'').

\begin{figure*} 
\noindent
\hglue0.7cm\centerline{\epsfxsize=14.4cm\epsffile{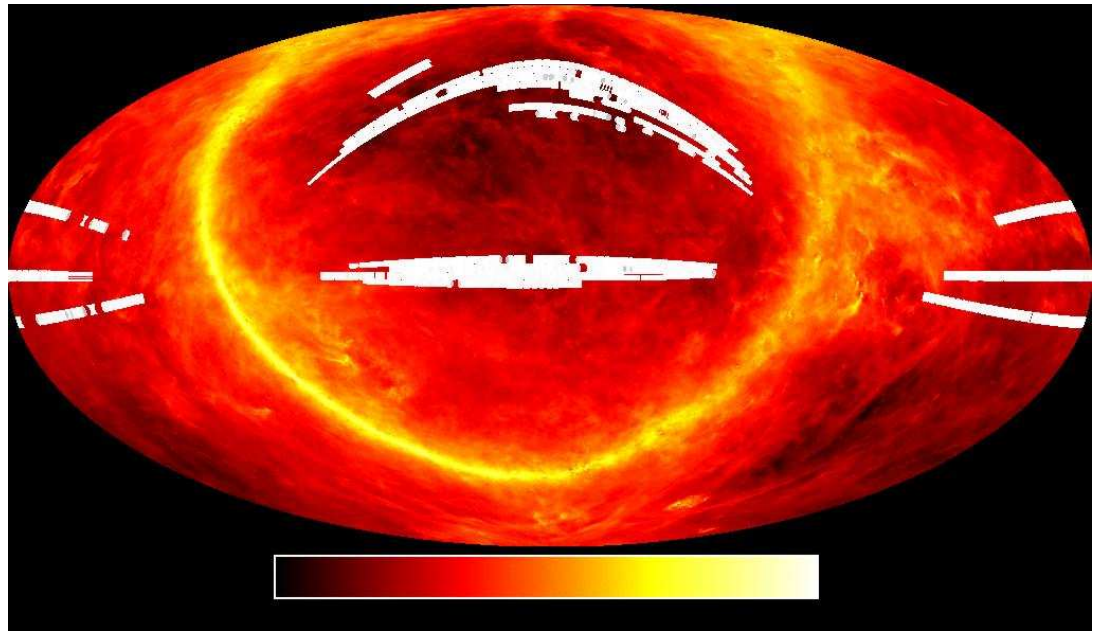}}
\hglue0.7cm\centerline{\epsfxsize=10cm\epsffile{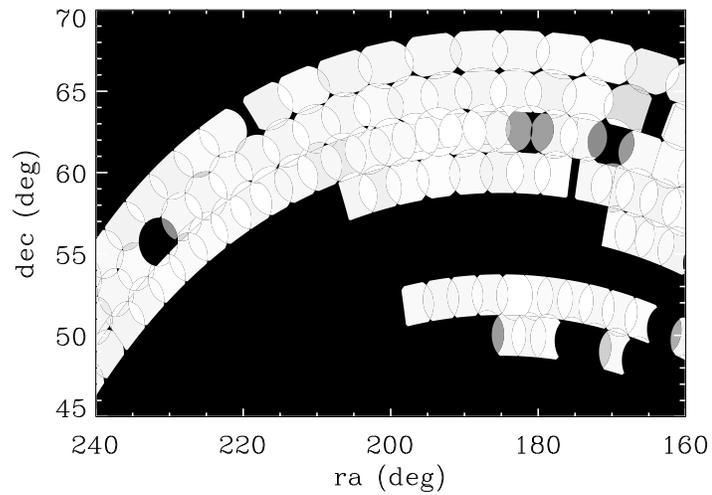}}
\noindent
\caption[1]{\label{aitoffFig}\footnotesize%
The upper panel shows the angular completeness map, 
the relative probabilities that galaxies in various directions
get included, in Hammer-Aitoff projection
in equatorial coordinates on a grayscale ranging from black (0) to white (1).
It is this completeness map that we expand in spherical harmonics.
The backdrop is the logarithm of the dust map from 
Schlegel, Finkbeiner, \& Davis (1998), indicating which sky regions
are most likely to be affected by extinction-related systematic errors.
The lower panel illustrates the complex nature of the completeness map 
and the high average completeness with a zoom of a small sky region.
}
\end{figure*}

\section{Data and data modeling}
\label{DataSec}

The SDSS uses a mosaic CCD camera (Gunn {\etal} 1998) to image the sky in five photometric
bandpasses denoted $u$, $g$, $r$, $i$, $z$\footnote{
The Fukugita {\etal} (1996) paper actually defines a slightly different system, 
denoted $u'$, $g'$, $r'$, $i'$, $z'$, but SDSS magnitudes are now referred to the native filter
system of the 2.5m survey telescope, for which the bandpass notation is unprimed.
} 
(Fukugita {\etal} 1996).
After astrometric calibration (Pier {\etal} 2003), 
photometric data reduction (Lupton {\etal} 2003, in preparation; see Lupton {\etal} 2001 and 
Stoughton {\etal} 2002 for summaries)
and photometric calibration (Hogg {\etal} 2001; Smith {\etal} 2002), 
galaxies are selected for spectroscopic observations using the algorithm described 
by Strauss {\etal} (2002). To a good approximation, the main galaxy sample consists of 
all galaxies with $r$-band apparent Petrosian magnitude $r<17.77$; see
Appendix~\ref{DataAppendix}. 
Galaxy spectra are also measured for a luminous red galaxy sample 
(Eisenstein {\etal} 2001), for which clustering results will be reported in a separate paper.
These targets are assigned to spectroscopic plates by an adaptive 
tiling algorithm (Blanton {\etal} 2003) and 
observed with a pair of fiber-fed CCD spectrographs
(Uomoto {\etal}, in preparation), after which the
spectroscopic data reduction and redshift determination are performed by 
automated pipelines
(Schlegel {\etal}, in preparation; Frieman {\etal}, in preparation).
The rms galaxy redshift errors are $\sim$ 30 km/s and hence negligible for the
purpose of the present paper.

Our analysis is based on SDSS {\tt sample11} (Blanton {\etal} 2003c), 
consisting of the 205{,}443 galaxies observed before 
July 2002, all of which will be included in the upcoming SDSS Data Release 2. 
From this basic sample, we produce a 
set of subsamples as specified in Table 1. 
The details of how this basic sample was processed, modeled and subdivided are given
in Appendix A. The bottom line is that each sample is completely specified by three entities:
\begin{enumerate}
\item The galaxy positions (a list of RA, Dec and comoving redshift
space distance $r$ for each galaxy)
\item The radial selection function $\nbar(r)$, which gives the expected (not
observed) number density of galaxies as a function of distance
\item The angular selection function $\nbar(\rhat)$, which gives the 
completeness as a function of direction in the sky
\end{enumerate}
Our samples are constructed so that 
their three-dimensional selection function is separable, \ie, simply the product 
$\nbar(\r)=\nbar(\rhat)\nbar(r)$ of an angular and a radial part; 
here $r\equiv |\r|$ and $\rhat \equiv \r/r$ are the comoving radial distance
and the unit vector corresponding to the position $\r$.
The conversion from redshift $z$ to comoving distance was made for a flat cosmological model with
a cosmological constant $\Omega_\Lambda=0.7$ --- below we will see that our results
are insensitive to this assumption.

\def\MM{M_{\band{0.1}r}}

\begin{table*}
\bigskip
\noindent
{\footnotesize {\bf Table 1} -- The table summarizes the various
galaxy samples used in our analysis, listing cuts made on
evolution-corrected absolute magnitude $M_{\band{0.1}r}$ (for $h=1$), apparent
magnitude $r$ and redshift $z$. 
$M_{\band{0.1}r}$ was computed from $r$ and $z$ assuming a flat cosmological model with
$\Ol=0.7$.  
\bigskip
\begin{center}   
{\footnotesize
\begin{tabular}{lrrrrr}
\hline
Sample name	&Abs.~mag  	&App.~mag   		&Redshift	&Sq. degrees	&Galaxies\\
\hline
{\tt all}	&All	  	&$r<17.5-17.77$ 	&All		&2417        	&205{,}443\\  
{\tt safe0}	&All		&$14.50<r<17.50$	&All		&2417		&157{,}389\\  
{\tt safe13}	&$-23<\MM<-18.5$&$14.50<r<17.50$	&$0.001<z<0.400$&2417		&146{,}633\\  
{\tt safe22}	&$-22<\MM<-19.0$&$14.50<r<17.50$	&$0.001<z<0.400$&2417		&134{,}674\\  
{\tt baseline}	&$-23<\MM<-18.5$&$14.50<r<17.50$	&$0.002<z<0.210$&2417           &143{,}314\\  
\hline
{\bf Angular subsamples:}\\
{\tt A1} (south)	&$-23<\MM<-18.5$&$14.50<r<17.50$        &$0.017<z<0.210$& 600	       & 35{,}782\\ 
{\tt A2} (north)	&$-23<\MM<-18.5$&$14.50<r<17.50$        &$0.017<z<0.210$&1817	       &107{,}532\\ 
{\tt A3} (north eq)	&$-23<\MM<-18.5$&$14.50<r<17.50$        &$0.017<z<0.210$& 809	       & 52{,}081\\ 
{\tt A4} (north rest)	&$-23<\MM<-18.5$&$14.50<r<17.50$        &$0.017<z<0.210$&1007	       & 55{,}451\\ 
\hline
{\bf Radial subsamples:}\\
{\tt R1} (near)	&$-23<\MM<-18.5$&$14.50<r<17.50$        &$0.017<z<0.078$&2417	      &47{,}954\\ 
{\tt R2} (mid)	&$-23<\MM<-18.5$&$14.50<r<17.50$        &$0.078<z<0.117$&2417	      &47{,}089\\ 
{\tt R3} (far)	&$-23<\MM<-18.5$&$14.50<r<17.50$        &$0.117<z<0.210$&2417	      &48{,}271\\ 
\hline
{\bf Luminosity (volume-limited) subsamples:}\\
{\tt L1}	&$-17<\MM<-16$	&$14.50<r<17.50$	&$0.008<z<0.017$&2417		&   455\\ 
{\tt L2}	&$-18<\MM<-17$	&$14.50<r<17.50$	&$0.011<z<0.027$&2417		& 1{,}736\\ 
{\tt L3}	&$-19<\MM<-18$	&$14.50<r<17.50$	&$0.017<z<0.042$&2417		& 5{,}191\\ 
{\tt L4}	&$-20<\MM<-19$	&$14.50<r<17.50$	&$0.027<z<0.065$&2417		&14{,}356\\ 
{\tt L5}	&$-21<\MM<-20$	&$14.50<r<17.50$	&$0.042<z<0.103$&2417		&31{,}026\\ 
{\tt L6}	&$-22<\MM<-21$	&$14.50<r<17.50$	&$0.065<z<0.157$&2417		&24{,}489\\ 
{\tt L7}	&$-23<\MM<-22$	&$14.50<r<17.50$	&$0.104<z<0.238$&2417		& 3{,}594\\ 
{\tt L8}	&$-24<\MM<-23$	&$14.50<r<17.50$	&$0.164<z<0.349$&2417		&    95\\ 
\hline
{\bf Mock samples:}\\
{\tt M1-M275} (PThalos)	&		& 	 	     	&$0.015<z<0.240$&1395	     	&108{,}300\\ 
{\tt V1-V10} (VIRGO)	&		& 	 	     	&$0.001<z<0.150$&1139	     	&103{,}400\\ 
\hline  
\end{tabular}
}
\end{center}     
} 
\end{table*}

\subsection{Angular selection function}

The angular selection function $\nbar(\rhat)$ is shown in
\fig{aitoffFig}.  For the baseline sample, it covers a sky area of
2499 square degrees.  The function $\nbar(\rhat)$ is defined to be the
completeness, \ie, the probability that a galaxy satisfying the sample
cuts actually gets assigned a redshift (including the 6\% of the total
which are determined based on the nearest neighbor redshift as
described in Appendix A).  
Therefore the completeness is a dimensionless number between zero and
one.  The effective area is $\int\nbar(\rhat)d\Omega\approx 2417$ square
degrees, corresponding to an average completeness of $96.7\%$.
As detailed in Appendix~\ref{MaskSec}, we model $\nbar(\rhat)$ as a piecewise constant function. 
We specify this function by giving its value in each of a large number of disjoint spherical polygons, 
within each of which it takes a constant value.  
There are 2914 such polygons for the baseline sample, encoding the geometric boundaries 
of spectroscopic tiles, holes and other relevant entities. 
\Fig{aitoffFig} shows that the sky coverage naturally separates into three fairly compact 
regions of comparable size: north of the Galactic plane (in the center of the figure), 
there is one region on the celestial equator and another at high declination; south of the Galactic plane, 
there is a set of three stripes near the equator. For the purpose of testing for systematic errors, 
we define angular subsamples A1, A3 and A4 corresponding to these regions (see Table 1), which have effective
areas of 809, 1007 and 600 square degrees, respectively.

\subsection{Radial selection function}

\begin{figure} 
\centerline{\epsfxsize=\figsize\epsffile{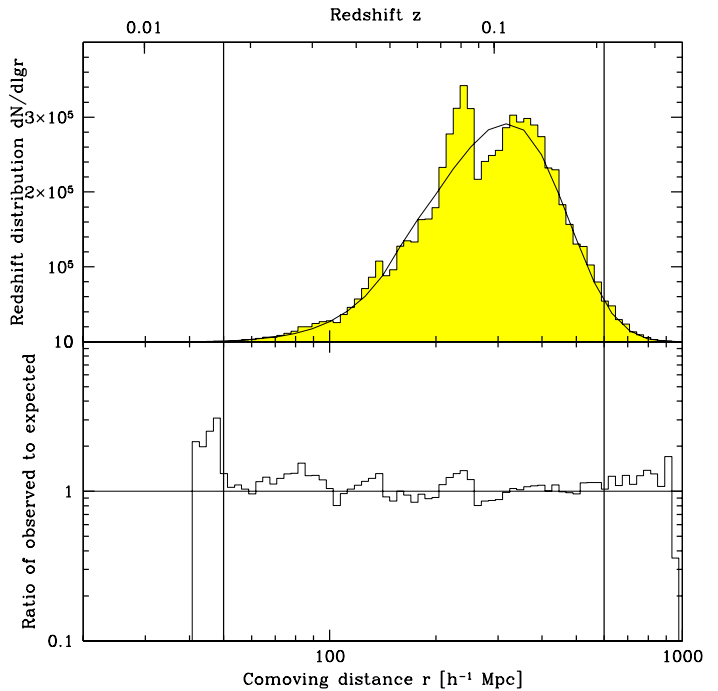}}
\caption[1]{\label{zhistFig}\footnotesize%
The redshift distribution of the galaxies
in sample {\tt safe13} is shown as a histogram and compared with 
the expected distribution in the absence of clustering, 
$\ln 10\nbar(r)r^3 d\Omega$ (solid curve)
in comoving coordinates assuming a flat $\Omega_\Lambda=0.7$ cosmology.
The bottom panel shows the ratio of observed and expected distributions.
The vertical lines indicate the redshift limits
($50\hMpc<r<600\hMpc$) employed in the baseline
analysis. This near cut removes only 22 galaxies, the far cut 3295.
}
\end{figure}

Our estimate of the radial selection function for the baseline sample is shown in 
\fig{zhistFig}, together with a histogram of the galaxy distances.  
The full details of the derivation of the radial
selection function can be found in Appendix~\ref{selfuncSec}, 
including both evolution and K-corrections. Our basic sample has magnitude limits
$r\ge 14.5$ at the bright end (since the survey becomes incomplete for bright galaxies with large angular size)
and $r\le 17.5$ at the faint end (Appendix~\ref{selfuncSec}).

\begin{figure} 
\centerline{\epsfxsize=\figsize\epsffile{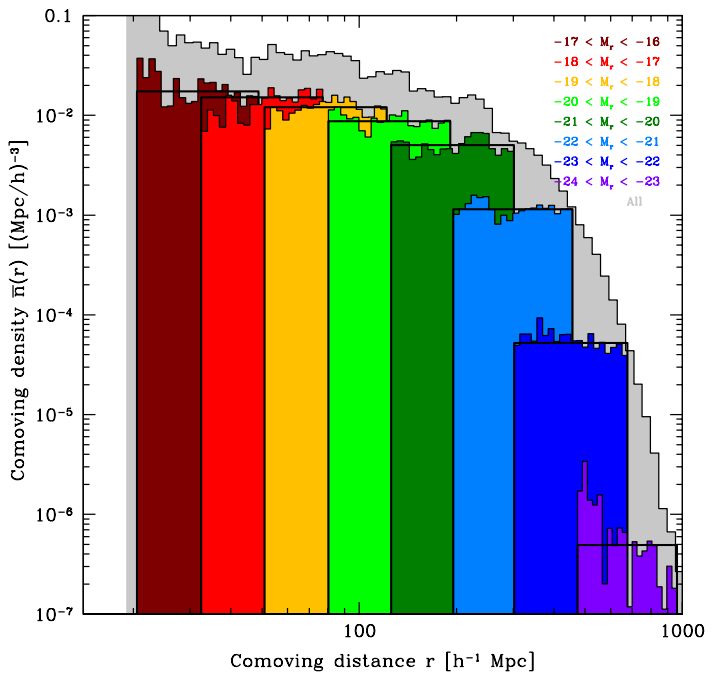}}
\caption[1]{
\label{vlim_nbarFig}\footnotesize%
Same as \fig{zhistFig} but plotted as comoving number density.
The grey (background) histogram shows the full flux-limited sample
and the others show the volume-limited subsamples, with
lines indicating their predicted constant selection functions.
}
\end{figure}

\begin{figure} 
\centerline{\epsfxsize=\figsize\epsffile{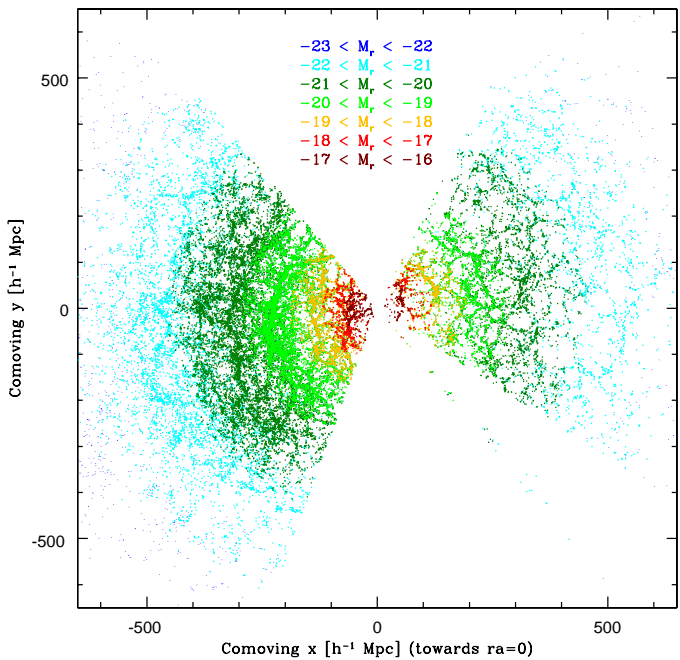}}
\caption[1]{\label{vlim_slice3Fig}\footnotesize%
The distribution of 67{,}676  
galaxies within $5^\circ$ of the Equatorial plane,
color coded by their absolute magnitudes. $M_r$ in the figure refers
to the absolute $r$-magnitude $K$-corrected to $z=0.1$. 
}
\end{figure}

\Fig{vlim_slice3Fig} shows all galaxies within $5^\circ$ of the equator in a pie diagram, 
color coded by their absolute magnitude, and illustrates 
one of the fundamental challenges for our project (and indeed for the analysis of 
any flux-limited sample): luminous galaxies dominate the sample
at large distances and dim ones dominate nearby.
A measurement of the power spectrum on very large scales is therefore statistically dominated by
luminous galaxies whereas a measurement on small scales is dominated by dim ones (since they
have much higher number density). Yet it is well-known that luminous galaxies cluster more
than dim ones (\eg, Davis {\etal} 1988; Hamilton 1988; Norberg {\etal} 2001;
Zehavi {\etal} 2002; Verde {\etal} 2003), so when comparing $P(k)$ on large and small 
scales we are in effect comparing apples with
oranges, and may mistakenly conclude that the power spectrum is red-tilted (with a spectral 
index $n=0.94$, say) even if the truth is $n=1$.
So far, no magnitude-limited galaxy power spectrum analysis has been
corrected for this effect. We do so in \sec{BiasSec}.
Although this effect is is not large in an absolute sense, 
we find that it must nonetheless be included for precision cosmology applications.

\begin{figure} 
\centerline{\epsfxsize=\figsize\epsffile{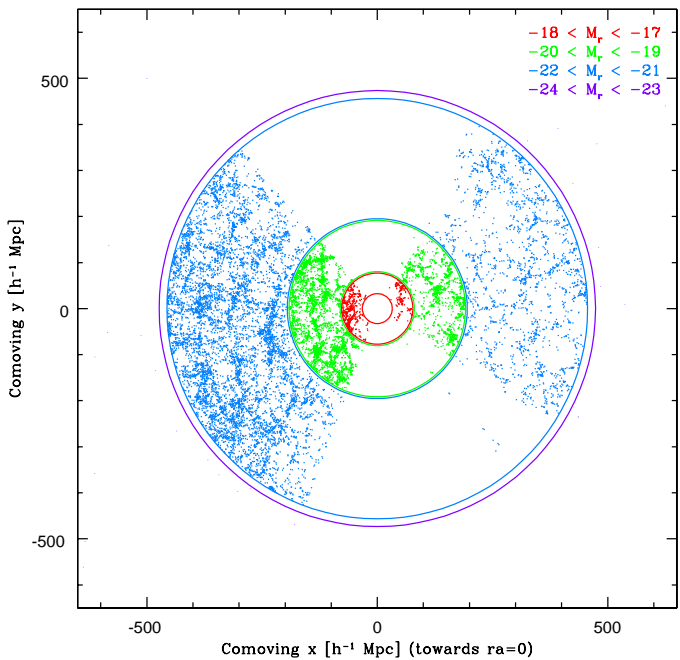}}
\caption[1]{\label{vlim_slice1Fig}\footnotesize%
The distribution of galaxies within $5^\circ$ of the equatorial plane
is shown for  the volume-limited subsamples L1, L3, L5 and L7 from
Table 1.}
\end{figure}

\begin{figure} 
\centerline{\epsfxsize=\figsize\epsffile{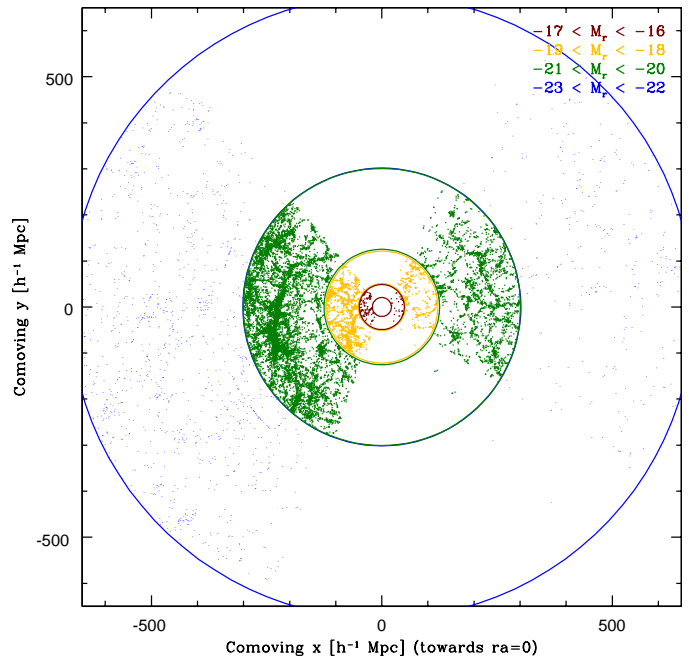}}
\caption[1]{\label{vlim_slice2Fig}\footnotesize%
The distribution of galaxies within $5^\circ$ of the equatorial plane
is shown for the remaining four volume-limited subsamples from Table 1, \ie,
L2, L4, L6 and L8.}
\end{figure}

The first step is to quantify the luminosity-dependence of bias. For this purpose, we define a
series of volume-limited samples as specified in Table 1, constructed
by discarding all galaxies that are too faint to be included at the
far limit or too bright to be included at the near limit.  This gives
a radial selection function (\fig{vlim_nbarFig}) that is constant
within the radial limits and zero elsewhere.  The radial limits are
set so that galaxies at the far (near) radial limit and the dim (luminous)
end of the absolute magnitude range in question 
have fluxes at the faint (bright) flux limits,
respectively.  Because the flux range $14.5<r<17.5$ spans exactly
three magnitudes, these subsamples overlap spatially only with
their nearest neighbor samples, and have a near limit that would be equal to the far
limit of the sample that is two notches more luminous if it were not for evolution and 
K-corrections. This is clear in
Figures~\ref{vlim_slice1Fig} and \ref{vlim_slice2Fig}.  These samples
have the advantage that the measured clustering is that of a
well-defined set of objects whose selection is redshift-independent.
Although we have not accounted for our surface
brightness limits in defining these samples, very few of even the
lowest luminosity galaxies in our sample are affected by the
surface-brightness limits of the survey (Blanton {\etal} 2002c; Strauss {\etal} 2002).

\bigskip\bigskip\bigskip 
\section{Method and basic analysis}
\label{AnalysisSec}

We now turn to our basic goal: accurately measuring the shape of the matter power spectrum $P(k)$
on large scales using the data described above, \ie, measuring a curve that equals the
large-scale matter power spectrum up to an unknown overall multiplicative bias factor that is
independent of scale.
This involves four basic challenges:
\begin{enumerate}
\item Accounting for the complicated survey geometry
\item Correcting for the effect of redshift-space distortions
\item Correcting for bias effects, which as described in \sec{BiasSec} cause
an artificial red-tilt in the power spectrum
\item Checking for potential systematic errors
\end{enumerate}
Before delving into detail, let us summarize each of these challenges and 
how we will tackle them. 

\subsection{Battle plan}

\subsubsection{Survey geometry and method of estimating power} 

It is well-known that since galaxies in a redshift survey 
probe the underlying density field only in a finite volume,
the power spectrum estimated with traditional Fourier methods
(\eg, Percival {\etal} 2001; Feldman, Kaiser \& Peacock 1994), 
is complicated to interpret: 
it corresponds to a smeared-out version of the true power spectrum,
can underestimate power on the largest scales due to the so-called
integral constraint (Peacock \& Nicholson 1991) and has correlated errors.
We therefore measure power spectra with an alternative, matrix-based 
approach which, although more numerically demanding, has several advantages on the
large scales that are the focus of this paper.
It facilitates tests for radial and angular systematic errors.
If galaxies were faithful tracers of mass, then it would
produce unbiased minimum-variance power spectrum measurements
with uncorrelated error bars that are smaller than those from traditional Fourier methods.
The power smearing is quantified by window functions that are both narrower than with 
traditional Fourier methods and 
can be computed without need for approximations or Monte-Carlo simulations.
(We do, however, use Monte Carlo simulations in \sec{zspaceSec} to verify that the method and software 
work as advertised.)

\subsubsection{Redshift space distortions}
 
Our basic input data consist of galaxy positions $\r$ in three-dimensional 
``redshift space'', where the comoving distance $r=|\r|$ is that which would explain the
observed redshift if the galaxy were merely following the Hubble flow of the expanding Universe.
The same gravitational forces that cause galaxies to cluster also cause them
to move relative to the Hubble flow,
and these so-called peculiar velocities make the clustering in redshift space
anisotropic (Kaiser 1987; Hamilton 1998).
Although this effect can be modeled and accounted for exactly on very large scales
on which the clustering is linear (Kaiser 1987), nonlinear corrections cannot be neglected 
on some of the scales of interest to us (Scoccimarro {\etal} 2001; Seljak 2001;
Cole {\etal} 1994; Hatton \& Cole 1998).
\Sec{zspaceSec} is devoted to dealing with this complication, going beyond
the Kaiser approximation with 
a three-pronged approach:
\begin{enumerate}
\item We precede our power spectrum analysis by a nonlinear  
``finger-of-God'' compression step with a tunable threshold, to quantify 
the sensitivity of our results to nonlinear galaxy groups and clusters.
\item We measure three power spectra (galaxy-galaxy, galaxy-velocity and velocity-velocity spectra)
rather than one, quantifying the clustering anisotropy
and allowing the real-space power to be reconstructed beyond linear order.
\item We analyze an extensive set of mock galaxy catalogs to quantify 
the accuracy of our results and measure how the 
non-linear correction factor grows toward smaller scales.
\end{enumerate}
Our mock analysis will also allow us to quantify the effects of
nonlinear clustering on the
error bars and band-power correlations.
Step 1 is optional, and we present results both with and without it.

\subsubsection{Bias}
\label{BiasPlanSec}

All we can ever measure with galaxy redshift surveys is galaxy clustering, whereas what we care about for constraining
cosmological models is the clustering of the underlying matter distribution. 
Our ability to do cosmology with the real-space galaxy power spectrum $\Pgg(k)$ 
is therefore only as good as our understanding of bias,
\ie, the relation of $\Pgg(k)$ to the matter power spectrum $P(k)$.
Pessimists have often argued that since we do not understand galaxy formation at high precision, 
we cannot understand bias accurately either, and so galaxy surveys will be relegated to a historical 
footnote, having no role to play in the precision cosmology era.
Optimists retort that no matter how complicated 
the gas-dynamical and radiative processes involved in galaxy formation may be, they 
have only a finite spatial range (a few $h^{-1}$Mpc, say), leading to a generic prediction that 
bias on much larger scales ($>20h^{-1}$Mpc, say) should be scale-independent for any particular type of galaxy
(Coles 1993;
Fry and Gazta\~naga 1993;
Scherrer \& Weinberg 1998;
Mann, Peacock \& Heavens 1998;
Coles, Melott \& Munshi 1999;
Heavens, Matarrese \& Verde 1999;
Blanton {\etal} 2000;
Narayanan {\etal} 2000).
This theoretical expectation is supported by visual inspection of the galaxy distribution.
Comparing early and late type galaxies in the 2dF galaxy redshift survey (Peacock 2003; Madgewick {\etal} 2003) 
shows that whereas the small-scale distribution differs (ellipticals display a more
``skeletal'' distribution than do cluster-shunning
spirals), their large-scale clustering patterns are indistinguishable. 

We will devote \sec{BiasSec} to the bias issue, 
arguing that both the pessimists and the optimists have turned out to be right:
yes, biasing is indeed 
complicated
on small scales (where the galaxy power spectrum will therefore teach us more
about galaxy formation than about cosmology) but no, this in no way prevents us from doing precision cosmology 
with the galaxy power spectrum on very large scales.
Our main tool will be analyzing our volume-limited magnitude subsamples, showing that their large-scale power
spectra are consistent with all having the same shape and differing merely in amplitude.

Although the argument above for scale-independent bias holds only for a volume-limited 
subsample, we wish to use our full galaxy sample over a broad range of redshifts, both to expand the range of $k$-scales
probed and to reduce shot noise.
We will therefore use our measured luminosity-dependence of bias to compute and 
remove the artificial red tilt in our full magnitude-limited baseline sample.

In future papers, we will constrain galaxy bias empirically using
clustering measurements on smaller scales (\eg, Zehavi {\etal} 2003),
which will allow us to calculate the effects of {\it scale}-dependent bias
on the power spectrum in the non-linear regime, and thus to extend
the measurement of the matter power spectrum shape to smaller scales.

\subsubsection{Systematic errors}

As the old saying goes, the devil you know poses a lesser threat than the devil you don't.
We will therefore devote \sec{SystematicsSec} to  testing for the sort of effects that are not 
included in our Monte Carlo simulations. This includes both radial modulations 
(due to mis-estimates of evolution or $K$-corrections)
and angular modulations (due to effects such as  
uncorrected dust extinction, variable observing conditions, photometric calibration errors and fiber collisions).
Our tests use two basic approaches:
\begin{enumerate}
\item Analyzing subsets of galaxies: we compare the power spectra from different parts of the
sky (subsamples A1-A4 from Table 1) and different distance ranges (subsamples R1-R3) looking
for inconsistencies.
\item Analyzing subsets of modes: we look for excess power in purely angular and purely radial 
modes of the density field, which act like lightning rods for angular and radial modulations 
such as those mentioned above.
\end{enumerate}

\subsection{Three Step Power Spectrum Estimation}

Our matrix-based power spectrum estimation approach 
is described in Tegmark {\etal} (1998).
It starts by expanding the galaxy density field in a set of
functions known as Pseudo-Karhunen-Lo\`eve eigenmodes.
This step compresses the data set into a much smaller size
(from hundreds of thousands of galaxy coordinates to a few thousand expansion
coefficients) while retaining the large-scale cosmological information
in which we are interested. It also reduces the power spectrum estimation 
problem to a mathematical form equivalent to that encountered in CMB
analysis, enabling us to take advantage of a powerful set of matrix-based tools
that have been fruitfully employed in the CMB field.
Our basic analysis in the remainder of this section therefore 
consists of the following three steps:
\begin{enumerate}
\item Finger-of-god compression to remove redshift-space distortions
due to virialized structures; \sec{fogSec}   
\item Pseudo-Karhunen-Lo\`eve eigenmode expansion; \S~\ref{PKLsec} 
\item Power spectrum estimation using quadratic estimators;
\S~\ref{ResultsSec}. 
\end{enumerate}
As mentioned, the third step measures not one but three power spectra, encoding
clustering anisotropy that contains information about redshift space distortions.
In this section, we merely present the basic measurement of these three curves, 
which involves no assumptions about linearity, the nature of biasing or anything else.
We then return to modeling and interpreting these curves in terms of real-space power in \sec{zspaceSec}
and to bias modeling in \sec{BiasSec}.

\begin{figure*}[t] 
\centerline{\hglue0.5cm\epsfxsize=19.5cm\epsffile{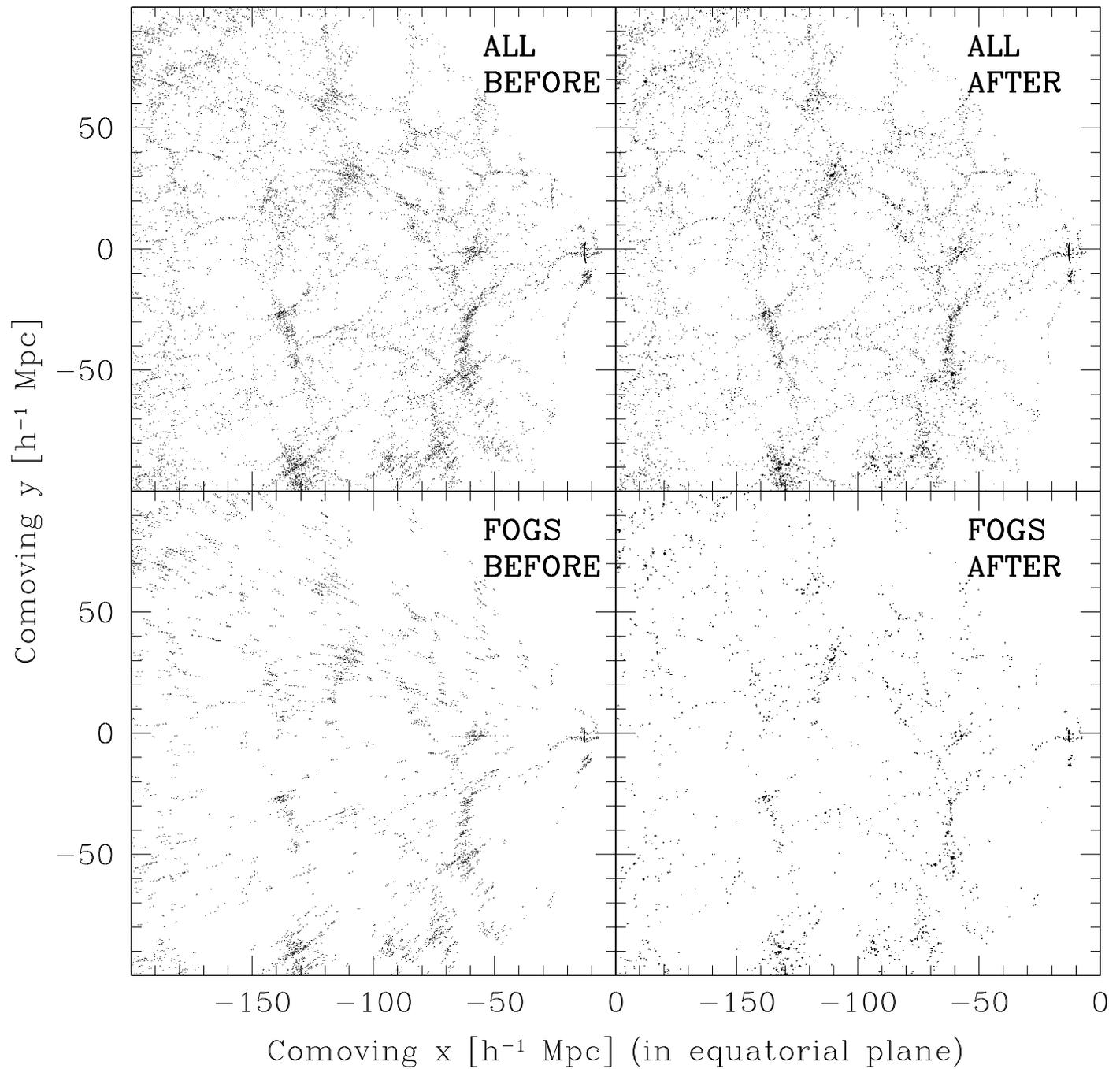}}
\caption[1]{\label{fogFig}\footnotesize%
The effect of our Finger-of-god (FOG) removal is shown in part of 
the equatorial slice $|\delta| < 5 ^\circ$ before (left) and after (right)
FOG compression. 
Top panels show all 67{,}626 galaxies in the slice, bottom panels show
the 28{,}255 that are identified as belonging to 
FOGs (with density threshold 200).
}
\end{figure*}

\subsection{Step 1: Finger-of-god compression}
\label{fogSec}

Since our analysis method is motivated by (although not limited to) 
the linear Kaiser approximation 
for redshift space distortions, it is crucial that we are able to empirically
quantify our sensitivity to the so-called finger-of-god (FOG) effect
whereby radial velocities in virialized clusters make them appear 
elongated along the line of sight. 
We therefore start our analysis by compressing (isotropizing) 
FOGs, as illustrated in \fig{fogFig}.
The FOG compression involves a tunable threshold density,
and in \sec{zspaceSec} we will study how the final 
results change as we gradually change this threshold to include 
lesser or greater numbers of FOGs.

We use a standard friends-of-friends algorithm,
in which two galaxies are considered friends, therefore
in the same cluster,
if the density windowed through an ellipse 10 times longer
in the radial than transverse directions,
centered on the pair,
exceeds a certain overdensity threshold.
To avoid linking well-separated galaxies in deep,
sparsely sampled parts of the survey,
we impose the additional constraint that friends
should be closer than
$r_{\perp\max} = 5 \, h^{-1} \Mpc$ in the transverse direction.
The two conditions are combined into the following
single criterion:
two galaxies separated by $r_\parallel$ in the radial
direction and by $r_\perp$ in the transverse direction
are considered friends if
\begin{equation}
\left[(r_\parallel / 10)^2 + r_\perp^2\right]^{1/2}
\le
\left[{4\over 3}\pi \nbar ( 1 + \delta_c ) + r_{\perp\max}^{-3} \right]^{-1/3}
\end{equation}
where $\nbar$ is the 3D selection function 
at the position of the pair,
and $\delta_c$ is an overdensity threshold.
Note that $\delta_c$ represents not the overdensity of the pair
as seen in redshift space, but rather the overdensity of the pair
after their radial separation has been reduced by a factor of 10.
Thus $\delta_c$ is intended to approximate the
threshold overdensity of a cluster in real space.
We have chosen $r_{\perp\max}$ somewhat larger than
the virial diameter of typical clusters to be conservative, 
minimizing the risk of missing FOGs --- for our baseline 
threshold $\delta_c=200$, our results are essentially unaffected 
by this choice of $r_{\perp\max}$.
Having identified a cluster by friends-of-friends in this fashion,
we measure the dispersion of galaxy positions about the center
of the cluster in both radial and transverse directions.
If the one-dimensional radial dispersion exceeds the transverse
dispersion, then the cluster is deemed a FOG,
and the FOG is then compressed radially so that the cluster becomes round,
that is, the transverse dispersion equals the radial dispersion.
We perform the entire analysis five times, using 
density cutoffs
$1{+}\delta_c=\infty$, 200, 100, 50 and 25, respectively; in our
analyses below, we will explore the sensitivity of our results to this
cutoff. 
The infinite threshold
$1{+}\delta_c=\infty$ corresponds to no compression at all.

\Fig{fogFig} illustrates FOG compression with threshold density
$1{+}\delta_c=200$, and 
unless explicitly stated otherwise, all results presented in this paper
are for this threshold density.
We make this choice to be on the safe side: Bryan \& Norman (1998) estimate that the
overdensity of a cluster at virialization is about 337 in a $\Lambda$CDM model, 
rising further as the Universe expands and the background density continues to drop.

\begin{figure} 
\vskip\smtopskip
\centerline{\epsfxsize=\figsiz\epsffile{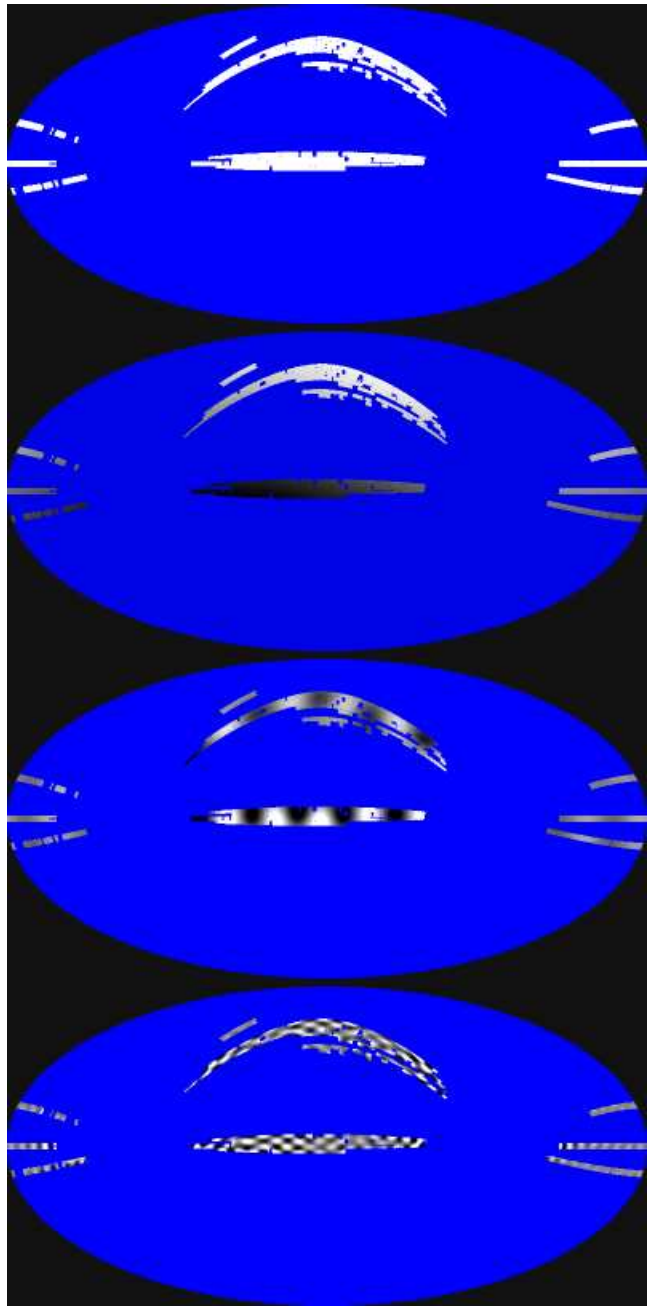}}
\caption[1]{\label{angularmodesFig}\footnotesize%
A sample of four angular pseudo-KL (PKL) modes are shown
in Hammer-Aitoff projection in equatorial coordinates, 
with grey representing zero weight,
and lighter/darker shades indicating positive/negative weight, respectively.
Uniform blue/grey areas are outside the observed region.
From top to bottom, they are
angular modes 1 (the mean mode), 3, 35 and 286, and
are seen to probe progressively smaller angular scales.
}
\end{figure}

\clearpage 
\subsection{Step 2: Pseudo-KL pixelization}

\label{PKLsec}

\begin{figure} 
\centerline{\epsfxsize=\figsiz\epsffile{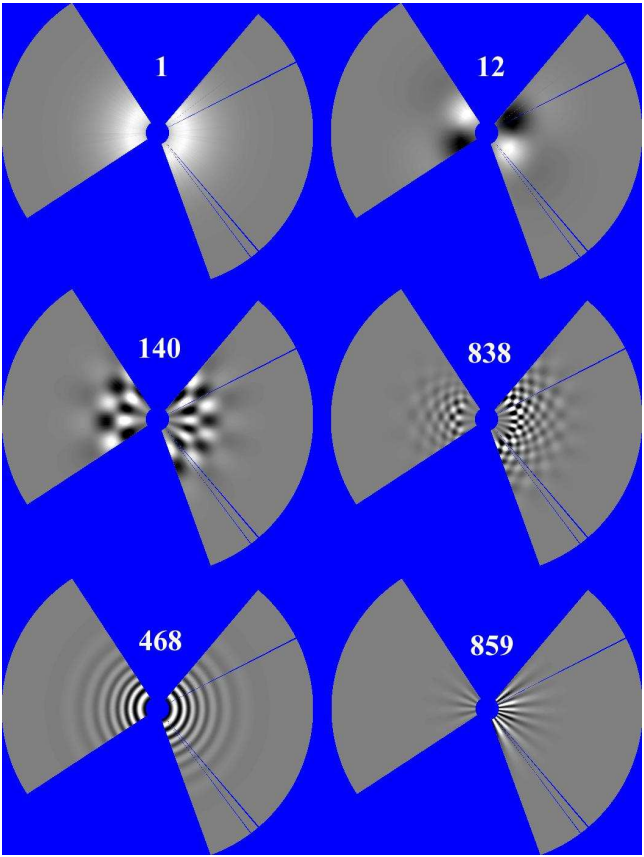}}
\caption[1]{\label{slicemodesFig}\footnotesize%
A sample of six pseudo-KL modes are shown
in the equatorial plane.
Grey represents zero weight,
and lighter/darker shades indicate positive/negative weight, respectively.
Uniform blue/grey areas are outside the volume used.
The bottom row gives examples of special modes, showing a 
purely radial mode (left) and a purely angular mode (right).
}
\end{figure}

\begin{figure} 
\centerline{\epsfxsize=\figsize\epsffile{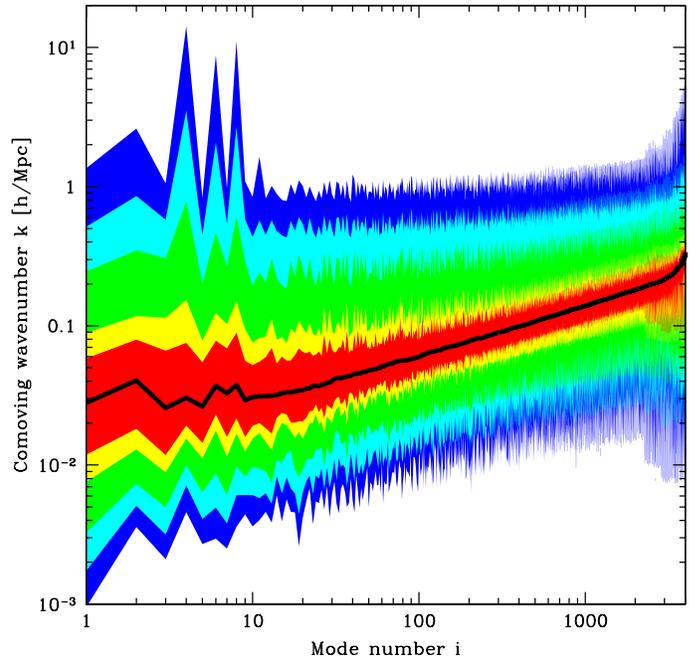}}
\caption[1]{\label{mode_keffFig}\footnotesize%
Relation between mode number $i$ and physical scale $k$.
The variance $\expec{x_i^2}$ gets non-negative contributions from
all $k$ as per \eq{gg_gv_vv_defEq}. 
$50\%$ of the contribution comes from below the solid black curve,
which we can interpret as the median $k$-value probed by the $\ith$ mode.
The shaded regions show percentiles of the contribution:
from outside in, they show 
the k-ranges giving $99.98\%$, $99.8\%$, $98\%$, $80\%$ and $60\%$
of the contribution, respectively.
Apart from the first 8 special modes, the modes are ordered by increasing 
median $k$-value.
}
\end{figure}

\begin{figure} 
\vskip\smtopskip
\centerline{\epsfxsize=\figsize\epsffile{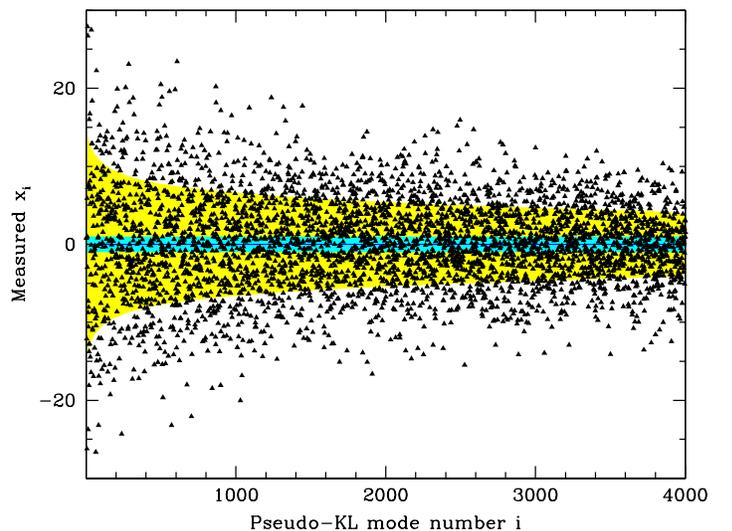}}
\vskip\smbotskip
\caption[1]{\label{xFig}\footnotesize%
The triangles show the 4000 elements $x_i$ of the data vector $\x$
(the pseudo-KL expansion coefficients) 
for the baseline galaxy sample.
If there were no clustering in the survey, merely shot noise, 
they would have unit variance, and about $68\%$ of them would 
lie within the blue/dark grey band.
If our prior power spectrum were correct,
then the standard deviation would be larger, as indicated by the 
shaded yellow/light grey band. 
To reduce clutter, the modes are (apart from the first 8 special modes), ordered by decreasing  
signal-to-noise ratio, which corresponds approximately to the ordering by scale in
Figure~\ref{mode_keffFig}.
}
\end{figure}

\begin{figure} 
\centerline{\epsfxsize=\figsize\epsffile{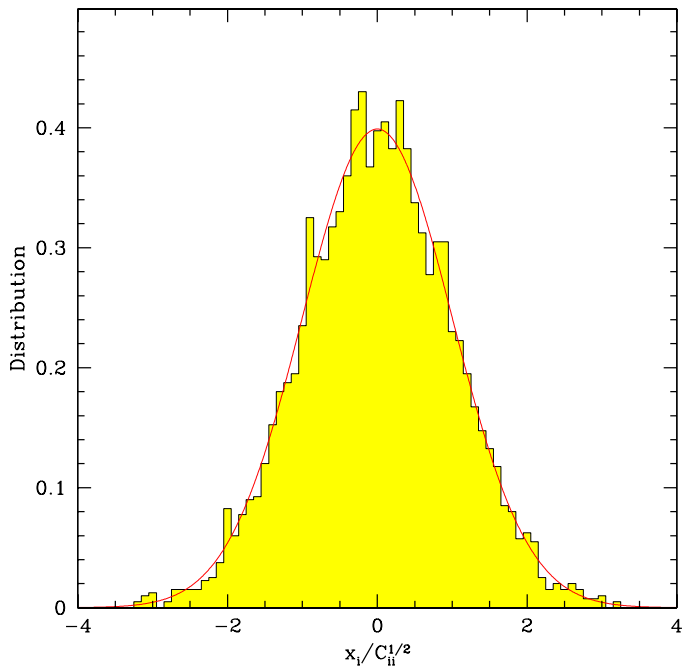}}
\caption[1]{\label{xhistFig}\footnotesize%
Gaussianity of fluctuations.
The histogram shows the distribution of the 4000
PKL coefficients $x_i$ from the previous figure after dividing by
their predicted standard deviation $\C_{ii}^{1/2}$, assuming our 
prior power spectrum.
The histogram has been normalized to have unit area.
The solid curve shows a Gaussian of unit variance, zero mean,
and unit area.
}
\end{figure}

The raw data consist of three-dimensional
vectors $\r_\alpha$, $\alpha=1,...,\ngal$, giving the
measured positions of each galaxy in redshift space,
with the number of galaxies $\ngal$ given in Table 1 for each sample.
As in Tegmark {\etal} (1998), we expand the observed three-dimensional density field 
in a basis of $\nx$ noise-orthonormal functions $\psi_i$, $i=1,...,\nx$,
\beq{xDefEq}
x_i \equiv\int_V {n(\r)\over\nbar(\r)}\psi_i(\r) d^3r
\eeq
and work with the $\nx$-dimensional data vector $\x$ of expansion coefficients instead of 
the $3\times\ngal$ numbers $\r_\alpha$.  Here, $\nbar$ is the three-dimensional selection function described in 
\sec{DataSec}, {\ie}, $\nbar(\r)dV$
is the expected (not the observed) number of galaxies in 
a volume $dV$ about $\r$ in the absence of clustering,
and the integration is carried out over the volume $V$ of the sample
where the selection function $\nbar(\r)$ is nonzero.
We will frequently refer to the functions $\psi_i$ as ``modes''.
As we will see below, these modes play a role quite analogous to
pixels in CMB maps, with the variance of $x_i$ depending linearly on the power
spectrum that we wish to measure.

Galaxies are
(from a cosmological perspective)
delta-functions in space,
so the integral in \eq{xDefEq} reduces to a discrete
sum over galaxies.
We do not rebin the galaxies spatially,
which would artificially degrade the resolution.
It is convenient to isolate the mean density into a single mode
$\psi_1(\r) = \nbar(\r)$,
with amplitude
\beq{meanDefEq1}
x_1=\int n(\r) d^3 r=\ngal,
\eeq
and to arrange for all other modes to have zero mean
\beq{meanDefEq2}
\expec{x_i} = \int\psi_i(\r) d^3r = 0
\quad(i \neq 1).
\eeq
The covariance matrix of the vector $\x$ of amplitudes is
a sum of noise and signal terms
\beq{xCovEq}
\expec{\Delta\x\Delta\x^t}=\C\equiv\N+\Sb.
\eeq
where
the shot noise covariance matrix is given by
\beq{NdefEq}
\N_{ij} = \int_V {\psi_i(\r)\psi_j(\r)\over\nbar(\r)}d^3 r
\eeq
and
the signal covariance matrix is 
\beq{SdefEq}
\Sb_{ij} = \int\psih_i(\k)\psih_j(\k)^*
\Pgg(k)\dV
\eeq
in the absence of redshift-space distortions (which will be included in \sec{3specSec}).
Here hats denote Fourier transforms and the star denotes complex
conjugation. 
$\Pgg(k)$ is the (real-space) galaxy power spectrum, which for a random field of density fluctuations
$\delta(\r)$ is defined by
$\expec{\deltah(\k)\deltah(\k')^*}=(2\pi)^3\delta_{\rm Dirac}(\k-\k')\discretionary{}{}{}\Pgg(k)$.
We take the functions $\psi_i$ to have units of inverse volume, so 
$\x$, $\N$, $\Sb$ and $\psih_i$ are all dimensionless.

Our method requires computing the signal covariance matrix $\Sb$, 
both to calculate power spectrum error bars
and to find the power spectrum estimator that minimizes them.
\Eq{SdefEq} shows that this requires assuming a power 
spectrum $\Pgg(k)$. For this spectrum, which we refer to as our {\it prior}, 
we use a simple two-parameter fit as described in \sec{PriorSec},
whose parameters are determined from our measurements by iterating the entire analysis.

As our functions $\psi_i(\r)$, we use the 
{\it pseudo-Karhunen-Lo\`eve (PKL) eigenmodes} defined in
Hamilton, Tegmark \& Padmanabhan (2000; hereafter ``HTP00'') and 
Tegmark, Hamilton \& Xu (2002; hereafter ``THX02'').
The construction of these PKL modes $\psi_i(\r)$
explicitly uses the three-dimensional selection function $\nbar(\r)$, but is 
model-independent since it does not 
depend on the power spectrum.

To provide an intuitive feel for the nature of these modes, 
a sample is plotted in \fig{angularmodesFig}
and \fig{slicemodesFig}. We use these modes because they have the
following desirable properties:
\begin{enumerate}

\item They form a complete set of basis functions probing 
successively smaller scales, so that 
a finite number of them (we use the first 4000, for the reasons given in
\sec{npixSec}) allow 
essentially all information about the density field on large scales 
to be distilled into the vector $\x$.

\item They are orthonormal with respect to the shot noise, 
\ie, such that \eq{NdefEq} gives $\N=\I$, the identity matrix.
The construction of the modes thus depends explicitly on the survey geometry 
as specified by $\nbar(\r)$, and $\psi_i(\r)=0$ in regions of 
space where $\nbar(\r)=0$.

\item They allow
the covariance matrix $\Sb$ to be  fairly rapidly computed.

\item They are the product of an angular and a radial part, \ie,
take the separable form $\psi_i(\r) = \psi_i(\rh)\psi_i(r)$,
which accelerates numerical computations and helps isolate radial and angular 
systematic problems.

\item A range of potential sources of systematic problems are isolated 
into special modes that are orthogonal to all other modes.
This means that we can test for the presence of such problems by looking 
for excess power in these modes, and immunize against their effects
by discarding these modes.

\end{enumerate}

  We have four types of such special modes:
  \begin{enumerate}
  \item 
   The very first mode is the mean density, 
   $\psi_1(\r)=\nbar(\r)$.
   The mean mode is used in determining the maximum likelihood
   normalization of the selection function,
   but is then discarded from the analysis,
   since it is impossible to measure the fluctuation of the mean mode.
   The idea of solving the so-called integral constraint problem 
   by making all modes orthogonal to the mean (Eq.~\ref{meanDefEq2})
   goes back to Fisher {\etal} (1993). 
   \item 
   Modes 2-8 are associated with the motion of the Local Group
   through the Cosmic Microwave Background
   at 622~km/s towards (B1950 FK4) RA = $162^\circ$, Dec = $-27^\circ$
   (Lineweaver {\etal} 1996;
   Courteau \& van den Bergh 1999).
   To first order, these modes are the only modes affected
   by mis-estimates of the motion of the Local Group.
   \item 
   Purely radial modes (for example mode 468 in \fig{slicemodesFig})
   are to first order the only ones affected by mis-estimates of
   the radial selection function $\nbar(r)$.
   \item 
   Purely angular modes (for example mode 859 in \fig{slicemodesFig})
   are to first order the only ones affected by mis-estimates of
   the angular selection function $\nbar(\rh)$, as may result from
   inadequate corrections for, \eg, extinction, the variable magnitude limit,
   the variable magnitude completeness or photometric zero-point offsets.
   \end{enumerate}
The computation of the modes in practice is described in detail in
THX02 and 
in even more detail in 
Hamilton \& Tegmark (2003).
  
The pixelized data vector $\x$ is shown in 
\fig{xFig}. This data compression step has thus distilled 
the large-scale information about the galaxy density field
from $3\times\ngal=429{,}942$ galaxy coordinates into 
4000 PKL-coefficients.  The order of these coefficients is one of
decreasing scale (increasing $k$) as is shown in Figure~\ref{mode_keffFig}.
If there were no cosmological density fluctuations
in the survey, merely Poisson fluctuations, 
the PKL-coefficients $x_i$ would be uncorrelated with unit variance (since $\N=\I$),
so about $68\%$ of them would be
expected to lie within the blue/dark grey band.
\Fig{xFig} shows that the fluctuations are considerably larger
than Poisson, especially for the largest-scale modes (to the left),
demonstrating the obvious fact that cosmological density fluctuations are
present, as expected.

\subsection{What we wish to measure: three power spectra, not one}
\label{3specSec}

Following HTP00 and THX02,
we will measure three separate power spectra,
whose ratios encode information about 
clustering anisotropy due to redshift space distortions.
Let us now give their definition and some intuition 
for how to interpret them.

On large scales where redshift space distortions can be treated 
in the linear approximation
(Kaiser 1987), the signal covariance matrix $\Sb$
in \eq{xCovEq} can be generalized from \eq{SdefEq} and written in the form
\beq{gg_gv_vv_defEq}
\Sb = \int_0^\infty\left[\Sgg(k)\Pgg(k) + \Sgv(k)\Pgv(k) + \Svv(k)\Pvv(k)\right]{k^2 dk\over (2\pi)^3},
\eeq
where $\Pgg(k)$, $\Pgv(k)$ and $\Pvv(k)$ are three power spectra 
defined in real space (as opposed to redshift space) and
$\Sgg(k)$, $\Sgv(k)$ and $\Svv(k)$ are known dimensionless matrix-valued functions.
We will refer to these three power spectra as the
galaxy-galaxy power, the galaxy-velocity power and the velocity-velocity power,
respectively, or $gg$, $gv$ and $vv$ for short.
Specifically, 
$\Pgg(k)$ is the real-space galaxy power spectrum,
$\Pvv(k)$ is the velocity power spectrum and 
$\Pgv(k)$ is the cross-power between galaxies and velocity.
More rigorously, `velocity' here refers to minus the velocity divergence $\nabla\cdot\v$,
which in linear theory is related to the mass (not galaxy)
overdensity $\delta$ by
$f\delta + \nabla\cdot\v = 0$,
where $\nabla$ denotes the comoving gradient in velocity units.
Here $f\approx\Omega_{\rm m}^{0.6}$ is the dimensionless growth rate
for linear density perturbations (see Hamilton 2001). 
The three matrix-valued functions are determined 
directly from the modes $\psi_i$, \ie, by geometry alone:
\beqa{SEq}
{(\Sgg)}_{ij} &=& \int\psih_i(\k)\psih_j(\k)^*d\Omega_k
,
\\
{(\Sgv)}_{ij} &=& \int[\psih_i(\k)\zetah_j(\k)^* + \zetah_i(\k)\psih_j(\k)^*] d\Omega_k
,
\\
{(\Svv)}_{ij} &=& \int\zetah_i(\k)\zetah_j(\k)^*d\Omega_k
,
\eeqa
in which the velocity mode $\zetah_i(\k)$
is related to the position mode $\psih_i(\k)$ by
(Fisher, Scharf \& Lahav 1994; Heavens \& Taylor 1995; Hamilton 1998 eq.~8.13)
\beqa{zetaEq}
\zeta_i = \M^\dagger \psi_i
\eeqa
where $\M^\dagger$ is the Hermitian conjugate of
the velocity part of the linear redshift distortion operator.
In the small-angle, or distant observer, approximation,
the operator $\M$ takes the familiar Kaiser (1987) form,
a diagonal operator in Fourier space
\beq{MplaneEq}
\M \approx \mu_\k^2
\eeq
where $\mu_\k \equiv \khat\cdot\z$ is the cosine of the angle between the
wavevector $\k$ and the line of sight $\z$.
Here however we do not assume the small-angle approximation,
but rather take into account the full radial nature of redshift distortions.
Radial redshift distortions destroy translation invariance,
so that the radial redshift distortion operator is no longer diagonal
in Fourier space,
as it is in the small-angle approximation;
indeed, the radial redshift distortion operator takes
a rather complicated form in Fourier space
(Hamilton 1998, eq.~4.37).
The radial redshift distortion operator takes a simpler form in real space,
where $\M$, expressed in the frame of the Local Group,
can be written as the integro-differential operator
(Hamilton 1998 eq.~4.46)
\beq{MEq}
  \M = \left[
    {\partial^2 \over \partial r^2} + {\alpha(\r) \rhat \over r} \, .
    \left( {\partial \over \partial\r}
    - \left. {\partial \over \partial\r} \right|_{\r = 0} \right)
    \right] \nabla^{-2}
\eeq
with $\alpha(\r)$ the logarithmic derivative of $r^2$ times the
selection function $\nbar(\r)$,
\beq{alphaEq}
\alpha(\r) \equiv {\partial \ln[r^2 \nbar(\r)] \over \partial \ln r}
.
\eeq
The $\left. \partial/\partial\r \right|_{\r = 0}$
term inside parentheses in eq.~(\ref{MEq}),
which subtracts from the first term $\partial/\partial\r$
its value at the position $\r = 0$ of the Local Group,
is the term that arises from the motion of the Local Group.
The Hermitian conjugate $\M^\dagger$
which enters equation~(\ref{zetaEq}) for the velocity mode $\zeta_i$
can be written
(Hamilton 1998 eq.~4.50)
\beq{MDagEq}
  \M^\dagger =
    \nabla^{-2} r^{-2} {\partial \over \partial r}
    \left( {\partial \over \partial r} - {\alpha(\r) \over r} \right) r^2
    - {\rhat \over r^2} . \left. {\partial \over \partial\r} \right|_{\r = 0}
    \nabla^{-2} \alpha(\r) r
\eeq
in which the last term
is again the term arising from the motion of the Local Group.

Although the definition of these three power spectra
assumes that redshift distortions conform to the linear Kaiser model,
they measure useful information from the data
even if the linear model fails.
In the small-angle (distant observer) approximation,
they reduce to simple linear combinations of the
monopole, quadrupole and hexadecapole power spectra in redshift space
(Cole, Fisher \& Weinberg 1994; Hamilton 1998):
\beq{multipole2flavorEq}
\left(\nskip\begin{tabular}{c}
$\Pgg(k)$\\[4pt]
$\Pgv(k)$\\[4pt]
$\Pvv(k)$
\end{tabular}\nskip\right)
=
\left(\begin{tabular}{ccc}
$1$	&$-{1\over 2}$	&${3\over 8}$\\[4pt]
$0$	&${3\over 4}$	&$-{15\over 8}$\\[4pt]
$0$	&$0$	&${35\over 8}$
\end{tabular}\right)
\left(\nskip\begin{tabular}{c}
$\Pmono(k)$\\[4pt]
$\Pquad(k)$\\[4pt]
$\Phexa(k)$
\end{tabular}\nskip\right).
\eeq 
Whereas the vector on the right hand side is closer to the measurements
(and also more familiar in the literature),
the vector on the left hand side is closer to the physics
of linear redshift distortions.
Indeed, inverting \eq{multipole2flavorEq},
\beq{flavor2multipoleEq}
\left(\nskip\begin{tabular}{c}
$\Pmono(k)$\\[4pt]
$\Pquad(k)$\\[4pt]
$\Phexa(k)$
\end{tabular}\nskip\right)
=
\left(\begin{tabular}{ccc}
$1$	&${2\over 3}$	&${1\over 5}$\\[4pt]
$0$	&${4\over 3}$	&${4\over 7}$\\[4pt]
$0$	&$0$	&${8\over 35}$
\end{tabular}\right)
\left(\nskip\begin{tabular}{c}
$\Pgg(k)$\\[4pt]
$\Pgv(k)$\\[4pt]
$\Pvv(k)$
\end{tabular}\nskip\right),
\eeq
we see that we can use this last equation as an improved {\it definition}
of monopole, quadrupole and hexadecapole, remaining valid even 
in the regime where the small-angle approximation fails.
For the reader more used to thinking in terms of the multipole formalism, 
the bottom line is that our main measurement $\Pgg(k)$ 
is basically the monopole power minus half the quadrupole power
plus three eights of the hexadecapole power, as per \eq{multipole2flavorEq}.

Because redshift distortions displace galaxies only along the line of sight,
the transverse, or angular, power spectrum is completely unaffected by
redshift distortions,
a point emphasized by Hamilton \& Tegmark (2002).
In the small-angle approximation,
the galaxy-galaxy power spectrum equals the
redshift space power spectrum in the transverse direction,
\beq{PggEq}
\Pgg(k) = P^{\rm s}(\mu_\k = 0)
= \sum_{\l ~ \rm{even}} {(-1)^{\l/2} (\l-1)!! \over \l!!} P^{\rm s}_\l(k)
,
\eeq
which is true in all circumstances, linear or nonlinear,
regardless of the character of redshift distortions.
The coefficients of the expansion~(\ref{PggEq}) are
the values of Legendre polynomials
${\cal P}_\l(\mu_\k)$
in the transverse direction $\mu_\k = 0$.
The first few terms of the expansion~(\ref{PggEq}) are
\beq{PggexpEq}
\Pgg(k) =
P^{\rm s}_0(k) - \frac{1}{2} P^{\rm s}_2(k) + \frac{3}{8} P^{\rm s}_4(k)
-
\frac{15}{48} P^{\rm s}_6(k)
+
\ldots
,
\eeq
which shows that our linear estimate of $\Pgg(k)$
is effectively the expansion~(\ref{PggEq})
truncated at the $\l = 4$ harmonic,
as predicted by linear theory (Kaiser 1987).
We expect on general grounds that the linear estimate of $\Pgg(k)$
will underestimate the true galaxy-galaxy power spectrum at nonlinear scales
(Hamilton \& Tegmark 2002),
although this underestimate should be mitigated by FOG compression.

In \sec{zspaceSec}, we will demonstrate with Monte Carlo simulations that 
$\Pgg$ faithfully recovers the true real-space galaxy power spectrum on large scales, and we will
quantify what constitutes ``large'', finding accurate recovery on substantially smaller
scales that those where the Kaiser approximation is valid.

A wide range of approximations for $\Pgv$ and $\Pvv$ have been introduced in the literature.
Using our notation, the Kaiser (1987) approximation becomes simply
\beqa{KaiserLimitEq1}
\Pgv(k) &=& \beta(k)r(k)\Pgg(k),\\
\Pvv(k) &=& \beta(k)^2\Pgg(k),\label{KaiserLimitEq2}
\eeqa
where $\beta(k)\equiv f/b(k)$, $b(k)$ is the bias factor,
$r(k)$ is the dimensionless correlation coefficient between
the galaxy and matter density (Dekel \& Lahav 1999; Pen 1998; Tegmark \& Peebles 1998)
and $f\approx\Omega_{\rm m}^{0.6}$ was defined above. 
Since both $b$ and $r$ can in principle depend on scale, we 
have two unknown functions $\beta(k)$ and $r(k)$ that can in principle
be determined uniquely from the two measured ratios
$\Pgv(k)/\Pgg(k)$ and $\Pvv(k)/\Pgg(k)$ in the Kaiser approximation.
Further popular approximations in the literature 
are that both $b$ and $r$ are constant, and most workers also assume $r=1$
despite some observational (Tegmark \& Bromley 1999; Blanton 2000) and
simulational (Blanton {\etal} 1999; Cen \& Ostriker 2000; Somerville {\etal} 2001) 
evidence that $r$ may be of the order of 0.9 for some galaxies\footnote{
Although $r<1$ is normally referred to as {\it stochastic bias}, this 
does of course not imply any randomness in
the galaxy formation process, merely that additional factors
besides the present-day dark matter 
density may be important (gas temperature, say). 
The evidence for $r<1$ thus far comes from scales smaller than 
those that are the focus of this paper.
More details on the
relationship between our three power spectra and the stochastic bias
formalism are given in Section 3.4 of THX02. 
}.

We will not make {\it any} of these approximations in our basic data analysis,
simply reporting measurements of $\Pgg(k)$, $\Pgv(k)$ and $\Pvv(k)$ from the SDSS 
data. In \sec{zspaceSec} we use the approximation that $\l>4$ anisotropies are negligible,
assessing its accuracy 
with Monte Carlo simulations and tunable finger-of-god compression, but 
the reader wishing to avoid approximations can simply fit better simulations directly to
our three measured curves.
Specifically, using an axis of a periodic cube as the line-of-sight
direction, so that the distant observer approximation holds perfectly,
one can compute the monopole, quadrupole, and hexadecapole components
of the redshift-space power spectrum and transform them to
$\Pgg$, $\Pgv$, and $\Pvv$ via \eq{multipole2flavorEq}.

\subsection{Step 3: Power spectrum estimation}
\label{ResultsSec}

All our information about the SDSS density field is encoded in the 4000-dimensional vector 
$\x$ plotted in \fig{xFig}, and \eq{gg_gv_vv_defEq} shows
that the covariance matrix of $\x$ depends linearly on the three power spectra that we want to measure.
We wish to invert equations\eqn{xCovEq} and\eqn{gg_gv_vv_defEq} to estimate the power spectra from the data vector.
This problem is mathematically equivalent to that of measuring the power spectrum from a CMB map,
and can be solved optimally with so-called quadratic estimators (Tegmark 1997b; Bond {\etal} 2000).
We describe our analysis method in full detail in Appendix B. However,  
since it is important for the interpretation, let us briefly review here
how the measurements are computed from the input data $\x$.

\begin{figure} 
\vskip-0.5cm
\epsfxsize=17cm\hglue-2mm\epsffile{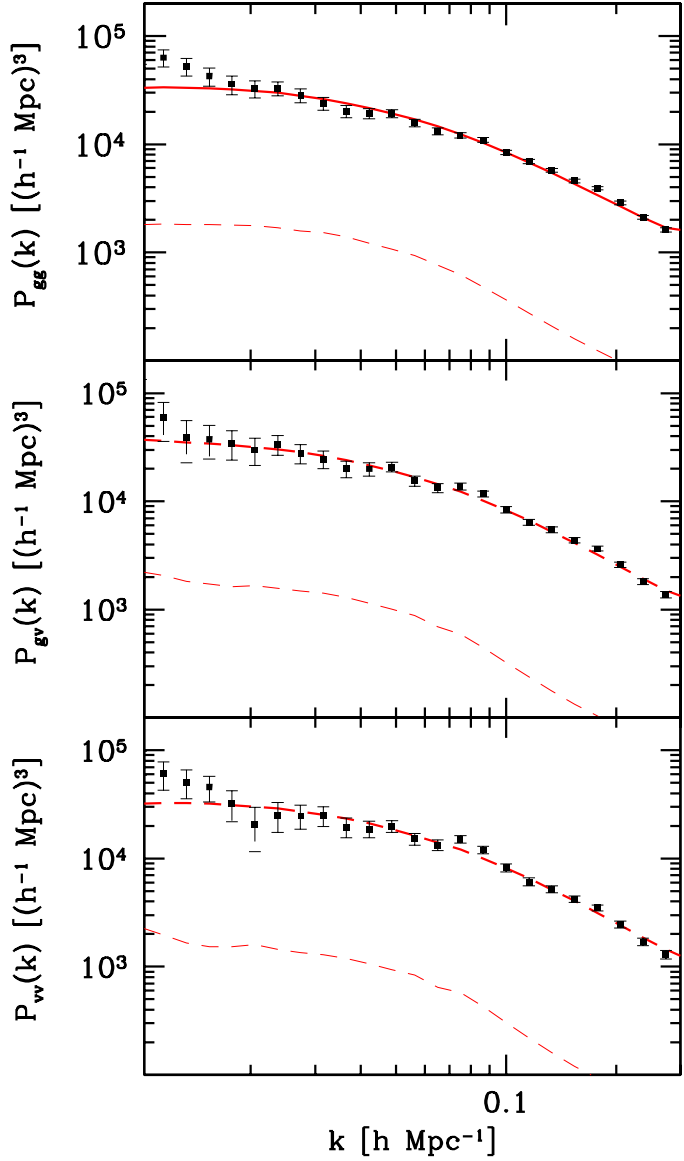}
\vskip-0.5cm
\caption[1]{\label{power_all3_correlatedFig}\footnotesize%
The minimum-variance quadratic estimators $\ph_i$ nominally measuring the 
galaxy-galaxy power spectrum (top), the galaxy-velocity power spectrum (middle)
and the velocity-velocity power spectrum (bottom) for the baseline 
galaxy sample.
They {\it cannot} be directly interpreted as power spectrum measurements,
since each point probes a linear combination of 
all three power spectra over a broad range of scales,
typically centered at a $k$-value different than the nominal 
$k$ where it is plotted. Moreover, nearby points are 
strongly correlated, causing this plot to 
overrepresent the amount of information present in the data.
The solid curves show the window-convolved prior power spectrum $\W\p$,
and the dashed curve shows the shot noise contribution subtracted out.
}
\end{figure}

\begin{figure} 
\vskip-0.5cm
\epsfxsize=17cm\hglue-2mm\epsffile{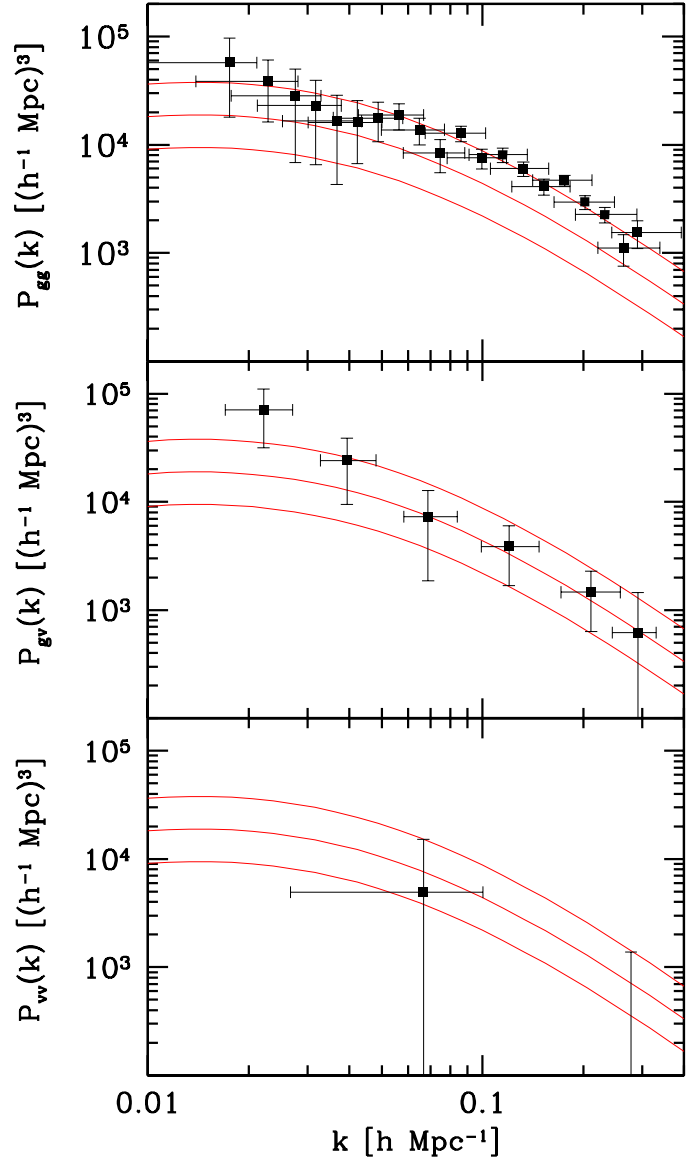}
\vskip-0.4cm
\caption[1]{\label{power_all3_binnedFig}\footnotesize%
Decorrelated and disentangled measurements of the galaxy-galaxy
power spectrum (top), the galaxy-velocity power spectrum (middle)
and the velocity-velocity power spectrum (bottom) for the baseline 
galaxy sample. 
Each point is plotted at the $k$-value that is the 
median of its window function, and the 
horizontal bars range from the 20th to the 80th percentile of
the window function.  The values of $\Pgg(k)$ are given in
Table~3.
From top to bottom, the three curves shows our prior for
$\Pgg(k)$, $\Pgv(k)$ and $\Pvv(k)$.
Note that most of the information in the survey is in 
the galaxy-galaxy spectrum. 
Band-power measurements with very low information 
content have been binned into fewer (still uncorrelated) bands.
All these measurements are for our baseline FOG compression threshold of 200.
Unlike \fig{power_all3_correlatedFig}, these points are
uncorrelated (affecting all three panels)
and the leakage between $gg$, $gv$ and $vv$ 
has been removed (affecting mostly the lower panels), giving much
larger (and easier to interpret) error bars.
}
\end{figure}

We parameterize our three power spectra by their amplitudes in 97 separate logarithmically spaced 
$k$-bands as detailed in Appendix B, 
so our goal is to measure $3\times 97=291$ band powers $p_i$, $i=1,...,291$.
Quadratic estimators $\ph_i$ are simply quadratic 
functions of the data vector $\x$,
and the most general unbiased case can be written as
\beq{phDefEq}
\ph_i\equiv\x^t\Q_i\x - \tr[\N\Q_i],
\eeq
for some symmetric $N\times N$-dimensional matrices $\Q_i$;
the second term merely subtracts off the expected contribution from the shot noise.

The basic idea behind quadratic estimators is that each matrix 
$\Q_i$ can be chosen to 
effectively Fourier transform the density field, square the Fourier modes
in the $\ith$ power spectrum band and average the results together,
thereby probing the power spectrum on that scale.
Grouping the parameters 
$p_i$ and the estimators $\ph_i$ into  
vectors denoted $\p$ and $\phat$, the expectation value and covariance of the measurements is given by
\beqa{pMomentsEq1a}
\expec{\phat} &=      &\W\p,\\
{\rm cov}(\phat)     &\equiv &\expec{\phat\phat^t}-\expec{\phat}\expec{\phat}^t =\SS,\label{pMomentsEq1b}
\eeqa
where the matrices $\W$ and $\SS$ can be computed 
from the $\Q_i$-matrices and the geometry alone via 
equations\eqn{pMomentsEq2a} and\eqn{pMomentsEq2b} in Appendix B.

As detailed in Appendix B, there are several attractive choices of 
$\Q$-matrices, each giving different desirable properties to the matrices $\W$ and $\SS$.
\Fig{power_all3_correlatedFig} shows the power spectrum 
measurements $\phat$ for the 
baseline galaxy sample using
the choice of $\Q_i$ that gives the smallest error bars,
and \fig{power_all3_binnedFig} shows them using a better choice described below.

\begin{figure} 
\centerline{\epsfxsize=\figsize\epsffile{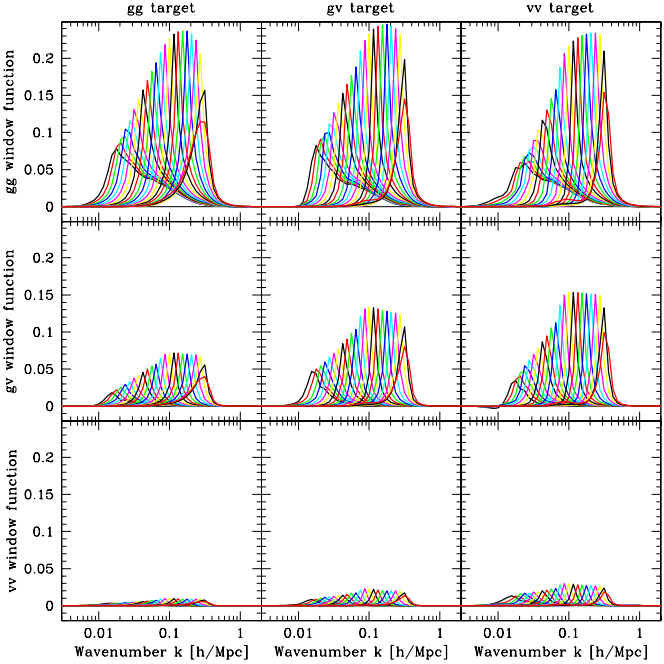}}
\caption[1]{\label{W0fig}\footnotesize%
Using the correlated minimum-variance estimators, the window functions are 
shown for those $k$-bands with non-negligible information content.
The $\ith$ row of $\W$ typically peaks at the
$\ith$ band, the scale $k$ that the band power 
estimator $\ph_i$ was designed to probe.
The three rows correspond to the estimators of 
$gg$, $gv$ and $vv$ power and the three columns to their 
sensitivity to $gg$, $gv$ and $vv$ power.
For example, the window function of the quadratic estimator targeting $\Pgv(k)$ 
at $k=0.1h/$Mpc is given by the three curves in the middle column
peaking at $k=0.1h/$Mpc (top, middle and bottom panels), and the normalization is 
such that  sum of the areas under these three curves is unity.
}
\end{figure}

Although \fig{power_all3_correlatedFig} looks 
impressive, it fails to convey two important complications.
The first is that the
error bars are strongly correlated between neighboring bands, \ie, 
the covariance matrix $\SS$ is far from diagonal. The second complication 
involves the matrix $\W$, known as the {\it window matrix}.
The $\Q$-matrices are normalized so that each row of the window matrix sums to unity. 
\Eq{pMomentsEq1a} shows that this normalization enables us to interpret
each band power measurement $\ph_i$ as a weighted average of
the true power spectrum $p_j$, the elements of
the $\ith$ row of $\W$ giving the weights (the so-called window function).
In short, the window functions 
connect our measurements $\phat$ to the underlying
power spectrum parameters $\p$.
The windows are plotted in 
\fig{W0fig}, and we see that they are complicated in two different ways,
making \fig{power_all3_correlatedFig} hard to interpret:
\begin{enumerate}
\item {\bf Smearing:} They have a non-zero width $\Delta k$, 
so that our estimate of the power on some scale $k$ is really the weighted average
of the power over a range of scales around $k$. In other words, 
\fig{power_all3_correlatedFig} shows a measurement of the true power spectrum
that has been smoothed, convolved with rather broad window functions.
\item {\bf Leakage:} They mix the $gg$, $gv$ and $vv$ power spectra, so that
a nominal estimate of $gv$, say, is really a weighted average of $gg$, $gv$ and $vv$ power.
This is why the signal-to-noise ratio of $gv$ and $vv$ appear so high in \fig{power_all3_correlatedFig}.
\end{enumerate}

\begin{figure} 
\centerline{\epsfxsize=\figsize\epsffile{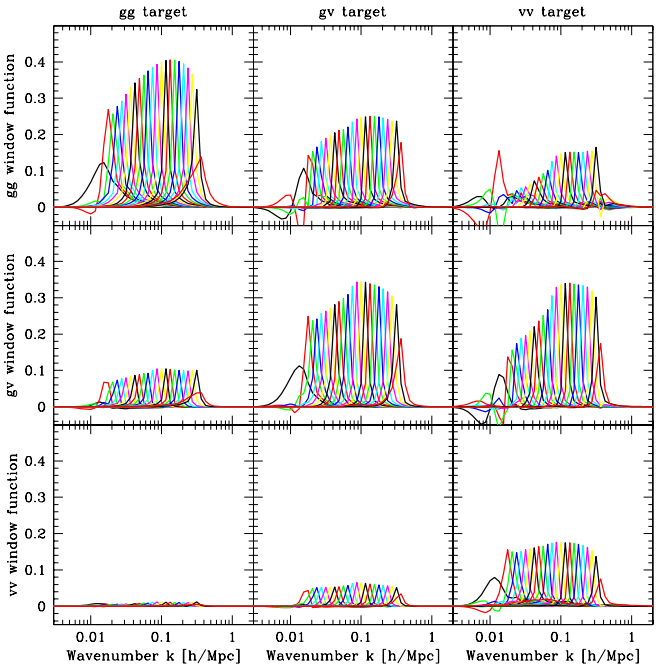}}
\caption[1]{\label{W1fig}\footnotesize%
Same as \fig{W0fig}, but 
using decorrelated estimators (before disentanglement).
Comparison with \fig{W0fig} shows that
decorrelation makes all windows substantially narrower.
}
\end{figure}

\begin{figure} 
\centerline{\epsfxsize=\figsize\epsffile{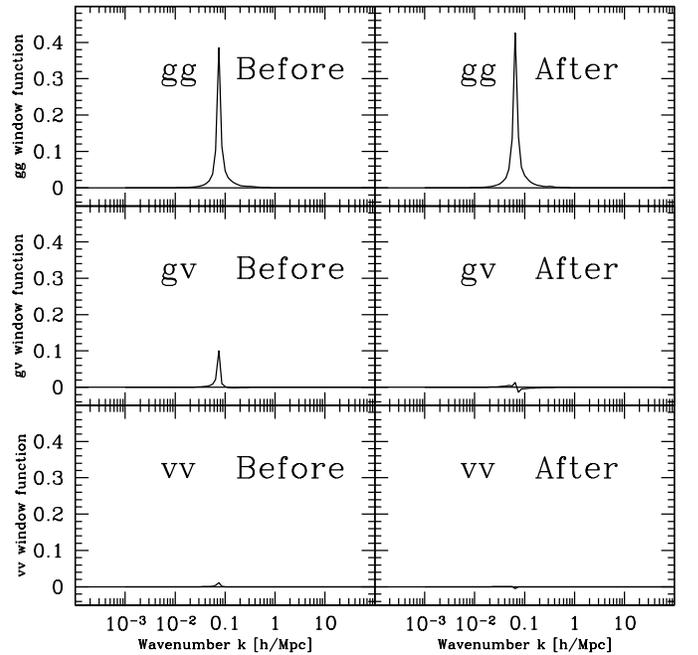}}
\caption[1]{\label{disentanglementFig}\footnotesize%
The window function for our measurement of the 49th
band of the galaxy-galaxy power is shown before
(left) and after (right) disentanglement.
Whereas unwanted leakage of $gv$ and $vv$ power 
is present initially, these unwanted
window functions both average to zero afterward.
The success of this method hinges on the fact that 
since the three initial functions (left) have similar shape,
it is possible to take linear combinations of
them that almost vanish (right). 
This procedure is repeated separately for each $k$.
}
\end{figure}

As described in Appendix B, both the correlation problem and the 
smearing problem can be tackled in one fell swoop with a better 
choice of quadratic estimators that give uncorrelated error bars
and narrower window functions, shown in \fig{W1fig}.
This choice makes the covariance matrix $\SS$ 
for the measured vector $\phat$ diagonal
(combining shot noise and sample variance errors), so it is completely
characterized by its diagonal elements, given by the error bars in 
Table 2 and \fig{power_all3_binnedFig}. 
A clearer and less cluttered view of a sample window function
for this uncorrelated case is given in \fig{disentanglementFig}
(top left panel).
We see that such a window is almost never negative, and tends to be sharply peaked around 
the $k$-value that it is designed to probe\footnote{ 
Its characteristic 
width $\Delta k$ corresponds roughly to the inverse width of the 
survey volume in its narrowest direction (Tegmark 1995),
so the windows will get narrower as the SDSS becomes more complete and
the thin sky slices seen in \fig{aitoffFig} thicken and merge.
Windows further to the left are slightly 
narrower (when plotted on a linear $k$-scale as opposed to the logarithmic
scale used here), since they probe more distant galaxies and hence a larger
effective volume. However, since our sample contains very few galaxies 
with $z\gg 0.2$, the window width $\Delta k$ approaches a constant 
as we keep moving to the left in \fig{W1fig}, causing the windows to
look wider on our logarithmic axis.
}.

\begin{figure} 
\centerline{\epsfxsize=\figsize\epsffile{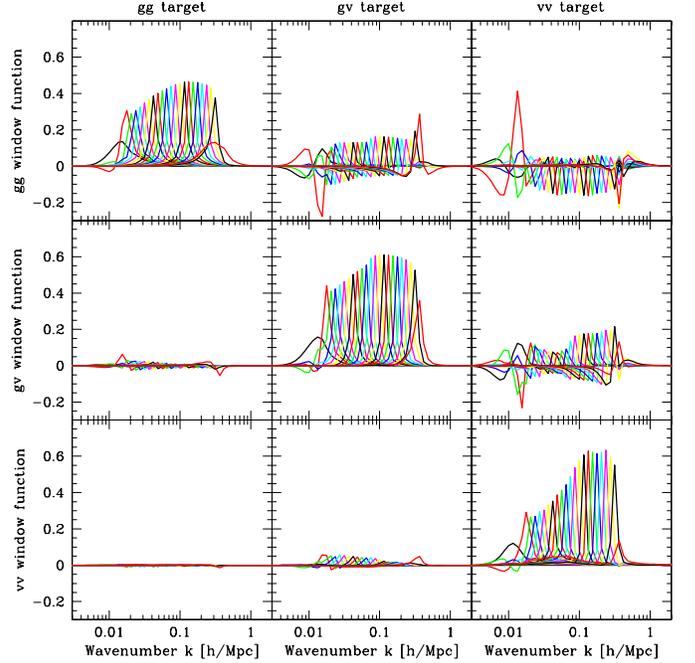}}
\caption[1]{\label{W2fig}\footnotesize%
Same as \fig{W1fig}, but decorrelated and disentangled 
estimators. Comparison with \fig{W1fig} shows that
disentanglement gives curves in the off-diagonal panels a vanishing 
average, and almost completely eliminates leakage of $gv$ and $vv$ power into
estimators of $gg$ power (two bottom panels in left column).
}
\end{figure}

\begin{figure} 
\centerline{\epsfxsize=\figsize\epsffile{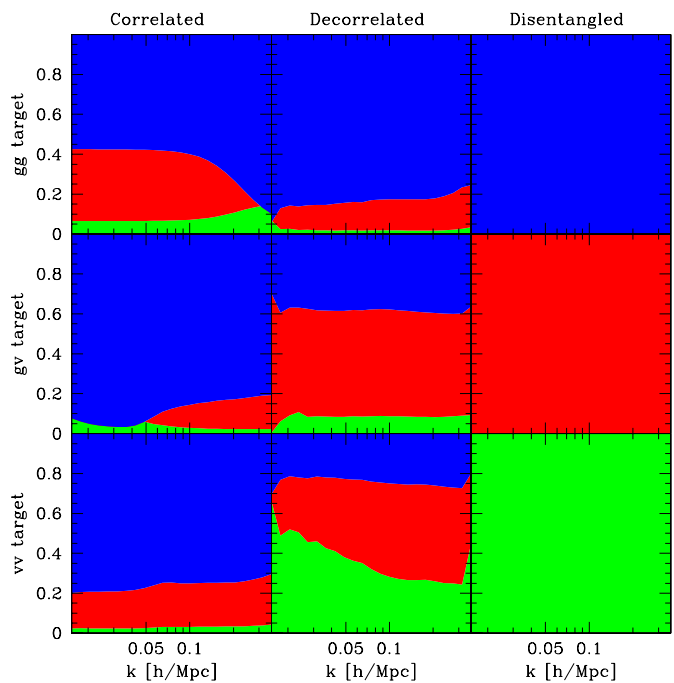}}
\caption[1]{\label{leakageFig}\footnotesize%
The leakage is shown for the correlated (left column), 
decorrelated (middle column) and disentangled (right column) methods.
From top to bottom within each box, the three bands
(blue/dark grey, red/grey and green/light grey) show the relative contributions
from $gg$, $gv$ and $vv$ power, respectively.
From top to bottom, the three rows are for estimators are for 
$gg$, $gv$ and $vv$ power, respectively.
}
\end{figure}

Let us now turn to the remaining problem: leakage.
The leakage results from a combination of two effects:
difficulties in separating the monopole, quadrupole and hexadecapole power
given the complicated survey geometry, and the mixing of these three multipoles
given by \eq{multipole2flavorEq}.
Figures \ref{W0fig}, \ref{W1fig} and \ref{W2fig} show the
$\ith$ window function (the $\ith$ row of the $\W$-matrix) as three 
curves plotted in the top, middle and bottom panels, 
giving the sensitivity of the estimator $\ph_i$ to 
$gg$, $gv$ and $vv$ power, respectively. If there were no leakage, then
all curves in the six off-diagonal panels in these figures would
be identically zero.  This is not the case, but the {\em area} under
the curves is much reduced in the off-diagonal panels, as is shown by
comparing the left-most and middle panels of \fig{leakageFig}. 
Switching to uncorrelated quadratic estimators 
causes a substantial leakage reduction as a side benefit, but that leakage is
still non-negligible: for instance, an estimate of the $gg$ power
is seen to give about 15\% 
weight to $\Pgv$
and about 2\% 
weight to $\Pvv$, with these percentages depending only weakly on $k$.

As detailed in Appendix B, we can eliminate the leakage problem and
measure one power spectrum, 
say $\Pgg(k)$, without any assumptions about the other two by 
effectively marginalizing over their amplitudes separately for each 
$k$-band. This procedure is equivalent to yet another choice of the 
$\Q$-matrices, which we refer to as {\it disentangled}.
As seen in \fig{W2fig} and the right panels of \fig{leakageFig}, it eliminates leakage
completely in the sense that all unwanted (off-diagonal) window functions 
have zero area.
The basic idea of the disentanglement procedure is illustrated in 
\fig{disentanglementFig}: since the $gg$, $gv$ and $vv$ components of
the window function have very similar shape, differing essentially
only in amplitude, it is possible to form linear combinations of them that 
for all practical purposes vanish.  In forming these linear
combinations, we do introduce statistical correlations between
$\Pgg(k)$, $\Pgv(k)$ and $\Pvv(k)$ at a given value of $k$; the values
at different values of $k$ remain uncorrelated.

In summary, we have measured the three power spectra
$\Pgg(k)$, $\Pgv(k)$ and $\Pvv(k)$, obtaining the results 
shown in 
\fig{power_all3_binnedFig}. These basic measurements are given in 
Table 2 and are
available at\\
{\it http://www.hep.upenn.edu/$\sim$max/sdss.html}
together
with their window matrix and likelihood calculation software, incorporating the bias correction described in 
\sec{BiasSec}.
The measurements make no assumptions whatsoever about redshift-space
distortions, and the issue of whether the density fluctuations are Gaussian
affects only the error bars, not the measurements themselves.

In the next section, we will model the effect of redshift distortions
and make what we argue is a more accurate estimate
of $\Pgg(k)$.
However, the conservative reader trusting only her/his own modeling 
can in principle stop right here and fit simulations directly to
our measurements from \fig{power_all3_binnedFig}, which are given in Table 2.

\begin{table}
\bigskip
\noindent
{\footnotesize {\bf Table 2} -- The real-space power spectrum 
$\Pgg(k)$ (top),  $\Pgv(k)$ (middle) and  $\Pvv(k)$ (bottom)  
measured with the disentanglement method.  The units of the power are
$(h^{-1}\Mpc)^3$. $\Delta P$ is the $1\sigma$ error.
These values have been corrected for luminosity-dependent
bias by dividing by the square of the last column, and thus refer to the clustering of $L_*$ galaxies. The  $k$-column gives 
the median of the window function and its $20^{th}$ and $80^{th}$ percentiles;
the exact window functions from 
{\it http://www.hep.upenn.edu/$\sim$max/sdss.html}.
should
be used for any quantitative analysis. The $\Pgg$ errors are uncorrelated with one another,
but are correlated with the $\Pgv$ and $\Pvv$ errors.
We recommend using column 2 for basic analysis. 
Column 3 is without FOG removal (i.e., with threshold
$\delta_c=\infty$) and is therefore
easier to compare against numerical simulations.
\bigskip
\begin{center}   
{\footnotesize
\begin{tabular}{|rrrrr|}
\hline
$k\>[h/$Mpc]			&$P$	&$P$ ($\perp$FOG)	&$\Delta P$	&$b$\\
\hline
$ 0.018^{+0.008}_{-0.004}$	&42098	&41081	&28850	&1.167\\ 
$ 0.023^{+0.009}_{-0.005}$	&28260	&28924	&16394	&1.167\\ 
$ 0.028^{+0.010}_{-0.005}$	&20880	&20508	&15849	&1.166\\ 
$ 0.032^{+0.010}_{-0.006}$	&16903	&17097	&12079	&1.165\\ 
$ 0.037^{+0.011}_{-0.007}$	&12178	&12119	& 9004	&1.164\\ 
$ 0.042^{+0.012}_{-0.008}$	&11887	&11996	& 6944	&1.163\\ 
$ 0.049^{+0.013}_{-0.009}$	&13098	&13094	& 5188	&1.161\\ 
$ 0.056^{+0.014}_{-0.010}$	&13996	&14003	& 3847	&1.159\\ 
$ 0.065^{+0.015}_{-0.012}$	&10273	&10333	& 2847	&1.157\\ 
$ 0.075^{+0.017}_{-0.014}$	& 6296	& 6366	& 2130	&1.153\\ 
$ 0.086^{+0.019}_{-0.016}$	& 9653	& 9687	& 1594	&1.149\\ 
$ 0.100^{+0.021}_{-0.018}$	& 5763	& 5814	& 1205	&1.144\\ 
$ 0.115^{+0.024}_{-0.021}$	& 6229	& 6273	&  921	&1.139\\ 
$ 0.132^{+0.027}_{-0.025}$	& 4693	& 4711	&  712	&1.132\\ 
$ 0.153^{+0.031}_{-0.030}$	& 3263	& 3321	&  554	&1.123\\ 
$ 0.176^{+0.035}_{-0.037}$	& 3778	& 3811	&  437	&1.114\\ 
$ 0.202^{+0.039}_{-0.045}$	& 2423	& 2428	&  356	&1.104\\ 
$ 0.232^{+0.043}_{-0.058}$	& 1891	& 1892	&  312	&1.093\\ 
$ 0.264^{+0.043}_{-0.075}$	&  952	&  947	&  304	&1.082\\ 
$ 0.290^{+0.047}_{-0.102}$	& 1340	& 1385	&  386	&1.074\\ 
\hline
$ 0.022^{+0.005}_{-0.005}$	&52115	&51536	&28798	&1.167\\ 
$ 0.039^{+0.007}_{-0.009}$	&17843	&17716	&10870	&1.164\\ 
$ 0.069^{+0.011}_{-0.015}$	& 5451	& 5233	& 4057	&1.155\\ 
$ 0.120^{+0.021}_{-0.028}$	& 2991	& 2746	& 1684	&1.137\\ 
$ 0.211^{+0.040}_{-0.047}$	& 1207	&  902	&  684	&1.101\\ 
$ 0.291^{+0.047}_{-0.039}$	&  537	&  319	&  720	&1.074\\ 
\hline
$ 0.067^{+0.040}_{-0.034}$	& 3700	& 3583	& 7712	&1.156\\ 
$ 0.278^{+0.128}_{-0.184}$	&    7	&  -81	& 1185	&1.078\\ 
\hline  
\end{tabular}
}
\end{center}     
} 
\end{table}

\clearpage 

\begin{table}
\bigskip
\noindent
{\footnotesize {\bf Table 3} -- The real-space power spectrum $\Pgg(k)$ in 
units of $(h^{-1}\Mpc)^3$ measured with
the modeling method. $\Delta\Pgg$ is the $1\sigma$ error, uncorrelated between bands.
These values have been corrected for luminosity-dependent
bias by dividing by the square of the last column (see \sec{BiasSec}),
and thus refer to the clustering of $L_*$ galaxies. 
The  $k$-column gives 
the median of the window function and its $20^{th}$ and $80^{th}$ percentiles;
the exact window functions from 
{\it http://www.hep.upenn.edu/$\sim$max/sdss.html}.
should
be used for any quantitative analysis.
\bigskip
\begin{center}   
{\footnotesize
\begin{tabular}{|rrrr|}
\hline
$k\>[h/$Mpc]		&$\Pgg$	&$\Delta\Pgg$	&$b$\\
\hline
$ 0.016^{+0.006}_{-0.003}$	&21573	&33320	&1.168\\ 
$ 0.018^{+0.006}_{-0.003}$	&33255	&24573	&1.167\\ 
$ 0.021^{+0.007}_{-0.004}$	&13846	&17712	&1.167\\ 
$ 0.024^{+0.007}_{-0.004}$	&38361	&13320	&1.167\\ 
$ 0.028^{+0.008}_{-0.005}$	&24143	&10047	&1.166\\ 
$ 0.032^{+0.008}_{-0.005}$	&19709	& 7414	&1.165\\ 
$ 0.037^{+0.009}_{-0.006}$	&12596	& 5486	&1.164\\ 
$ 0.043^{+0.010}_{-0.007}$	&13559	& 4078	&1.163\\ 
$ 0.049^{+0.011}_{-0.008}$	&18311	& 2974	&1.161\\ 
$ 0.057^{+0.012}_{-0.008}$	&12081	& 2140	&1.159\\ 
$ 0.065^{+0.013}_{-0.010}$	& 9217	& 1580	&1.156\\ 
$ 0.075^{+0.015}_{-0.011}$	& 9751	& 1128	&1.153\\ 
$ 0.087^{+0.017}_{-0.012}$	& 9530	&  818	&1.149\\ 
$ 0.100^{+0.019}_{-0.014}$	& 6385	&  602	&1.144\\ 
$ 0.116^{+0.021}_{-0.016}$	& 5295	&  447	&1.138\\ 
$ 0.134^{+0.024}_{-0.019}$	& 4630	&  335	&1.131\\ 
$ 0.154^{+0.027}_{-0.022}$	& 3574	&  254	&1.123\\ 
$ 0.178^{+0.031}_{-0.027}$	& 3394	&  195	&1.114\\ 
$ 0.205^{+0.036}_{-0.032}$	& 2298	&  153	&1.103\\ 
$ 0.236^{+0.041}_{-0.039}$	& 1597	&  124	&1.092\\ 
$ 0.271^{+0.043}_{-0.048}$	& 1105	&  107	&1.080\\ 
$ 0.306^{+0.042}_{-0.075}$	& 1013	&  110	&1.069\\ 
\hline  
\end{tabular}
}
\end{center}     
} 
\end{table}

\begin{figure} 
\vskip-0.5cm
\epsfxsize=17cm\hglue-2mm\epsffile{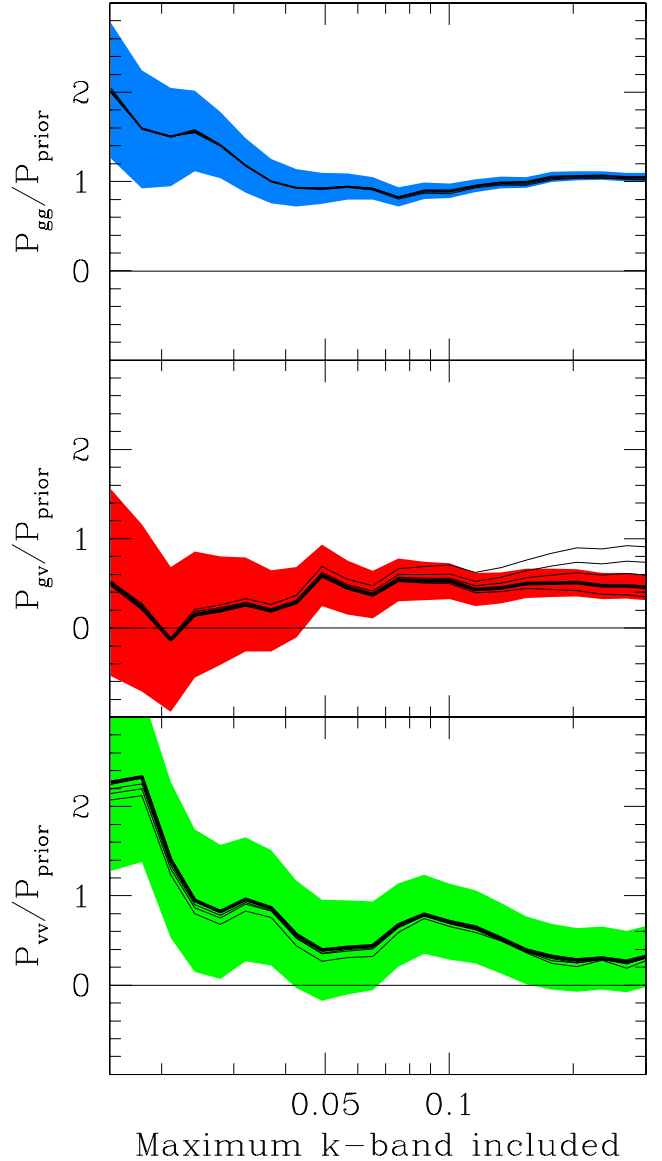}
\vskip-0.5cm
\caption[1]{\label{chi2Fig}\footnotesize%
The blue, red and green bands show the $1\sigma$ allowed ranges for
the amplitude of the $gg$, $gv$ and $vv$ power, respectively, relative to the prior $gg$ spectrum,
as a function of the maximum wavenumber included in the fit.
The sets of five black curves show the best fit values 
using FOG compression with the five density thresholds 
$1{+}\delta_c=\infty$ (no FOG compression),
200 (our baseline; heavy curve), 100, 50 and 25 (successive curves 
go higher for $gv$ and lower for $gg$ and $vv$).
}
\end{figure}

\section{Accounting for redshift space distortions}
\label{zspaceSec}

So far, we have measured the SDSS galaxy power spectrum and its redshift-space anisotropy,
encoded in the three functions $\Pgg(k)$, $\Pgv(k)$ and $\Pvv(k)$.
In the present section, we focus on this anisotropy to model, quantify and correct for 
the effects of redshift-space distortions, producing an estimate of the 
true real-space galaxy power spectrum, $\Pggtrue(k)$.
We will use Monte-Carlo simulations to assess the accuracy of 
two alternative approaches: 
\begin{enumerate}
\item {\bf Disentanglement approach:} perform FOG compression, 
      then simply use $\Pgg(k)$ from \fig{power_all3_binnedFig} as the estimator
      of $\Pggtrue(k)$.
\item {\bf Modeling approach:} perform FOG compression, then make the Kaiser approximation 
      of equations\eqn{KaiserLimitEq1}
      and\eqn{KaiserLimitEq2} with the best-fit constant values of $\beta$ and $r$
      to eliminate $\Pgv(k)$ and $\Pvv(k)$ from the problem, 
      solving for the 97 decorrelated measurements of $\Pggtrue(k)$ 
      (see Appendix B and THX02 for details).
\end{enumerate}
The difference between the two approaches is essentially between marginalizing over 
the other two power spectra ($\Pgv(k)$ and $\Pvv(k)$) and modeling them. 
Both approaches break down on small scales, so we focus on quantifying
their accuracy as a function of $k$.
We will see that although the disentanglement approach is more robust to nonlinearities,
the modeling approach has the advantage of roughly halving the error bars, 
corresponding to quadrupling the Fisher information.
The gain comes from using rather than discarding measured information about   
the amplitudes of the $gv$ and $vv$ power spectra.
We will argue that the disentanglement method is overly conservative,
especially on extremely large scales
like $k < 0.05h/$Mpc where we need all the statistical power that we can get.

There are two separate sources of statistical bias in our measurement that we will quantify below.
The first is that $\Pgg(k)$ will only equal the true real-space power
on scales on which the Kaiser approximation holds, generally underestimating it on smaller scales.
The second occurs only in the modeling approach, which produces a 
biased measurement of $\Pgg(k)$ if the model parameters $\beta$ and $r$ are
incorrect.

\begin{figure} 
\centerline{\epsfxsize=\figsize\epsffile{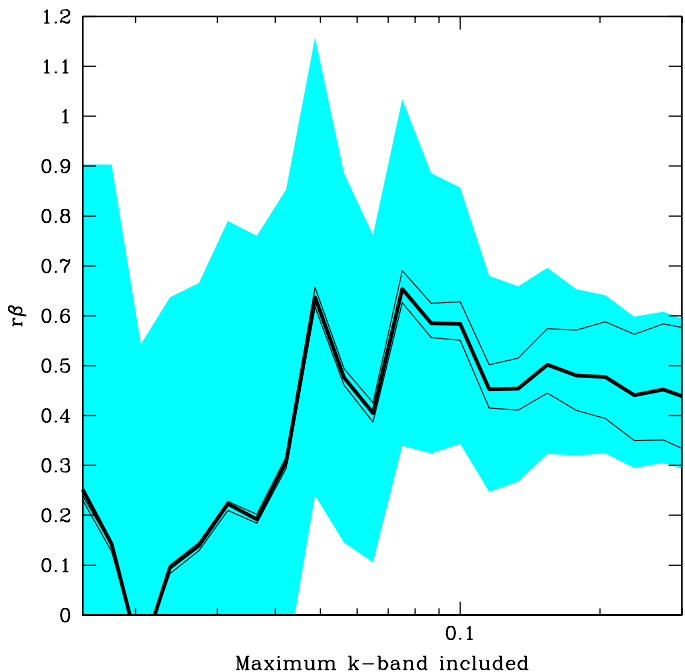}}
\caption[1]{\label{betaFig}\footnotesize%
The blue/grey band shows the $1\sigma$ allowed range
for $r\beta$, assuming the {\it shape} of the prior $\protect\Pgg(k)$
but marginalizing over the power spectrum normalization,
using FOG compression with our baseline density threshold $1{+}\delta_c=200$.
From bottom to top, the three curves show the best fit $\beta$ for
FOG thresholds $1{+}\delta_c=\infty$ (no FOG compression),
200 (our baseline; heavy curve) and 100. 
}
\end{figure}

Let us begin our investigation by studying the real data, then 
turn to Monte Carlo simulations to better understand and quantify the 
results.

\begin{figure*} 
\vskip-0.4cm
\centerline{\epsfxsize=18cm\epsffile{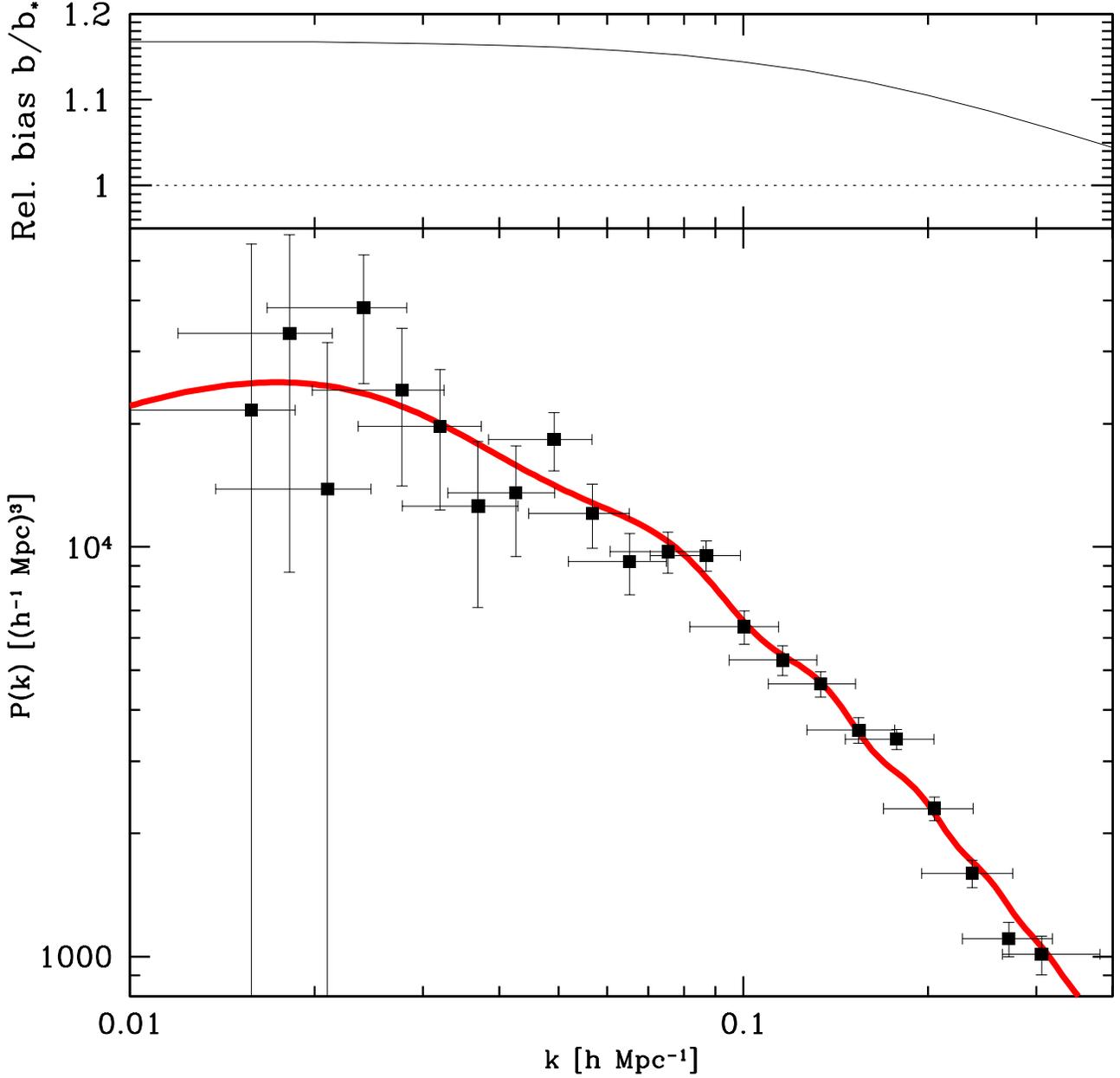}}
\caption[1]{\label{power_ggFig}\footnotesize%
The decorrelated real-space galaxy-galaxy power spectrum using the
modeling method is shown (bottom panel) for the 
baseline galaxy sample
assuming $\beta=0.5$ and $r=1$. 
As discussed in the text, 
uncertainty in $\beta$ and $r$ contribute to an overall 
calibration uncertainty of order $4\%$ which is
not included in these error bars.
To remove scale-dependent bias caused by luminosity-dependent clustering, 
the measurements have been divided by the square of the curve in the top panel,
which shows the bias relative to $L_*$ galaxies.
This means that the points in the lower panel 
can be interpreted as the power spectrum of $L_*$ galaxies.
The solid curve (bottom) is the best fit linear $\Lambda$CDM model of \sec{BiasSec}.
}
\end{figure*}

\subsection{Results based on the data}

\Fig{power_all3_binnedFig} shows that whereas we have precise measurements of 
$\Pgg(k)$, we have rather limited information about $\Pgv(k)$ and close to 
no information about $\Pvv(k)$.
\Fig{chi2Fig} shows a slightly less noisy rendition of the same information.
Here we have taken all three curves to have the shape of the prior power spectrum
and plotted their best-fit amplitudes relative to the prior.
These fits are performed cumulatively, using all measurements for all wavenumbers $\le k$.
The three bands give the $1\sigma$ allowed ranges for $gg$, $gv$ and $vv$, respectively.
It is well-known that as $k$ increases, nonlinearities become more important and start
reducing $gv$, eventually driving it negative. This is because the FOG effect 
has the opposite sign of the linear Kaiser infall, causing less rather than more radial power
(or larger rather than smaller radial correlations, for the reader preferring real space over
Fourier space).
The fact that \fig{chi2Fig} does not show this effect is therefore a first encouraging indication 
that nonlinearities have only a minor impact on our results over the range
of scales that we consider.
Since we have used only 4000 PKL modes, most of the information from 
scales $k\gg 0.1 h/$Mpc is excluded from our analysis (cf.,
Figure~\ref{mode_keffFig}). 
The bands in the figure therefore stop getting thinner for $k \gg 0.1 h/$Mpc.
In other words, the information contained in our data vector $\x$ 
describes only a highly smoothed version of the density field,
where nonlinear effects are small.

The five thin lines in \fig{chi2Fig} correspond to our 
five FOG compression thresholds, and show several noteworthy things.
First of all, changing the FOG threshold is seen to have a strong effect on 
$gv$ but almost no effect on $gg$ (the quantity that we really care about in this paper),
providing another encouraging indication that virialized structures do not pose an unsurmountable 
problem for us.
Second, more aggressive FOG removal is seen to raise the $gv$ amplitude. This is the
expected sign of the effect, since it removes (and eventually over-corrects for) the FOG effect
which suppresses $gv$.
Third, the $gg$ and $gv$ curve pentuplets are seen to diverge as $k$ increases, as nonlinearities
become more important. For $gv$, the spread between the baseline threshold $1+\delta_c=200$ 
and the rather extreme neighboring thresholds ($100$ and $\infty$)
equals the error bar for $k\sim 0.3 h/\Mpc$, suggesting that
nonlinearity-related uncertainties become comparable to statistical 
uncertainties on this scale when trying to measure the redshift distortion parameter $\beta$.
For $gg$, on the other hand, the statistical uncertainties dominate on all scales 
to which we are sensitive.
The optimal FOG compression threshold should be expected to lie somewhere between
our options $200$ and $\infty$, since
$1{+}\delta_c=200$ is the approximate overdensity of a cluster
that has just formed, and many FOG systems will have formed earlier and hence have
higher overdensities.
The other thresholds plotted, \ie, 100, 50, and 25, are thus more extreme
and eventually unphysical --- we have used and plotted them merely to
exaggerate and illustrate the effect of FOG removal more clearly. 

Since $vv$ is so noisy, our main constraint on redshift space distortion
comes from the ratio of $gv$ to $gg$ power, \ie, on $r\beta=\Pgg/\Pgv$.
\Fig{betaFig} shows our $1-\sigma$ constraints on $r\beta$ as a function of the maximal
$k$-band included in a cumulative fit, 
discarding the $vv$ information to be conservative. 
The effect of FOG removal is seen to be smaller than the statistical
errors for all scales that we consider.
Our (loose) constraints agree well with a previous $\beta$-measurement
from earlier SDSS data (Zehavi {\etal} 2002) and also with 
measurements from the 2dFGRS (Peacock {\etal} 2001; THX02)
assuming that the bias does not differ dramatically between the
$r$-band selected SDSS galaxies and $B$-band selected 2dF galaxies.
We are unable to break their near degeneracy and 
place strong constraints on $\beta$ and $r$ separately,
but a joint likelihood analysis marginally favors $r\sim 1$.

\begin{figure} 
\vskip-0.8cm
\centerline{\epsfxsize=\figsize\epsffile{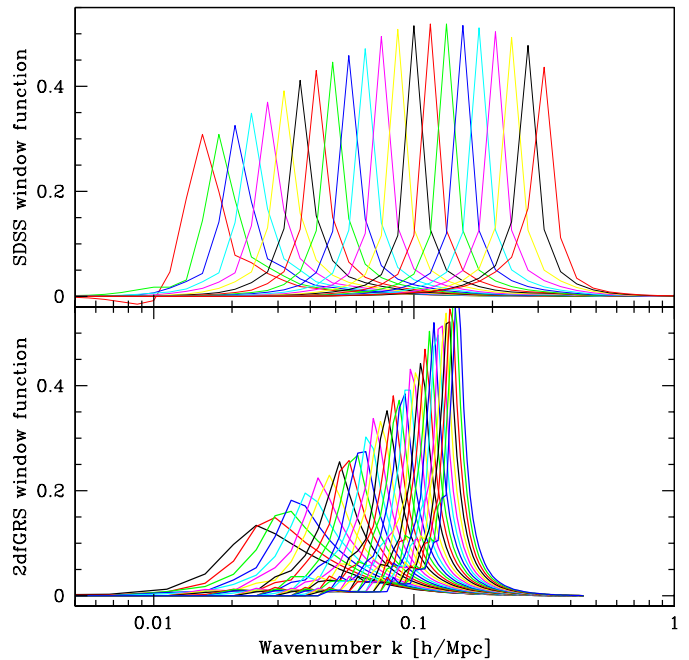}}
\caption[1]{\label{Wfig}\footnotesize%
The SDSS window functions (rows of $\W$) are shown (top panel)
using decorrelated estimators for the modeling method.
The $\ith$ row of $\W$ typically peaks at the
$\ith$ band, the scale $k$ that the band power 
estimator $\ph_i$ was designed to probe.
All curves have been normalized to have the same area, 
so the highest peaks indicate the scales where the window 
functions are narrowest.
The turnover in the envelope at $k\sim 0.1\hperMpc$
reflects our running out of information due to 
omission of modes probing smaller scales.
The 32 2dFGRS window functions estimated by Percival {\etal} (2001) are shown 
for comparison, plotted with the exact same conventions. 
They correspond to correlated rather than uncorrelated measurements.
Their shapes and widths is seem to agree well with the SDSS windows 
around $k\sim 0.1h/$Mpc, becoming substantially wider on larger
scales; this is a key advantage of our analysis method.
}
\end{figure}

Our estimate of the real-space galaxy power spectrum from the disentanglement approach
is simply the top panel of \fig{power_all3_binnedFig}.
The corresponding estimate using the model approach is shown in 
\fig{power_ggFig}; the values are tabulated in Table 3. Here we use 
$\beta=0.5$ and $r=1$, which provides a good fit to our data. 
The measurements are also tabulated in Tables 2 and 3. 
Changing these two parameters within their measurement uncertainty 
causes an uncertainty of 4\% 
in the overall normalization of the recovered $gg$ power spectrum
(which is, of course, degenerate with a 2\% 
change in the galaxy bias).
The corresponding window functions are shown in \fig{Wfig}; compare
with Figure~\ref{W2fig}.

\begin{figure} 
\vskip-0.4cm
\centerline{\epsfxsize=\figsize\epsffile{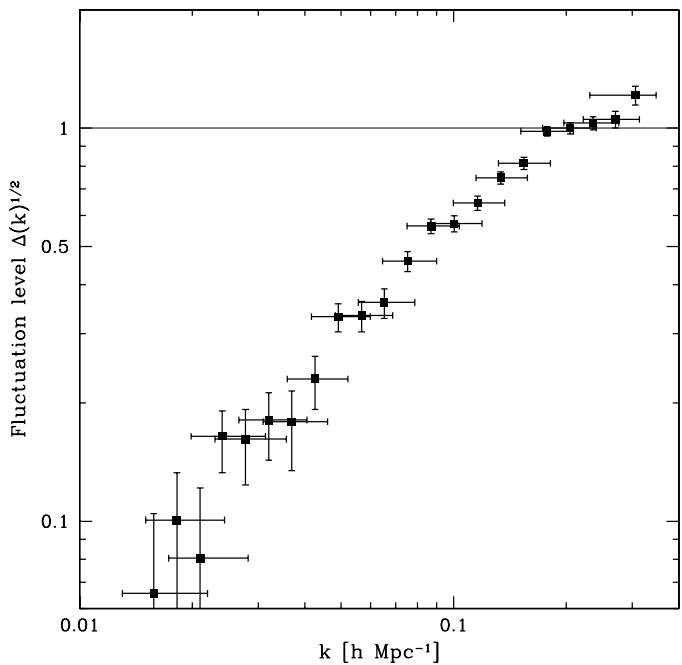}}
\caption[1]{\label{Delta_ggFig}\footnotesize%
The rms density 
fluctuation amplitude $\Delta(k)^{1/2}$ derived from the modeling
method power spectrum of the previous figure, where
$\Delta(k)\equiv 4\pi\Pgg(k)[k/(2\pi)]^3$ and $k$ is the effective
$k$-value from Table 2. 
}
\end{figure}

To indicate how linear the fluctuations are on various scales, 
\fig{Delta_ggFig} shows the square root of the corresponding dimensionless power
spectrum, which can be crudely interpreted as the rms fluctuation on that scale.
This fluctuation level is seen to drop below 10\% 
on the largest scales, $k\simgt 0.02 h/$Mpc, with the curve being strikingly different from 
a power law (more clearly seen in \fig{power_ggFig})\footnote{
To make this more quantitative, we fit the measurements to a parabola
in $(\log k,\log P)$, obtaining a curvature 
$d\log P/d\log k = -1.28\pm 0.49$.
For a Markov Chain with $10^6$ models, 99.9\% had $\alpha<0$, 
thereby driving yet another nail into the coffin of the fractal Universe hypothesis
and any other models predicting a power law power spectrum ($\alpha=0$).
}.
The nonlinearity transition $\Delta\sim 1$ is seen to occur
around $k\sim 0.2h$/Mpc, but this is a crude estimate since what matters
is of course the fluctuation level of the matter, not of the galaxies, and
the two differ by the bias factor.
As detailed in \sec{DiscussionSec}, our $L^*$ galaxies have $\sigma_8\approx 0.93$,
so if $\sigma_8\approx 0.8$ for the matter as indicated by many recent
CMB, lensing and cluster studies
(Lahav {\etal} 2002; Wang {\etal} 2002; Bennett {\etal} 2003),
the fluctuations are slightly more
linear than \fig{Delta_ggFig} indicates.

\begin{figure} 
\centerline{\epsfxsize=\figsize\epsffile{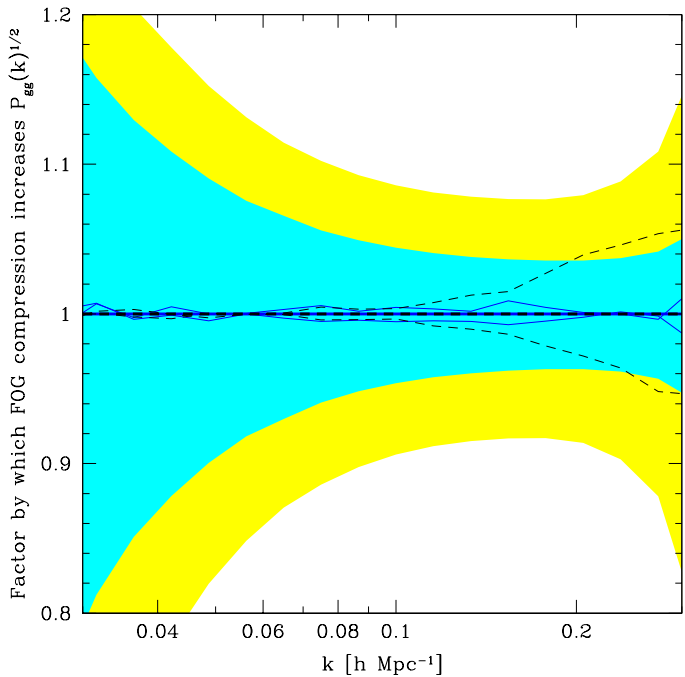}} 
\caption[1]{\label{defog_powerFig}\footnotesize%
Effect of finger-of-god (FOG) compression on the measured power spectrum.
From bottom to top, the three solid curves show 
the factor by which the measured fluctuation amplitude $\Pgg(k)^{1/2}$ is increased
by FOG compression with overdensity thresholds
$\infty$ (no compression), 200 (our baseline; horizontal line) 
and 100, respectively, using the disentanglement method.
The dashed curves show the same ratios for the modeling method.
To place these effects in context, the relative $1\sigma$ measurement errors on 
the power spectrum amplitude are indicated by the yellow/light grey band for the 
disentanglement method and by the cyan/grey band for the modeling
method. 
}
\end{figure}

To quantify the FOG effect on our recovered real-space power spectrum,
\fig{defog_powerFig} shows the ratio of the measured power spectrum amplitude
to its value with our baseline FOG compression. 
Just as we saw in \fig{chi2Fig}, nonlinearities become progressively more
important toward smaller scales.
Quantitatively, the disentanglement method is seen to 
be almost unaffected by FOG-compression.
Over the range $0.1h/\Mpc < k < 0.2h/\Mpc$ where the error bars are smallest,
changing the FOG compression threshold within the rather extreme
range $100-\infty$ changes the measured fluctuation amplitude by only about 1\%,
which should be compared to statistical error bars of 8\% 
or more.
The sensitivity of the modeling method to these nonlinear effects is slightly greater: 1\% 
at $k\sim 0.1h/\Mpc$ and 4\%
at $k\sim 0.2h/\Mpc$, again letting the FOG threshold vary across the rather
extreme range $100-\infty$.

\subsection{Results from Monte Carlo simulations}

We need to quantify how accurately what we measure, $\Pgg(k)$, recovers
what we really care about, \ie, the real space matter power spectrum $P(k)$.
Nonlinear clustering {\it per se} would not bias quadratic estimators of the power spectrum, but 
how much do non-linearities in the redshift-space distortions affect the results?
\Fig{defog_powerFig} shows
that the sensitivity of the $\Pgg(k)$-measurement to 
FOG nonlinearities is around $1\%$ at $k\sim 0.1h/\Mpc$, \ie, 
negligible compared to our statistical measurement errors.
Although fingers of god are perhaps the most important way 
in which nonlinear redshift distortions manifest themselves, 
mildly nonlinear effects on larger scales are also important
(Scoccimarro {\etal} 2001; Scoccimarro 2003).
To be prudent, we therefore complement the above-mentioned tests with a
Monte Carlo analysis in which the true $P(k)$ is known and we can directly determine
how well we recover it.

We use two suites of Monte Carlo simulations as summarized in Table 1.
The first consists of 275 simulations constructed with the PThalos code (Scoccimarro \& Sheth 2002), 
covering 1395 square degrees with an angular completeness map corresponding to the northern 
part of SDSS ({\tt sample9}, an earlier version of {\tt sample11}
discussed in Appendix~\ref{DataAppendix}. 
In short, this code is
a fast approximate method to build non-Gaussian density fields with the halo model.
It produces realistic correlation functions and includes non-trivial galaxy biasing
by placing galaxies within dark matter halos with a prescribed
halo occupation number as a function of halo mass.
The second suite of simulated surveys is based on the 
Hubble volume $\Lambda$CDM $n$-body simulation (Frenk {\etal} 2000; Evrard {\etal} 2002).
10 mock surveys were produced by sparse-sampling different parts of the simulation 
cube to reproduce the three-dimensional selection function $\nbar(\r)$ for
SDSS {\tt sample8},
so although these mock surveys include fully nonlinear gravitational clustering, 
they have trivial light-to-mass bias with $b=r=1$
(the ``galaxies'' are simply a random subset of the dark matter particles).

\begin{figure} 
\centerline{\epsfxsize=\figsize\epsffile{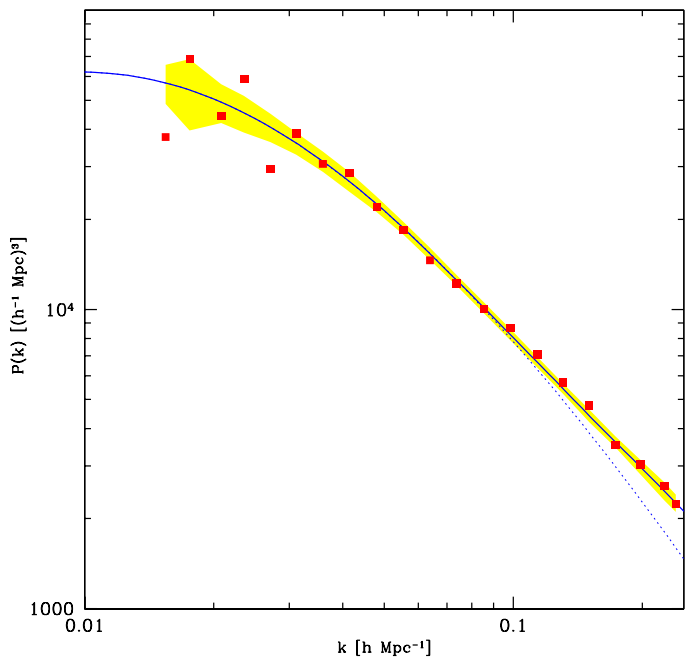}} 
\caption[1]{\label{roman_meanFig}\footnotesize%
Squares show the mean power spectrum $\Pgg(k)$ recovered from 62 PThalos simulations
by our analysis pipeline, using the disentanglement method without FOG compression.
If these squares faithfully measure the average real-space matter power spectrum of the simulations
(solid line), then about 68\% 
of them should lie in the shaded band around this curve, whose width is
given by the $1\sigma$ errors computed by our pipeline divided by the square root of the number of 
simulations.
For comparison, the dotted curve shows the linear power spectrum used as input for the simulations.
}
\end{figure}

\Fig{roman_meanFig} shows that the average $\Pgg(k)$ recovered using the
methodology described in this paper from the PThalos simulations
agrees with the matter $P(k)$ on all relevant scales to within the sensitivity we can test,
as expected given the above indications that the effect of nonlinearities on $\Pgg$ is small.
It also confirms that the analysis pipeline produces unbiased results (this was also 
demonstrated with extensive Monte Carlo simulations in THX02).
The mock surveys based on the Hubble Volume simulation give similar agreement, although 
with larger noise since there are only ten of them.

\begin{figure} 
\centerline{\epsfxsize=\figsize\epsffile{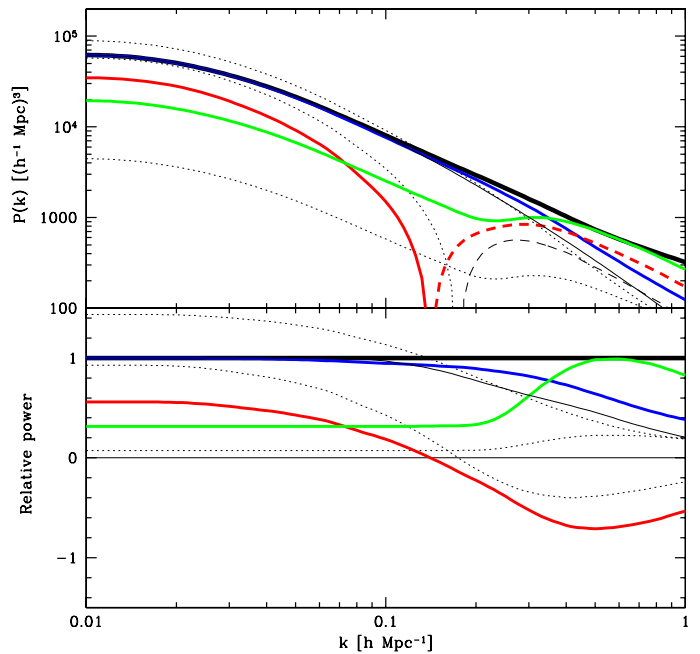}}
\caption[1]{\label{multipolesFig}\footnotesize%
Power spectra from 275 PThalos mock catalogs, 
quantifying how accurately our $\Pgg(k)$ statistic recovers the true power spectrum.
The thick black curve in the upper panel shows the real-space power spectrum that we wish to 
recover, on a logarithmic scale. The lower panel,
on a linear scale, shows the same curves as the upper panel, but divided by 
this reference model.
The thin solid black curve is the linear power spectrum that was taken
as input for the simulations.
The thick blue/dark grey, red/grey and green/light grey curves show the 
recovered $gg$, $gv$ and $vv$ power
spectra and the three dotted curves show the redshift space 
monopole, quadrupole and hexadecapole power from top to bottom on left hand side (see
\eq{multipole2flavorEq}), all curves being averages 
from 275 simulations using about $10^6$ galaxies each in the full 
simulation cube. Where they are negative, the $gv$ and quadrupole curves are plotted positive
and dashed in the upper panel.

The velocity dispersion is higher than in the 
real SDSS data and no FOG compression has been performed, 
so this should be viewed as a worst-case scenario.
Nonetheless, the figure shows that $\Pgg(k)$ (blue) recovers the true power spectrum (thick black)
to within a few percent at $k=0.1h$/Mpc even though strong departures from 
the Kaiser approximation (which predicts all curves being horizontal in the lower panel) 
are evident in the $\Pgv(k)$ curve (thick red/grey) on these scales.
The reason that our method works so well is that $\Pgg(k)$ recovers the transverse power
(which is unaffected by redshift distortions) beyond the Kaiser approximation,
requiring merely that $\l\ge 6$ anisotropies are negligible.
}
\end{figure}

Since the possible biases that we wish to quantify are so small (at the percent level), it is desirable to
have still more statistical testing power than these numerical
experiments provide.  In particular, we wish to test the breakdown of
the Kaiser approximation as a function of scale; here we are not
concerned with the effects of the survey geometry. 
For the 275 PThalos simulations, we therefore measure the various power spectra using 
all of the roughly $10^6$ galaxies in each the full simulation cubes.
(Since $\nbar(\r)$ is now constant and the boundary conditions are periodic, we do this 
by simply using fast Fourier transforms, matching to the Kaiser limit as $k\to 0$ to 
reduce sample variance; see Scoccimarro \& Sheth 2002).  No FOG
correction was applied to these simulations.    

The results are shown in \fig{multipolesFig}.  The upper panel (on a
logarithmic scale) shows the input power spectrum, the quantities
$\Pgg, Pgv$, and $Pvv$, as well as the monopole, 
quadrupole, and hexadecapole $P_0^s$, $P_2^s$, and $P_4^s$ as dotted
lines.  The lower panel shows (on a linear scale) the ratio of each of these
quantities to the input power spectrum.  In the absence of nonlinear
clustering and bias, each of these lines would be horizontal. 
We see that $\Pgg(k)$ agrees well with the real-space matter power spectrum $P(k)$ 
on large scales and progressively underestimates it more and more as $k$ increases.
Quantitatively, it is off by $4\%$ at $k\sim 0.1h/\Mpc$ and $6\%$ at $k\sim 0.2h/\Mpc$,
corresponding to $2\%$ and $3\%$ in fluctuation amplitude, respectively.
These numbers are thus in the same ballpark as those we found from varying the
FOG compression threshold above.

While the agreement is impressive, we note that PTHalos code may not
have a fully accurate radial distribution of galaxies inside halos, nor is 
the halo occupation number as a function of mass uniquely determined from 
the observations. For these reasons one should exercise caution when using these 
results in the nonlinear regime ($k\simgt 0.2h/\Mpc$), bearing in mind that different galaxy
distribution models may lead to larger differences between the nonlinear matter power
spectrum and $P_{gg}$.

Comparing \fig{multipolesFig} with figures~\ref{power_all3_binnedFig} and~\ref{chi2Fig},
it is striking that the simulations display stronger nonlinearity than the real data.
The simulations show $\Pgv(k)$ going negative for $k\simgt 0.14h/\Mpc$, whereas
the data
show no statistically significant detection of negative $\Pgv$ power on any scale probed.
This difference reflects the fact that the small-scale velocity dispersion in the PThalos simulations
is larger than those actually observed. In other words, our PThalos results should not 
be interpreted as our best estimate of how large the nonlinear problems are, but 
rather more as a worst-case scenario for the importance of nonlinear corrections.

In the Kaiser approximation, all curves in the lower panel of \fig{multipolesFig}
would be horizontal lines.
It is noteworthy that although the strong nonlinearities in the simulations cause 
the Kaiser approximation for $\Pmono(k)$ and $\Pquad(k)$ (dotted lines
in Figure~\ref{multipolesFig})
to break down on very large scales, $k\simgt 0.02 h/$Mpc,
the combination that represents $\Pgg(k)$ remains an accurate estimate of 
$\Pggtrue(k)$ down to much smaller scales.
We obtain similar results using the analytic halo model approach of Seljak (2001)
in place of our simulations. 
Scoccimarro (2003) shows that this is in fact a generic result: as long as the
wavenumber $k$ times the $rms$ pairwise velocity dispersion is smaller than
the Hubble parameter $H$, 
$\Pggtrue(k)$ is accurately approximated by 
\eq{multipole2flavorEq} even if the coefficients in this expansion are not
well approximated by the Kaiser formula\footnote{
Unfortunately, this useful result does not hold
for $\Pgv(k)$ or $\Pvv(k)$, so these two functions cannot be interpreted as simply the 
galaxy-velocity and velocity-velocity power when nonlinearities are important.
}.
This can be intuitively understood from the fact that 
$\Pgg(k)$ is equal to transverse power under all circumstances, linear or nonlinear,
as exploited in Hamilton \& Tegmark (2002).
As long as redshift distortions can be reasonably approximated by quadrupole
and hexadecapole distortions, then the arbitrary functions $\Pgv(k)$
and $\Pvv(k)$ contain enough freedom to model distortions completely,
even if they do not conform to the Kaiser model.

A third and final piece of evidence that nonlinearities have no major effect on our
measurement of the large-scale real-space power comes from a direct comparison of 
$\Pgg(k)$ recovered with our disentanglement and modeling methods. Although
the former has about twice as much scatter as the latter, the two measurements show excellent
agreement. There is no hint of systematic differences between the two on any scale.

The bottom line of this section is that although 
estimates of the redshift space distortions (estimates of $\beta$, 
the $gv/gg$ ratio, the quadrupole-to-monopole ratio, \etc)
are very sensitive to nonlinear effects, 
our estimates of the real-space matter power are {\it not}.
We have argued that any scale-dependent statistical bias in our $\Pgg(k)$ results 
due to nonlinear redshift distortions (or errors in our code)
is smaller than a few percent for $k<0.1h/\Mpc$
\ie, that the systematic errors associated with this are 
negligible compared with the statistical errors.

\clearpage

\section{Accounting for luminosity-dependent bias}
\label{BiasSec}

We have now measured the real-space power spectrum $\Pgg(k)$ of the SDSS galaxies, obtaining the
results shown in \fig{power_ggFig}. The goal of this section is to compute and apply a 
small ($\sim 10\%$)
scale dependent bias correction, producing a curve proportional to the
underlying matter power spectrum and usable for cosmological parameter estimation.  

As discussed in \sec{BiasPlanSec}, there is good reason to believe that bias 
is complicated on small scales, yet simple and essentially scale-independent on 
the extremely large scales $\lambda=2\pi/k\simgt 60 h^{-1}\Mpc$ that are the 
focus of this paper\footnote{On large scales, bias can also introduce an additive (as opposed to 
multiplicative) constant, related to halo 
shot noise, thereby affecting the shape of the power spectrum on
scales larger than the turnover
(Scherrer \& Weinberg 1998; Seljak 2001; Durrer {\etal} 2003).
Although this effect is negligible for $k\simgt 0.003h^{-1}\Mpc$, and
is therefore unimportant for the present paper, 
it may be important for the upcoming analysis of the SDSS luminous red galaxy (LRG) sample,
both because the halo shot noise effect is larger for such rare objects and because 
LRGs probe $P(k)$ out to larger scales than does the main SDSS galaxy sample analyzed here.
}.
However, since this scale-independent bias factor depends on
luminosity (among other galaxy properties),
we should expect to introduce an artificial scale-dependence of bias
from the magnitude-limited nature of our sample.

It is easy to understand how luminosity dependent clustering can masquerade as 
scale-dependent bias. 
Since luminous galaxies dominate the sample
at large distances and dim ones dominate nearby,
a measurement of $\Pgg(k)$ on very large scales is statistically dominated by
luminous galaxies whereas a measurement on small scales is dominated by dim ones 
(which have much higher number density). 
Since luminous galaxies cluster more
than dim ones, the measured power spectrum will therefore be redder than the 
matter power spectrum, with a lower ratio of small-scale to large-scale power.

Below we will quantify and correct for this effect. We emphasize that this is not intended to be
the mother of all bias treatments and the final word on the subject.
Rather, this artificial red-tilt is a small ($\sim 10\%$) effect
which has never previously been quantified, and we simply wish to make a
first order 
estimate of it. We start by measuring the luminosity dependence of bias using our volume-limited
subsamples in the next section, then use this to compute the scale-dependent effect.

It has been long known (Davis \& Geller 1976; Dressler 1980) that
galaxy bias depends on other galaxy properties as well, {\eg}, 
morphological type, color and environment.
Fortunately, the only intrinsic property which determines whether a galaxy gets included in 
our baseline sample is its luminosity, so we can ignore dependence on all
other properties for our present purposes (type dependence of clustering is 
of course a fascinating subject of its own, and will be the topic of 
future SDSS papers).

\begin{figure} 
\centerline{\epsfxsize=\figsize\epsffile{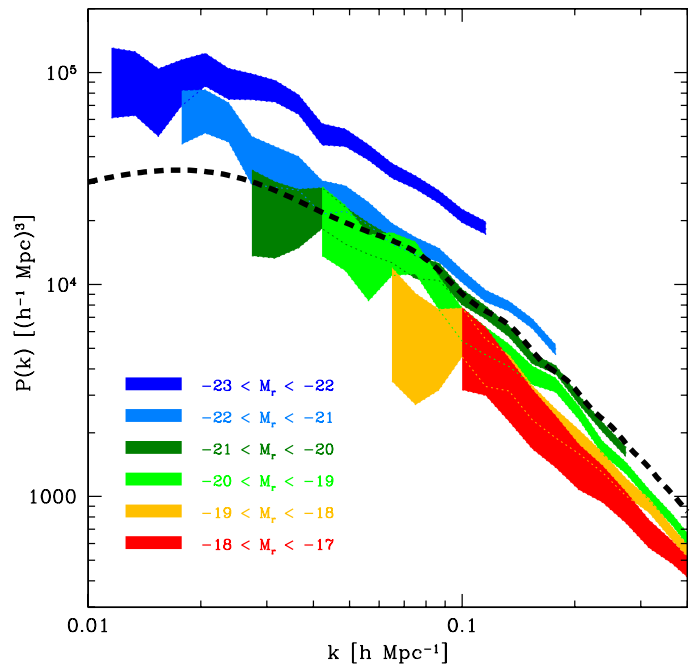}}
\caption[1]{\label{subsample_Pfig}\footnotesize%
The real-space power spectrum $\Pgg(k)$ is shown for galaxies in six bins of
absolute magnitude $\MM$ detailed in Table 1, with the shading indicating
$1-\sigma$ uncertainty.
All power spectra have roughly the same shape, increasing in 
amplitude as the galaxies become more luminous. The dashed curve is the best fit
linear $\Lambda$CDM model (see text) normalized to $\sigma_8=1$.  
}
\end{figure}

\begin{figure} 
\centerline{\epsfxsize=\figsize\epsffile{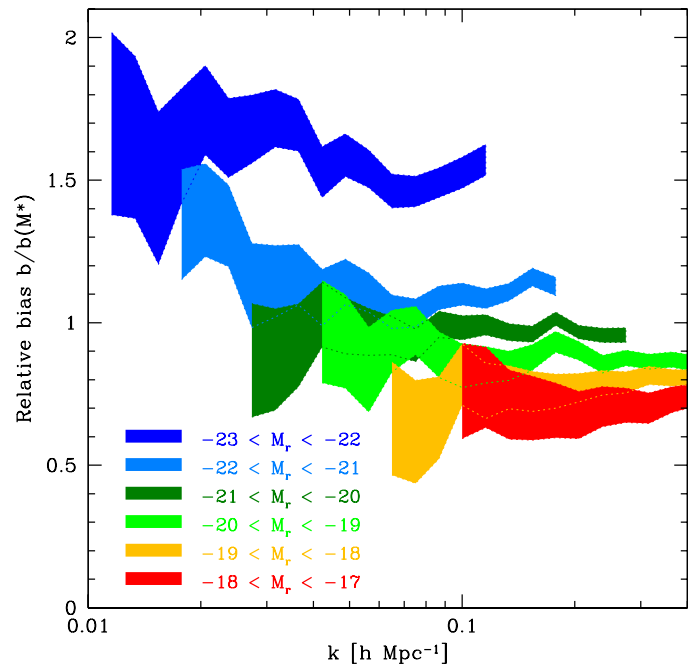}}
\caption[1]{\label{bias_allFig}\footnotesize%
The bias relative to the linear $\Lambda$CDM
model of the previous figure is shown 
for galaxies in six bins of
absolute magnitude $\MM$.
All six curves are consistent with being scale-independent,
the bias merely increasing in 
amplitude as the galaxies become more luminous.
}
\end{figure}

\subsection{Measurement of the luminosity-dependence of bias}

To quantify the luminosity-dependence of bias for the SDSS galaxies,
we repeat our entire analysis for each of the 
volume-limited samples L2-L7 specified in Table 1 and plotted in \sec{DataSec}
(samples L1 and L8 contain too few galaxies to be useful).
The resulting power spectra are shown in \fig{subsample_Pfig}.
To avoid excessive clutter in this figure, we plot the minimum variance 
power spectrum estimate described in Appendix~\ref{Msec}. 
To indicate that the measurement errors are correlated between $k$-bins, we show the
measurements as a shaded band rather than as separate points, with the vertical thickness of the band
corresponding to the $1\sigma$ uncertainty. (The bias fitting below 
uses the full covariance matrix and is of course independent 
of what plotting convention is used, as is $\chi^2$ computed with \eq{chi2Eq}.)

\Fig{subsample_Pfig} shows that all power spectra have roughly the same shape, 
increasing in amplitude as the galaxies become more luminous. 
This is seen more clearly in \fig{bias_allFig}, where we have divided them all by 
the linear power spectrum of the simple $\Lambda$CDM reference model described below.

\begin{figure} 
\centerline{\epsfxsize=\figsize\epsffile{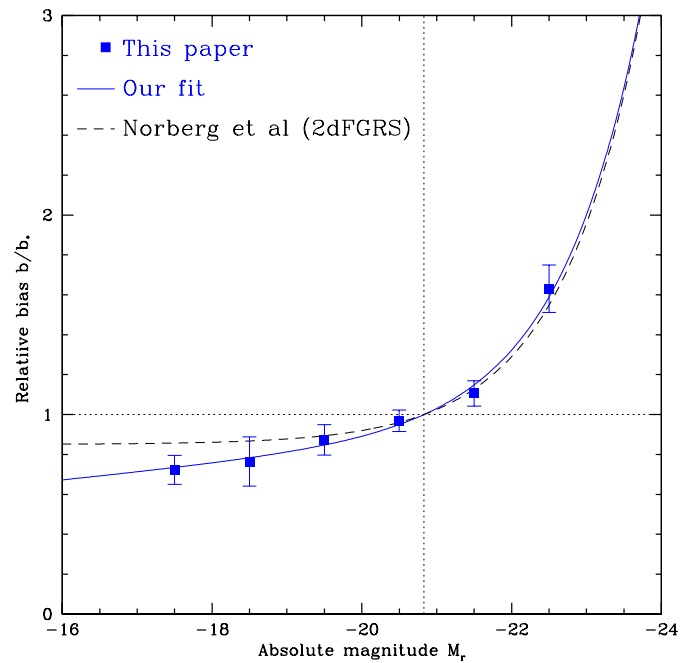}}
\caption[1]{\label{MbiasFig}\footnotesize%
The squares show the measured bias of our SDSS galaxies as a function of magnitude
relative to $\bstar$, the bias at $M*=-20.83$ (vertical dotted line).
The solid curve shows the fit to our measurements described in the text,
$b(M)/\bstar =  0.895 + 0.150 L/L_* - 0.040 (M-M_*)$,
and the 
dashed curve shows the corresponding fit of Norberg {\etal} based on 2dFGRS data,
$b(M)/\bstar =  0.85 + 0.15 L/L*$.
}
\end{figure}

To quantify this similarity of shapes, we fit each of the measured power spectra to the reference 
$\Lambda$CDM curve with the
amplitude freely adjustable. 
All six cases produce acceptable fits with reduced $\chi^2$ of order unity, and the
corresponding best-fit normalizations are shown in \fig{MbiasFig}.

We want our reference model to provide an accurate empirical characterization of the
SDSS data with as few parameters as possible.
We choose it to be a flat scale-invariant $\Lambda$CDM model with the baryon density 
$h^2\Ob=0.024$
preferred by WMAP (Bennett {\etal} 2003) 
and the Hubble parameter $h=0.72$ preferred by the HST key project
leaving $\Om$ as the only free parameter determining its shape.
We determine $\Om$ by the following iterative procedure: 
\begin{enumerate}
\item Given an $\Om$-value, we compute the reference model $P(k)$ normalized to $\sigma_8=1$.

\item Given the reference model, we fit for the six bias factors plotted in \fig{MbiasFig}.

\item We fit these bias factors to a smooth curve 
$b(M)/b_* =  A + B (L/L*) + C(M-M_*)$
shown in \fig{MbiasFig}
given by the three parameters $(A,B,C)$.

\item We compute the correction $\beff(k)$ for scale-dependent bias shown in 
\fig{power_ggFig} (top) as described below.

\item We compute the value of $\Om$ that best fits the bias-corrected measurements in 
\fig{power_ggFig} (bottom).

\end{enumerate}
This procedure converges to within floating-point numerical precision in merely a few
iterations for starting values anywhere in the range $0.1<\Om<1.0$, 
yielding $\Om=0.300$ and $(A,B,C)=(0.895, 0.150, -0.040)$.
The basic reason for this robustness is that changing the
shape of the fiducial model changes $\beff(k)^2$ by a much smaller amount,
because of the smearing by the integrals below.

\subsection{Correcting for the luminosity-dependence of bias}

\begin{figure} 
\centerline{\epsfxsize=\figsize\epsffile{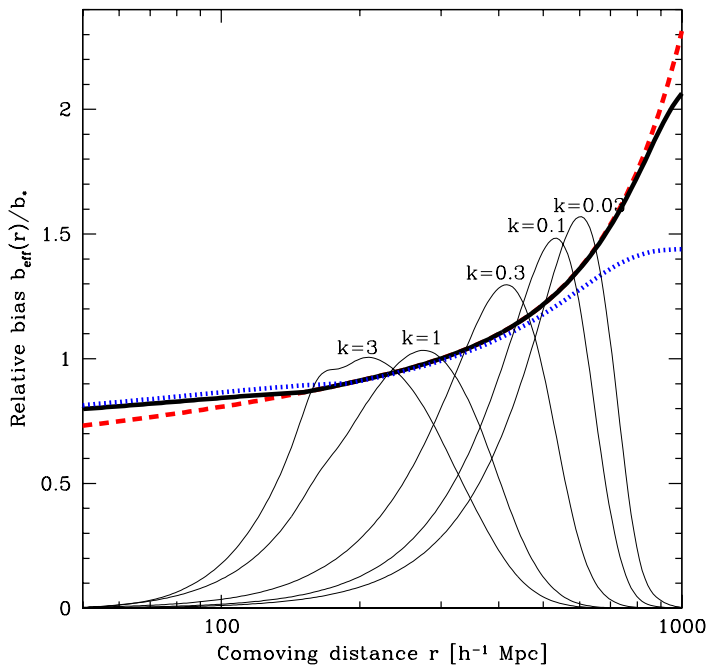}}
\caption[1]{\label{rbiasFig}\footnotesize%
The effective bias (equation~\ref{rbiasEq}; heavy curves) is seen to increase with distance, 
reflecting the fact that more distant galaxies are on average more luminous.
The curves become shallower as the range of absolute magnitudes
in the sample is cut, going from {\tt safe0} (dashed; no cuts) to 
{\tt safe13} (solid; $-23<\MM<-18.5$) to {\tt safe22} (dotted; $-22<\MM<-18$).
Our volume-limited samples simply have $\beff(r)$ constant.
The five thin curves show the relative weights given to different distances
when measuring $P(k)$ for $k=0.03$, $0.1$, $0.3$, $1$ and $3h/\Mpc$
using the {\tt safe13} radial selection function.
}
\end{figure}

\begin{figure} 
\centerline{\epsfxsize=\figsize\epsffile{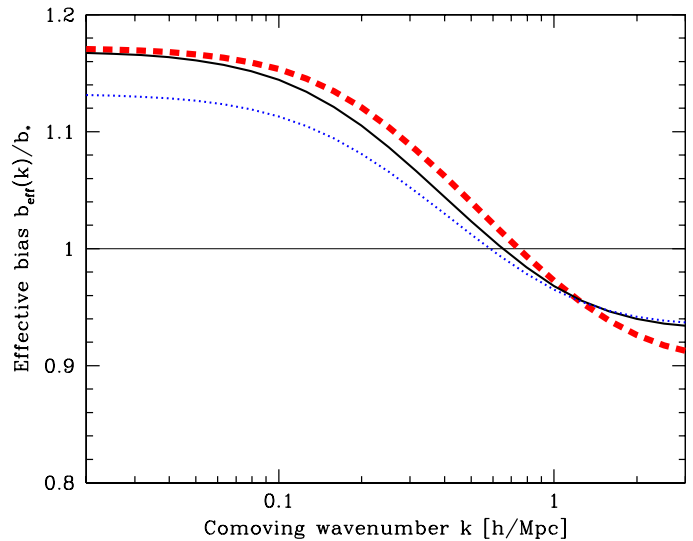}}
\caption[1]{\label{kbiasFig}\footnotesize%
The effective bias in the power spectrum measurement,
\eq{kbiasEq3}, is seen to
decrease with wavenumber $k$,  
reflecting the fact that low-$k$ measurements are dominated by
more distant and luminous galaxies.
The curves become shallower as the range of absolute magnitudes
in the sample is cut, going from {\tt safe0} (thick dashed; no cuts) to 
{\tt safe13} (solid; $-23<\MM<-18.5$) to {\tt safe22} (dotted; $-22<\MM<-18$).
}
\end{figure}

Above we quantified the well-known fact that the density field 
$\delta_M(\r)$ of galaxies of absolute magnitude $M$ is more strongly clustered
for larger luminosity (smaller absolute magnitude $M$). 
Let us consider the simple bias model
\beq{biasModelEq}
\delta_M(\r) = b(M)\delta(\r),
\eeq
where $\delta(\r)$ is the field of matter fluctuations 
and $b(M)$ is the luminosity-dependent bias factor 
proportional to what is plotted in \fig{MbiasFig}.
Since our observed galaxy sample mixes galaxies of various absolute magnitudes
with some probability distribution $f_M(M;r)$,
our observed density field can be written 
\beq{deltaobsEq}
\deltaobs(\r) = \int f_M(M;r)\delta_M(\r) dM.
\eeq
This probability distribution $f_M(M;r)$ is simply proportional 
to the galaxy luminosity function $\Phi(M)$ 
over the absolute magnitude range $\Mbri(r)<M<\Mdim(r)$ where the galaxy is observable at 
comoving distance $r$, zero otherwise, and normalized to integrate to unity.
$\Mbri(r)$ and $\Mdim(r)$ are given by \eq{eqn:mminmax} on inserting the appropriate
absolute magnitude cuts from Table 1
(for instance, the sample {\tt  safe13} has 
$\Mmin=-23$, $\Mmax=-18.5$).
Substituting \eq{biasModelEq} into \eq{deltaobsEq}, we obtain
\beq{deltaobsEq2}
\deltaobs(\r) = \beff(r)\delta(\r),
\eeq
where
\beqa{rbiasEq}
\beff(r) &=& \int f_M(M;r)b(M) dM\nonumber\\
         &=& {\int_{\Mdim(r)}^{\Mbri(r)} \Phi(M)b(M)dM \over \int_{\Mdim(r)}^{\Mbri(r)} \Phi(M)dM}.
\eeqa
We evaluate this expression using the SDSS luminosity function measured in Blanton {\etal} (2002).
The results are plotted in \fig{rbiasFig}, and the 
effective bias is seen to increase with distance as expected.
We see that the curve $\beff(r)$ become shallower as the range of absolute magnitudes
in the sample is cut, so the samples {\tt safe0}, {\tt safe13} and {\tt safe22} are progressively 
less affected. Our volume-limited samples by construction simply have
$\beff(r) = {\rm constant}$.

Bias is expected to depend not only on luminosity but also 
on time (Fry 1996; Tegmark \& Peebles 1998; Giavalisco {\etal} 1998;
Cen \& Ostriker 2000; Blanton {\etal} 2000).  In addition, the intrinsic matter clustering should increase over time.
Since the net result of these two counteracting effects is likely to be smaller than
the luminosity effect at the low redshifts ($z\simlt 0.2$) that we are considering, this is
difficult to measure separately. The same applies to the small effect from the time-dependence 
of the redshift-space distortion parameter $\beta$ caused by the time-dependence of 
$\Om$ and $\Ol$.
Indeed, since typical luminosity grows monotonically with distance,
the distance-dependent bias $\beff(r)$ plotted in \fig{rbiasFig} should be expected to approximately 
include the combination of all four effects, so that our analysis will to first order be corrected for 
all of them.

Let us now estimate how $\beff(r)$ translates into $k$-dependent bias in our measured power spectrum.
We do this in the FKP approximation (Feldman, Kaiser \& Peacock 1994). 
Here the density field $\deltaobs(r)$ is multiplied by a weight function
\beq{FKPweightEq}
\phi(r)\propto{\nbar(r)P(k)\over 1+\nbar(r)P(k)}
\eeq
and then Fourier transformed, giving 
\beq{FKPfourierEq}
\int \phi(r)\deltaobs(\r)e^{-i\k\cdot\r} d^3r  = 
\int \phi(r)\beff(r) \delta(\r) e^{-i\k\cdot\r} d^3r
\eeq
because of \eq{deltaobsEq2}, so we see that we are simply measuring 
the power spectrum using a modified effective weight function,
replacing $\phi$ by $\phi\beff$.
It is well-known (see Tegmark {\etal} 1998 for a review conforming to the present notation)
that the FKP estimate $\Phat(\k)$ of the three-dimensional power spectrum $P(\k)$ satisfies
\beq{FKPexpecEq}
\expec{\Phat(\k)}=(W\star P)(\k),
\eeq
\ie, that it 
probes the true power spectrum convolved with a window function $W(\k)$.
This window function is the square modulus of the Fourier transform of the weight function,
so it is given by 
\beq{W0eq}
W_0(\k)\equiv|\widehat{\phi}(\k)|^2, \quad W_1(\k)\equiv|\widehat{\phi\beff}(\k)|^2,
\eeq
where $W_0$ applies if we ignore bias and $W_1$ applies if we take bias into account.
According to \eq{FKPexpecEq}, galaxy bias therefore increases the measured power by a factor 
\beq{kbiasEq}
\beff(k)^2 = {(W_1\star P)(\k)\over (W_0\star P)(\k)}
= {\int W_1(\k')P(\k-\k')d^3k'\over \int W_0(\k')P(\k-\k')d^3k'}.
\eeq
The window function is normally narrower than the scale on which the power spectrum varies appreciably,
so we can make the approximation of taking it out of the convolution integral, obtaining simply
\beq{kbiasEq2}
\beff(k)^2 = {\int W_1(\k') d^3k'\over \int W_0(\k') d^3k'}
           = {\int \phi(r)^2\beff(r)^2 d^3r \over \int \phi(r)^2 d^3r},
\eeq
where we used Parseval's theorem 
\beq{ParsevalEq}
\int |\widehat{\phi}(\k)|^2 d^3k = (2\pi)^3 \int |\phi(r)|^2 d^3r
\eeq 
in the last step.
In summary, substituting \eq{FKPweightEq}, we have shown that the effective
bias as a function of wavenumber is 
\beq{kbiasEq3}
\beff(k) \approx \left[{
\int\left({\nbar(r)P(k)\over 1+\nbar(r)P(k)}\right)^2\beff(r)^2 r^3 d\ln r
\over
\int\left({\nbar(r)P(k)\over 1+\nbar(r)P(k)}\right)^2 r^3 d\ln r
}\right]^{1/2}.
\eeq
The resulting curves $\beff(k)$ are plotted in 
\fig{kbiasFig}. As expected, 
the effective bias is seen to decrease with wavenumber $k$, 
reflecting the fact that low-$k$ measurements are dominated by
more distant and luminous galaxies. Just as in the previous figure, 
the curves become shallower going from {\tt safe0} to {\tt safe13} to {\tt safe22},
as the range of absolute magnitudes
shrinks.

Note that if one treats $P(k)$ as a constant in \eq{kbiasEq3}, then
$\beff(k)$ becomes a constant independent of $k$.
Of the magnitude-limited galaxy survey analyses of the last decade, essentially
the only one using such a constant weighting was the 2dFGRS analysis by Percival {\etal} (2001).
Thus one can minimize the luminosity bias at the expense of increased statistical errors
due to suboptimal galaxy weighting (as shown by Feldman, Kaiser \& Peacock 1994, 
such uniform weighting is desirable only for volume-limited surveys where the
galaxy number density is constant). 
In this paper we have instead used minimum-variance methods to measure the luminosity bias 
and power spectrum jointly.
For a detailed discussion of these issues which appeared after the original version
of this paper was submitted, see Percival {\etal} (2003).

The bottom line of this section is that although luminosity-dependent bias
has only a small tilting effect on our measured SDSS power spectrum,
we can and should correct for it when doing precision cosmology, 
by simply dividing the measured power spectrum by
the square of the curve in the top panel of \fig{power_ggFig}.
This correction curve has a slope around
$-10\%$ per decade at $k\sim 0.2h/$Mpc. 
This means that fitting for cosmological parameters
ignoring this effect would give noticeably biased results.
For instance, assuming a power law primordial power spectrum of the form
$k^\ns$, this would correspond to shifting the best fit spectral index $\ns$
by an amount $\Delta\ns\approx -(2/\ln 10)\times 10\% \approx -0.1$,
and a more careful analysis in \sec{InterpretationSec} shows the net effect to be $-0.06$.
To place this in context, the popular stochastic eternal inflation model (Linde 1994) predicts 
$\ns\approx 0.96$, \ie, a substantially smaller 
departure from the $\ns=1$ scale-invariant case.

\section{Tests for systematic errors in the data}
\label{SystematicsSec}

The Monte Carlo experiments described in \sec{zspaceSec} provided an end-to-end
validation of our method and our software.
In this section, we turn to potential systematic errors in the SDSS data themselves.
Examples of such effects 
include radial modulations 
(due to mis-estimates of evolution or $K$-corrections)
and angular modulations (due to effects such as  
uncorrected dust extinction, variable observing conditions, 
photometric calibration errors and fiber collisions) of the density field.

\begin{figure} 
\centerline{\epsfxsize=\figsize\epsffile{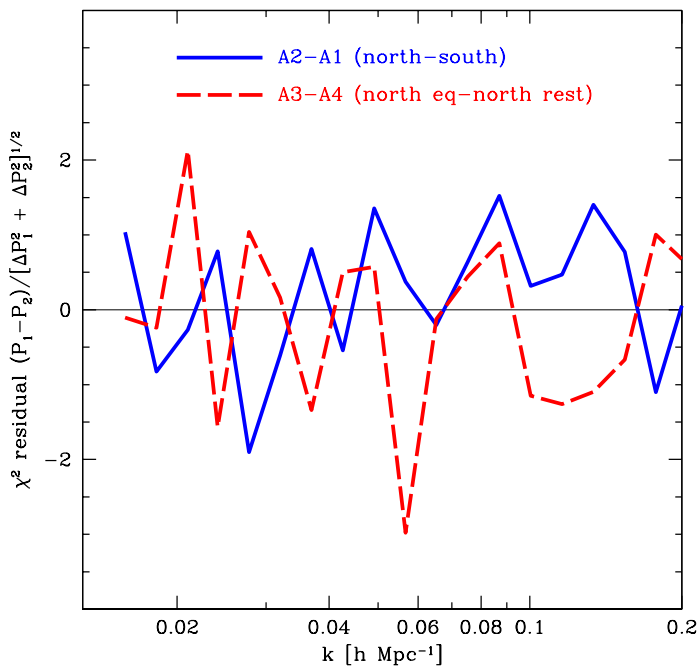}}
\caption[1]{\label{submask_diffFig}\footnotesize%
Comparison of power in different parts of the sky defined by the angular 
subsamples A1-A4 in Table 1.
Each curve shows the difference of two power spectra divided by the 
error bar on this quantity, so $\chi^2$ is simply the sum of the 
square of what is plotted. $\chi^2$ per degree of freedom is
0.86 for A2-A1 (north versus south) and
1.39 for A3-A4 (the two separate northern regions).
}
\end{figure}

\subsection{Analysis of subsets of galaxies}

Since such effects would be expected to vary across the sky (depending 
on, say, reddening, seasonally variable baseline offsets or 
observing conditions such as seeing and sky brightness), 
we repeat our entire analysis for four different angular subsets of the sky
(subsamples A1-A4 from Table 1) in search of inconsistencies.
We subtract the power spectrum measured south of the Galactic plane (A1)
from the power spectrum measured north of the Galactic plane (A2) for $k<0.2h/$Mpc
using the modeling method, 
and obtain a residual $\chi^2\approx 16$ for 19 degrees of freedom. 
A similar comparison of the two disjoint northern regions (A3 and A4)
gives a residual $\chi^2\approx 26$, again for 19 degrees of freedom.
Under the null hypothesis that a such a pair of curves are independent measurements 
of the same underlying power spectrum, the mean and standard deviation 
is $\expec{\chi^2}=19$ and $\Delta\chi^2=(2\times 19)^{1/2}\approx 6$, so 
these residuals are $-0.4$ and $+1.4$ standard deviations away from the expectation,
respectively. In other words, there is no significant evidence for discrepancies
between the power spectra measured in different parts of the sky.

The actual residuals are shown in \fig{submask_diffFig}. 
Since our measurements in different $k$-bands are uncorrelated, $\chi^2$ is 
simply the sum of the square of what is plotted. 
The most notable discrepancy is at
$k\sim 0.05h/$Mpc, where there
is more power in A4 than in A3. This appears to be related to a
striking wall-like structure that is seen in the northern galaxy distribution 
around $z\sim 0.08$ (see \fig{vlim_slice1Fig}). 
Although this ``great blob'' may be the largest coherent structure 
yet observed, a first crude estimate suggests that it is not inconsistent 
with Gaussian fluctuations: visual inspection of the 275 PThalos simulations reveals
similar structures in more than $10\%$ 
of the cases.

A similar comparison of the power spectra in the radial subsamples R1-R3
is less useful, since this radial binning is largely degenerate with the luminosity binning
(\sec{BiasSec}),
so we test for radial systematics with mode subsets instead.

\begin{figure} 
\centerline{\epsfxsize=\figsize\epsffile{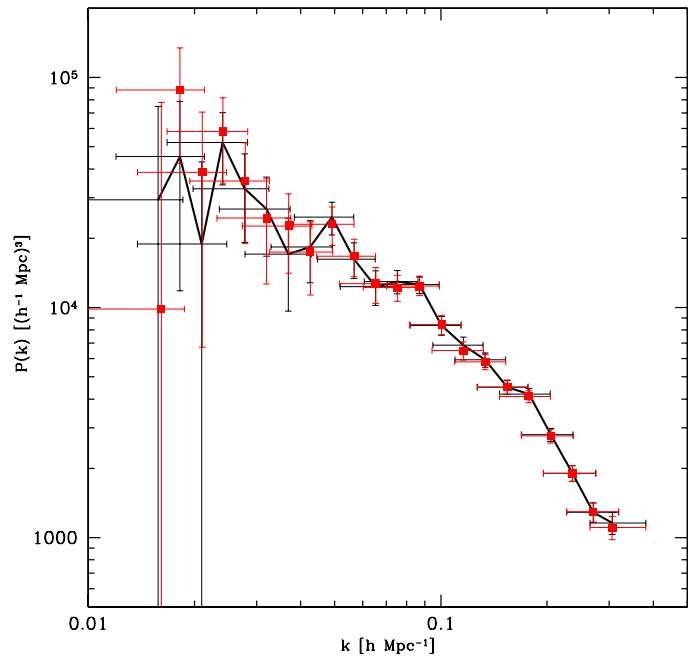}}
\caption[1]{\label{zapFig}\footnotesize%
Effect of removing special modes.
Black curve with associated error bars shows our measured power 
spectrum $\Pgg(k)$ from the modeling method using all 4000 PKL modes.
Red squares with error bars show the effect of removing the 234 
special modes corresponding to purely angular and purely
radial fluctuations as well as fluctuations associated with the motion of 
the Local Group relative to the CMB rest frame.
Any systematic errors adding power to these special modes would cause the 
curve to lie systematically {\it above} the squares. 
}
\end{figure}

\subsection{Analysis of subsets of modes}

Because of their angular or radial nature, all 
potential systematic errors discussed above
create excess power mainly in the radial and angular modes.
To quantify any such excess, we therefore repeat our entire analysis with 
all 233 special modes (27 radial modes, 199 angular modes and 7 Local Group modes) deleted.
The results of this test are shown in \fig{zapFig} and are very
encouraging; the differences are tiny.  
Any systematic errors adding power to these special modes would cause the 
squares to lie systematically {\it above} the crosses, yet the squares fall below 
the crosses for four out of the five leftmost bands, where such systematics would
be most important.  Thus there is no indication of excess radial or
angular power in the data. 

\Fig{zapFig} shows that removing the special modes 
results in a slight error bar increase 
on the largest scales and essentially no change on smaller scales.
This can be readily understood geometrically.
If we count the number of modes that probe mainly scales $k<k_*$,
then the number of purely radial, purely angular and arbitrary modes
will grow as $k_*$, $k_*^2$ and $k_*^3$, respectively.
This means that ``special'' modes (radial and angular) will make up a larger fraction
of the total pool on large scales (at small $k$), and that the purely 
radial ones will be outnumbered by the purely angular ones.

Our treatment of spectroscopic fiber collisions described in Appendix~\ref{DataAppendix}
is another source of potential angular/radial problems.
By assigning both members of some close pairs (separated by less than $55''$,
corresponding to $0.08 h^{-1}$Mpc at the mean depth of the survey)
at the same redshift, we {\it overestimate} the radial power on small scales.
Zehavi {\etal} (2002) perform extensive tests of this effect and show that
it is negligible on large scales considered in this paper.
As a further cross-check, we repeat our entire 
calculation with all galaxies with such assigned redshifts removed.
Since this second approach is guaranteed to {\it underestimate} the power, 
the two approaches will bracket the correct answer. As expected based on the
Zehavi {\etal} (2002) analysis, we 
find no evidence that fiber collisions are boosting our measured
power spectrum on the smallest scales we probe ($k\sim 0.3h/$Mpc).

\section{Discussion and conclusions}
\label{ConcSec}
\label{DiscussionSec}

\def\pgg{\Pgg(k)}
\def\pgv{\Pgv(k)}
\def\pvv{\Pvv(k)}
\def\pggtrue{P^{\rm true}_{\rm gg}(k)}

\subsection{Basic results}

We have measured the shape of the real-space power spectrum $P(k)$ on large scales
using the SDSS galaxy redshift catalog, paying particular attention to
quantifying and correcting for the effects of survey geometry, redshift space distortions
and luminosity-dependent bias.
Our principal results are the estimates $\Pgg(k)$ of the
real space galaxy power spectrum which are listed in Tables 2 and 3 for the 
disentanglement and modeling methods, respectively.
As discussed in \sec{zspaceSec},
the disentanglement method is more robust, but the modeling method (\fig{power_ggFig})
yields smaller statistical errors and appears in our tests to introduce
little systematic error. Table 3 lists results both for FOG compression with
$\delta_c=200$, which we consider the most reliable choice for
estimating the true real space power spectrum, and for no FOG compression,
which is the case easiest to model in detail. 
Our estimation procedure
yields uncorrelated error bars, so the reported errors in these
tables can be used as a diagonal covariance matrix when evaluating
the likelihood for model fits. For such fits, it is crucial to use the 
exact window functions, which are available at\\
{\it http://www.hep.upenn.edu/$\sim$max/sdss.html}
together with sample software for evaluating the SDSS likelihood function. 

As noted in \sec{zspaceSec},
uncertainties in the values of $\beta$ and $r$ leave a $4\%$ $1\sigma$
uncertainty in the overall normalization of $\Pgg(k)$ with the modeling method, in addition to
the error bars on individual points. There is no corresponding 
normalization uncertainty for the disentanglement method.
Our tabulated power measurements have all been corrected for the effect 
of luminosity-dependent bias as discussed in \sec{BiasSec}. The 
correction $b(k)$ used is given in Tables 2 and 3, and is normalized so that 
our quoted power measurements represent the power
spectrum of galaxies with absolute r-band magnitude $M_*\approx -20.83$;
the relative bias of galaxies as a function of luminosity can
be found in \fig{MbiasFig}.

\subsection{Using our results}
\label{RecipeSec}

There are several levels at which one might use our results, depending
on how conservative one wishes to be and how much energy one has for
theoretical modeling.  The mock catalog tests in Section 4.2 (\fig{multipolesFig} in
particular) suggest that our method is quite successful at correcting
for redshift space distortions to recover the real space galaxy
power spectrum $\pggtrue$.  However, there are notable departures from
perfect recovery at $k\simgt 0.15h/$Mpc, and the tests are in
any event carried out for a particular choice of cosmology and galaxy
bias model. The simplest and least conservative way to use our results
is to assume that we have indeed recovered $\pggtrue$ and to {\it further}
assume that on the scales of our measurement the galaxy power spectrum
is a scale-independent multiple of the linear theory matter power
spectrum, $\Pgg(k) = b_*^2 P(k)$ where $b_*$ represents the large scale
bias factor of $L_*$ galaxies.  The agreement of the lines representing
$\Pgg(k)$ and the linear $P(k)$ in \fig{multipolesFig} suggests that this approach
is probably safe for $k \simlt 0.1 h/{\rm Mpc}$, and that one can use the
more precise modeling estimates of $\Pgg(k)$ (Table 3) without incurring a
systematic error that is significant relative to the statistical
errors of the current measurement.  However, our tests are not exhaustive,
and it is possible that the agreement of $\Pgg(k)$ and linear $P(k)$ shapes
in \fig{multipolesFig} arises in part from a cancellation of non-linear gravitational
effects with errors in the redshift-space distortion correction.
This cancellation, in turn, might not hold for other cosmological or
galaxy bias models.

A second option is to assume that we have recovered $\pggtrue$ but not
assume that this has the same shape as the linear theory matter power
spectrum.  Here, for example, one could use N-body simulations or
analytic approximations to compute the non-linear, real space power spectrum,
incorporating galaxy bias based on semi-analytic galaxy formation calculations,
hydrodynamic simulations, or a ``halo occupation'' model constrained
by other measurements of galaxy clustering.
\Fig{multipolesFig} again suggests that this approach can be used safely for
$k \simlt 0.1 h/$Mpc (and perhaps a bit further) without systematic
errors that exceed the statistical errors.
One can also use the non-linear matter power spectrum and a constant $b$,
but there is good reason to expect scale-dependent bias
on scales where non-linearity is significant (Hamilton \& Tegmark 2002).
Finally, the most cumbersome but most reliable way to use our data
is to follow the course suggested at the end of \sec{zspaceSec}: create redshift-space
realizations using non-linear simulations with galaxy bias, compute
the monopole, quadrupole, and hexadecapole components of the redshift
space power spectrum in the distant observer approximation, and use
\eq{multipole2flavorEq} to convert them to $\Pgg(k)$.
These predictions should be compared directly to the disentanglement
estimates of $\Pgg(k)$, since the redshift-space distortions have been
incorporated into the model rather than removed from the data.
However, by focusing on a quantity $\Pgg(k)$ that responds mainly
to real space clustering (exactly so in the linear regime), such
a comparison will be insensitive to moderate errors in the redshift-space
distortions incorporated in the theoretical predictions.
This last approach is still much simpler than creating artificial
SDSS catalogs and reproducing our estimation method in its entirety,
but it should be equally valid.

\begin{figure} 
\centerline{\epsfxsize=\figsize\epsffile{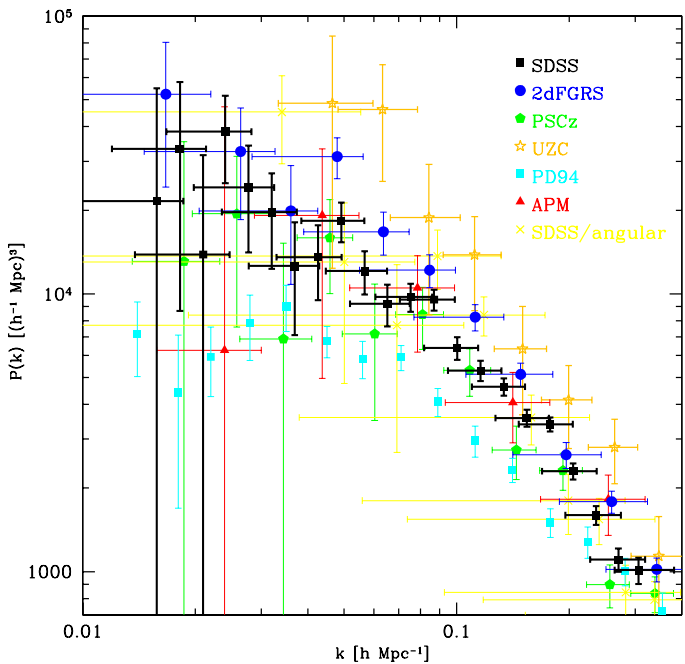}}
\caption[1]{\label{P1Fig}\footnotesize%
Comparison with other galaxy power spectrum measurements.
Numerous caveats must be borne in mind when interpreting this
figure. 
Our SDSS power spectrum measurements are those from \fig{power_ggFig},
corrected for the red-tilting effect of luminosity dependent bias.
The purely angular analyses of the 
APM survey (Efstathiou \& Moody 2001)
and the SDSS (the points are from Tegmark {\protect\etal} 2002
for galaxies in the magnitude range $21<r^*<22$  --- see also 
Dodelson {\protect\etal} 2002) should also be free of this effect,
but represent different mixtures of luminosities.
The 2dFGRS points are from the analysis of HTX02, and like the 
PSCz points (HTP00) and the UZC points (THX02) have not been corrected
for this effect, whereas the Percival {\etal} 2dFGRS analysis should
be unafflicted by such red-tilting. The influential PD94 points
(Table 1 from Peacock \& Dodds 1994),
summarizing the state-of-the-art a decade ago, are shown assuming IRAS bias of unity
and the then fashionable density parameter $\Omega_m=1$.
}
\end{figure}

\begin{figure} 
\centerline{\epsfxsize=\figsize\epsffile{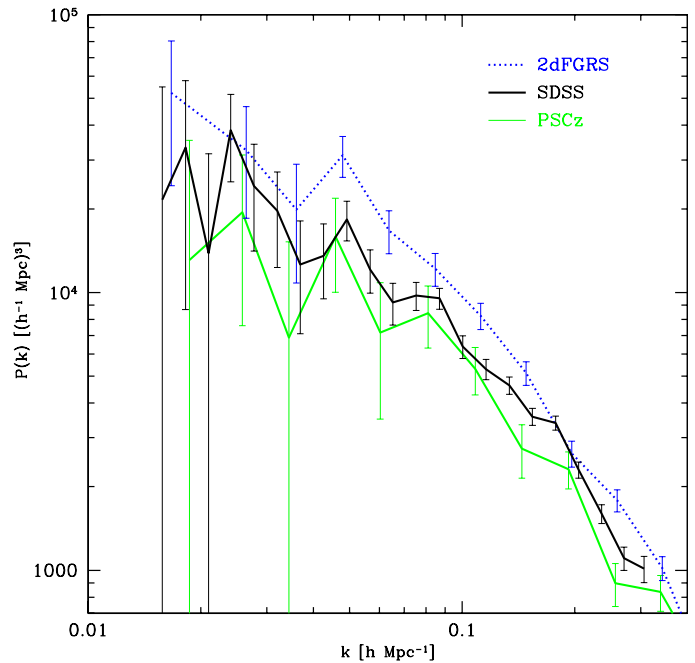}}
\caption[1]{\label{P3Fig}\footnotesize%
Same as Figure~\ref{P1Fig},
but restricted to a comparison of decorrelated power spectra,
those for SDSS, 2dFGRS and PSCz.
The similarity in the bumps and wiggles between the three power spectra
is intriguing.
}
\end{figure}

\begin{figure} 
\vskip\smtopskip
\bigskip
\centerline{\epsfxsize=\figsize\epsffile{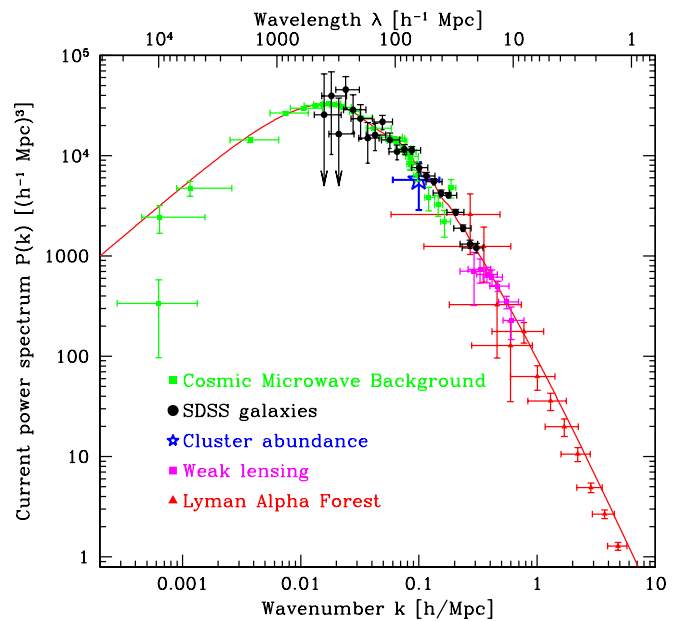}}
\vskip\smbotskip
\caption[1]{\label{pdataFig}\footnotesize%
Comparison of our results with other $P(k)$ constraints.
The location of CMB, cluster, lensing and Ly$\alpha$ forest points in this
plane depends on the cosmic matter budget (and, for the CMB,
on the reionization optical depth $\tau$), so requiring consistency
with SDSS constrains these cosmological parameters without
assumptions about the primordial power spectrum. This figure is for the case 
of a ``vanilla'' flat scalar scale-invariant model with $
\Om=0.28$, $h=0.72$ and $\Omega_b/\Omega_m=0.16$,
$\tau=0.17$ (Spergel {\etal} 2003; Verde {\etal} 2003, Tegmark {\etal} 2003b), 
assuming $b_*=0.92$ for the SDSS galaxies.
}
\end{figure}

\begin{figure} 
\centerline{\epsfxsize=\figsize\epsffile{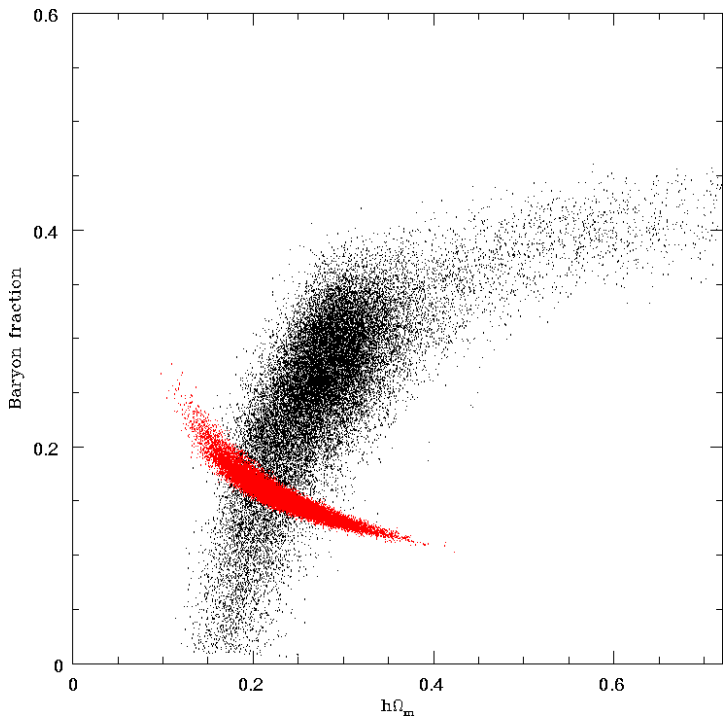}}
\caption[1]{\label{GammafbFig}\footnotesize%
Characterization of the SDSS power spectrum in terms on 
constraints on the ``shape parameter'' $h\Omega_m$ and the baryon 
fraction $f_b$. Black points show 100{,}000 Markov Chain steps for
SDSS, red/grey points are for WMAP data.
}
\end{figure}

\subsection{Comparison to other results}
\label{InterpretationSec}

Figure~\ref{P1Fig}
compares our results from Table 3 (modeling approach) 
with other measurements from galaxy surveys,
but must be interpreted with care. 
The UZC points may contain excess large-scale power due to selection
function effects (Padmanabhan {\etal} 2000; THX02), and the angular SDSS points measured
from the early data release sample are difficult to interpret
because of their extremely broad window functions.
Only the SDSS, APM and angular SDSS points can be interpreted as measuring the
large-scale matter power spectrum with constant bias, since the others have not been 
corrected for the red-tilting effect of luminosity-dependent bias. 
The Percival {\etal} (2001) 2dFGRS analysis 
unfortunately cannot be directly plotted in the
figure because of its complicated window functions.  

Figure~\ref{P3Fig}
is the same as Figure~\ref{P1Fig},
but restricted to a comparison of decorrelated power spectra,
those for SDSS, 2dFGRS and PSCz.
Because the power spectra are decorrelated,
it is fair to do ``chi-by-eye'' when examining this Figure.
The similarity in the bumps and wiggles between the three power spectra is quite striking.
Moreover, there is an interesting degree of similarity with
the power spectrum of the Abell/ACO cluster catalog (not shown) reported by
Miller \& Batuski (2001).
It is tempting to see hints of baryonic oscillations in these wiggles.
Indeed
Percival {\etal} (2001) in their analysis of the 2dFGRS,
and
Miller {\etal} (2001a,b; see also Miller {\etal} 2002)
in their analysis of
the Abell/ACO cluster (Miller \& Batuski (2001),
APM cluster (Tadros, Efstathiou \& Dalton 1998),
and PSCz surveys (HTP00),
already concluded that their data
mildly preferred model power spectra with baryonic oscillations
over those without.
However,
the oscillations at large scales evident in Figure~\ref{P3Fig}, notably
the dip at $k \sim 0.035 \, h \Mpc^{-1}$ and bump at $k \sim 0.05 \, h \Mpc^{-1}$,
are substantially larger than predicted by the standard $\Lambda$CDM concordance model
illustrated in Figure~\ref{power_ggFig};
if confirmed, such a feature
would challenge the $\Lambda$CDM concordance model
with scale-invariant initial conditions.
This preference for a large baryon fraction is also seen in \fig{GammafbFig}
(details below) which, however, shows that our SDSS data is nonetheless perfectly 
consistent with the concordance baryon fraction --- about one sixth of the 
100{,}000 points shown have a baryon fraction below the WMAP value of $17\%$.

\Fig{GammafbFig} also illustrates why galaxy clustering data is so complementary to
CMB measurements. The 100{,}000 red/grey points are from a Monte Carlo Markov Chain 
analysis of the WMAP for simple flat scalar adiabatic models parametrized by
the densities of dark energy, dark matter and baryronic matter, the spectral index and 
amplitude, and the reionization optical depth. 
As emphasized by Eisenstein {\etal} (1999) and Bridle {\etal} (2003), WMAP alone cannot 
determine $\Om$ to better than a factor of two or so because of a strong degeneracy with 
other parameters. Fortunately, the WMAP degeneracy banana in \fig{GammafbFig}
is seen to be almost orthogonal to the SDSS degeneracy, which means that combining the two 
measurements dramatically tightens the constraints on all the parameters involved in the
degeneracy --- notably $\Om$, $h$ and $\sigma_8$.

To place our SDSS results in a larger context, 
\fig{pdataFig} compares them with other 
measurements of the matter power spectrum $P(k)$. Here the CMB, galaxy cluster, lensing
and Ly$\alpha$ forest results have been mapped into $k$-space using the method of
Tegmark \& Zaldarriaga (2002), assuming the WMAP model given in the caption, and we have assumed 
an SDSS bias $b_*=1$. 
The CMB data combines the Boomerang, MAXIMA, DASI, CBI, VSA, ACBAR and WMAP data 
as in Wang {\etal} (2002) with the WMAP measurements (Hinshaw {\etal} 2003). 
The cluster point 
reflects the spread in the recent literature rather than any one quoted measurement.
The lensing data are from 
Hoekstra {\etal} (2002). The Ly$\alpha$ forest points are from the
Gnedin \& Hamilton (2002)
reanalysis of the Croft {\etal} (1999) data.

We leave detailed investigation of the implications of our measurement
to other papers (by ourselves in Paper II and, we hope, by others), since the primary goal of 
this paper is the measurement itself.
As a characterization of our data, we will
briefly indulge in the simplest of the interpretive approaches described in 
\sec{RecipeSec}.
For this purpose, we fit our 22 $\Pgg(k)$-measurements derived from the modeling method
with $\delta_c=200$ using the linear CDM power spectrum of
Eisenstein \& Hu (1999), fixing the baryon fraction
$\Omega_b/\Omega_m=0.17$ and the Hubble parameter $h=0.72$ as per the best fits from 
WMAP (Bennett {\etal} 2003; Spergel {\etal} 2003; Verde {\etal} 2003)
and no massive neutrino contribution.
If we further fix the
inflationary spectral index to $\ns=1$, then the shape of $P(k)$
is determined by the combination $h\Om$, and we find
$h\Om =0.201\pm 0.017$ at $1\sigma$, \ie, $\Om=0.300\pm 0.018$.

As discussed above, our modeling of nonlinear redshift space distortions is
only accurate on large scales, so we recommend not using the measurements 
with $k>0.2h/\Mpc$ for cosmological analysis.
In this spirit, we fit the 19 $\Pgg(k)$-measurements for $k<0.2h/\Mpc$ 
to the two-parameter model defined by the 
Smith {\etal} (2003) nonlinear power spectrum approximation using the 
Eisenstein \& Hu (1999) fit as above for the linear transfer function,
fixing the bias $b_*=1$.
This gives the shape parameter $h\Om =0.213\pm 0.0233$ at $1\sigma$, \ie, $\Om=0.295\pm 0.0323$.
This fit has $\chi^2=15.6$ for $19-2=17$ effective degrees of freedom, 
so this two-parameter model fit
can be regarded as an adequate representation of our results in
compact summary form. (Using a linear power spectrum model would increase this $\Om$-value by $0.043$.)
\Fig{GammafbFig}, which was commented on above, shows the result of repeating this same 
fit after adding the baryon fraction as a third free paramenter.
Fixing the best-fit shape $h\Om=0.213$, the power spectrum amplitude
corresponds to $\sigma_8=0.89\pm 0.02$ for $L^*$ galaxies after 
marginalizing over the redshift-space distortion parameters
$\beta$ and $r$. Note that this normalization is at the effective
redshift of the survey, not for $z=0$ galaxies.

If we fix $\Omega_m$ at the best-fit value $0.291$
and treat the spectral index as the free shape parameter,
then we find $\ns=0.995\pm 0.049$. 
Without our correction for luminosity-dependent bias, the corresponding numbers
are $\ns=0.933\pm 0.046$, so the statistical errors are now small enough for effects such as this 
to be important. Similarly, Table 7 of Paper II shows that ignoring this correction reduces the measured
value of $\Omega_m$ by $0.035$.

Paper II presents a thorough analysis of the cosmological constraints 
from our $P(k)$-measurement, finding them in good agreement with a 
``vanilla'' flat adiabatic $\Lambda$CDM model with neglibible  
tilt, running tilt, tensor modes or massive neutrinos.
Our $P(k)$ measurement provides a powerful confirmation of the results reported by the WMAP 
team, and more than halves the WMAP-only error bars on some parameters,
\eg, the matter density $\Om$ and the Hubble parameter $h$.
Paper II finds $\Om=0.30\pm 0.04$ from WMAP+SDSS when marginalizing over the 
other ``vanilla'' parameters. This is about 
$1\sigma$ higher that when using the 2dFGRS survey (which gave a slightly redder
$P(k)$ slope than we found) --- just the sort of statistical difference one would expect
from two completely independent samples.

\subsection{Outlook}

Let us conclude by looking ahead. 
Galaxy surveys have the potential to greatly improve the cosmological constraints
from the cosmic microwave background by breaking degeneracies and providing cross-checks,
so detailed joint analysis of our measurements with WMAP and other data sets will be worthwhile.
In particular, detecting the effect of baryons on $P(k)$ 
(Tegmark 1997a; Goldberg \& Strauss 1998) can provide powerful constraints
on the Hubble parameter (Eisenstein {\etal} 1998) and accurate 
determination of the shape of $P(k)$ can place strong constraints on 
neutrino masses (Hu {\etal} 1998; Spergel {\etal} 2003; Hannestad 2003) and
help pin down the primordial power spectrum. 
Deeper surveys can also provide interesting constraints on the evolution of clustering and dark energy, 
and the SDSS luminous red galaxy (LRG) sample and photometric redshift catalog will complement 
specialized deep redshift surveys such as DEEP (Davis {\etal} 2001) and VIRMOS 
(Le F\`evre {\etal} 2001) in this regard.

Prospects are also good for reducing systematic uncertainties involving both bias and redshift distortions.
A key virtue of having very large galaxy samples is it permits accurate measurements for 
large numbers of subsamples.
For instance, repeating our analysis for subsamples based on galaxy color or spectral type
will provide a powerful test of how scale-independent the bias is on large
scales.
Moreover, empirical constraints from SDSS on redshift-space distortions should improve substantially.
These constraints are currently rather weak 
because the survey geometry consists largely of thin wedges; we have
therefore focused simply on modeling distortions well enough
to remove their impact on the real space $P(k)$ estimate.
As the survey area fills in and becomes more contiguous, 
we expect to obtain interesting constraints on redshift distortions that can be used to 
test and refine theoretical and numerical models.

In addition to more careful modeling and combining with other
observational constraints, we anticipate several complementary
results from the SDSS in the near future, such as cosmological constraints 
directly from KL-modes (Pope {\etal} 2003),
real space clustering on small scales from the projected correlation function $w(r_p)$
(Zehavi {\etal} 2003b),
power spectrum measurements
on large scales using the luminous red galaxy sample and 
angular clustering measurements using photometric redshifts (Budavari {\etal} 2003).
This should help break degeneracies and provide cross-checks to 
{\em test} rather than {\em assume} the physics underlying the cosmological model,
thereby strengthening the weakest link in post-WMAP cosmology.

\bigskip
We wish to thank Adrian Jenkins and Carlton Baugh for providing Hubble Volume simulation results
and Scott Dodelson for useful suggestions.
MT thanks Ang{\e}lica de Oliveira-Costa for helpful comments and infinite patience.

Funding for the creation and distribution of the SDSS Archive has been provided
by the Alfred P. Sloan Foundation, the Participating Institutions, the National
Aeronautics and Space Administration, the National Science Foundation, the U.S.
Department of Energy, the Japanese Monbukagakusho, and the Max Planck Society. 
The SDSS Web site is http://www.sdss.org/. 

The SDSS is managed by the Astrophysical Research Consortium (ARC) for the
Participating Institutions.  The Participating Institutions are The University
of Chicago, Fermilab, the Institute for Advanced Study, the Japan Participation
Group, The Johns Hopkins University, Los Alamos National Laboratory, the
Max-Planck-Institute for Astronomy (MPIA), the Max-Planck-Institute for
Astrophysics (MPA), New Mexico State University, University of Pittsburgh, 
Princeton University, the United States Naval Observatory, and the University
of Washington.

MT was supported by NSF grants AST-0071213 \& AST-0134999, NASA grants
NAG5-9194 \& NAG5-11099 and fellowships from the David and Lucile
Packard Foundation and the Cottrell Foundation.  MAS acknowledges
support from NSF grant AST-007109,
and AJSH from NSF grant AST-0205981 and NASA grant NAG5-10763.
Our PThalos mock catalogs were
created on the NYU Beowulf cluster supported by NSF grants
PHY-0116590 and PHY-0101738, and NASA grant NAG5-12100.

\clearpage

\appendix

\section{Data and data modeling}
\label{DataAppendix}

In this Appendix, we provide a detailed description of how our basic galaxy sample
was processed, modeled and split into the subsamples used in our power spectrum analysis.

\subsection{The SDSS Galaxy Catalog}

The SDSS (York {\etal} 2000) is producing imaging and spectroscopic surveys
over about a quarter of the Celestial Sphere. A dedicated 2.5m
telescope at Apache Point Observatory, Sunspot, New Mexico, images the
sky in five bands between 3{,}000\Angstrom\ and 10{,}000\Angstrom\ ($u$, $g$, $r$, $i$,
$z$; Fukugita {\etal} 1996) using a drift-scanning, mosaic CCD camera
(Gunn {\etal} 1998), detecting objects to a flux limit of $r\sim 22.5$.  The
photometric quality of the observations are tracked using an automatic
photometricity monitor (Hogg {\etal} 2001).  One of the goals is to
spectroscopically observe 900{,}000 galaxies, (down to
$r_{\mathrm{lim}}\approx 17.77$; Strauss {\etal} 2002), 100{,}000 Luminous Red Galaxies
(LRGs; Eisenstein {\etal} 2001), and 100{,}000 QSOs (Richards {\etal} 2002).  
This spectroscopic follow-up uses two
digital spectrographs (Uomoto {\etal} 2003) on the same telescope as the
imaging camera. 
Other aspects of the survey are described in the Early Data
Release (EDR) paper (Stoughton {\etal} 2002).

The SDSS images are reduced and catalogs are produced by the SDSS
pipeline {\tt photo} (Lupton {\etal} 2001), which detects and measures
objects, the sky background, and the seeing conditions.  As described
in Smith {\etal} (2002), magnitudes are calibrated to a standard star
network approximately in the $AB$ system. The astrometric calibration
is also performed by an automatic pipeline which obtains absolute
positions to better than 0.1 arcsec rms per coordinate (Pier {\etal} 2003).

Object fluxes are determined in several different ways by {\tt photo},
as described in Stoughton {\etal} (2002). The primary measure of flux
used for galaxies is the SDSS Petrosian magnitude, a modified version
of the quantity proposed by Petrosian (1976). 
In the absence of seeing, Petrosian magnitudes measure a
constant fraction of a galaxy's light regardless of distance (or
size). They are described in greater detail by Blanton {\etal} (2001)
and Strauss {\etal} (2002). Another important measure of flux for
galaxies is the SDSS model magnitude, which is an estimate of the
magnitude using the better of a de Vaucouleurs and an exponential fit
to the image. Finally, the SDSS also measures the flux in each object
using the local PSF as a template, which is the highest
signal-to-noise ratio measurement of flux for point sources.

Main sample target selection (Strauss {\etal} 2002) involves 
star/galaxy separation,
application of the flux limit, application of the surface brightness limit
and application of the fiber magnitude limit. Expressed
quantitatively, the first three of these criteria are
\begin{eqnarray}
r_{\mathrm{PSF}} - r_{\mathrm{model}} &>& s_{\mathrm{limit}} \cr
r_{\mathrm{petro}} &<&\rdim, \mathrm{~and}\cr
\mu_{50} &<& \mu_{50,\mathrm{limit}}.
\end{eqnarray}
where $r_{\mathrm{petro}}$ is the dereddened Petrosian magnitude in
the $r$ band (using the dust maps of Schlegel, Finkbeiner \& Davis 1998),
$r_{\mathrm{model}}$ is the model magnitude, $r_{\mathrm{PSF}}$ is the
PSF magnitude, and $\mu_{50}$ is the Petrosian half-light surface
brightness of the object in the $r$-band. In practice, the values of
the target selection parameters vary across the survey in a
well-understood way, but for the bulk of the area, they are
$s_{\mathrm{limit}}=0.3$, $r_{\mathrm{limit}}=17.77$, and
$\mu_{50,\mathrm{limit}}=24.5$. 
We note here that objects near the
spectroscopic flux limit are nearly five magnitudes brighter than the
photometric limit; that is, the fluxes are measured at a signal-to-noise ratio
of a few hundred.

Fibers are assigned to a set of circular tiles with a field of view
$1.49^\circ$ in radius by an automatic tiling pipeline
(Blanton {\etal} 2003).  The targets are observed using a 640 fiber
spectrograph on the same telescope as the imaging camera. 
We extract
one-dimensional spectra from the two-dimensional spectrograms using a pipeline
({\tt idlspec2d}) created specifically for the SDSS instrumentation
(Schlegel {\etal} 2003); a further pipeline ({\tt specBS v4\_9}) fits for the redshift of each
spectrum. The official SDSS redshifts are obtained from a different
pipeline (Frieman {\etal} 2003). The two independent versions provide
a consistency check on the redshift determination. They are consistent
(for galaxies) at over the 99\% level.

Fibers on a single plate cannot be placed more closely than $55''$.
Thus, redshifts for both members of a close galaxy pair can only be
obtained in regions where tiles overlap. If we did not take fiber
collisions into account at all, we would systematically underestimate
correlations on all scales. We correct this bias by
assigning each galaxy pair member whose redshift was not obtained because of a
fiber collision the same redshift as the pair member whose redshift
was measured.
Thus, for 192{,}642 of the galaxies in the full sample, a
spectroscopic redshift is available, but for 12{,}801 ($\sim 6\%$) we
must assign redshifts according to this prescription.  Using the
overlaps of multiple tiles, where many of these pairs can be
recovered, Zehavi {\etal} (2002) have shown that this procedure works
well on large scales, and we confirm this conclusion with
additional tests in \sec{SystematicsSec}.

As of July 2002, the SDSS had imaged and targeted 2{,}873 deg$^2$ of sky
and taken 431{,}291 successful spectra (including 323{,}126 spectra of
galaxies) over $2{,}499$ deg$^2$ of that area. We
created a well-defined sample for calculating large-scale structure
and galaxy property statistics from these data, known as Large-Scale Structure {\tt
sample11}.  {\tt sample11} consists of all of the photometry for all
of the targets over that area (as extracted from the internal SDSS
operational database), all of the spectroscopic results (as output
from {\tt idlspec2d}), and a description of the
angular window function of the survey and the flux and surface
brightness limits used for galaxies in each area of the sky (discussed
more fully below).  For most of the area, we used the same version of the
analysis software that was used to create the target list.
However, for the area covered by the Early Data Release (EDR;
Stoughton {\etal} 2002) we used the version of the analysis software used for
that data release, since it was substantially better than the early
versions of the software used to target that area. For {\tt photo},
the most important piece of analysis software run on the data, the
versions used for the photometry range from {\tt v5\_0} to {\tt
v5\_2}.  The region covered by this sample is similar to, but somewhat
greater than, the region which will be released in the SDSS Data
Release 1 (DR1), (which will use a newer version of the software which
among other things improves the handling of large galaxies).
For all of the subsamples of {\tt sample11}
defined here, we define a bright limit of $r = 14.5$ (using
Petrosian, dereddened magnitudes) to exclude the spectroscopic bright
limits imposed in galaxy target selection as well as the possibility
of poor deblending of bright objects by {\tt photo} (a problem for
{\tt v5\_2} and previous, though less so for {\tt v5\_3}).

We measure galaxy magnitudes through a set of bandpasses that are 
constant in the observer frame.  These observer frame
magnitudes correspond to different rest-frame magnitudes depending on
the galaxy redshift.  In order to compare galaxies observed at
different redshifts, we convert all the magnitudes to a single fixed
set of bandpasses using the method of
Blanton {\etal} (2002b) ({\tt kcorrect v1\_11}). These routines fit a linear
combination of four spectral templates to each set of five magnitudes. The
coefficient $a_0$ of the first template is an estimate of the flux in
the rest-frame optical range ($3500\Angstrom < \lambda < 7500\Angstrom$), 
the fractional contribution
of the other coefficients $a_1/a_0$, $a_2/a_0$, and $a_3/a_0$
characterize the spectral energy distribution of the galaxy. The most
significant variation is along $a_3/a_0$.  Taking the sum of the
templates and projecting it onto filter responses, we can calculate
the $K$-corrections from the observed bandpass to any rest-frame
bandpass. In practice, for the rare galaxies in our sample around redshift $z\sim 0.3$ this
procedure is unstable. Thus, for this sample we fix the coefficients
for all galaxies at redshifts $z>0.28$ to the average value for
galaxies between redshifts $0.26 < z < 0.28$, corresponding to the typical
red luminous galaxy SED. 
The median redshift of the sample is 0.1, so in this paper, we quote absolute magnitude
restframe bandpasses shifted blueward from the 
observatory frame bandpasses by a factor 1.1 and denoted
$\band{0.1}{r}$ in $r$-band.  
This procedure minimizes the
uncertainties in the $K$-corrections, since galaxies near the median
redshift 
independent of their spectral energy distribution.
For a galaxy exactly at $z=0.1$, the $K$-corrections are trivial.

In the remainder of this section, we model the three-dimensional
selection function $\nbar(\r)$, which gives the expected number density of galaxies in the absence of clustering
as a function of three-dimensional position.
For a uniform
magnitude limit, our selection function is separable into the product
of an angular part and a radial part: \beq{SeparabilityEq} \nbar(\r) =
\nbar(\rh)\nbar(r), \eeq where $\r\equiv r\rh$ and $\rh$ is a unit vector.
The angular part $\nbar(\rh)$ may take any value between 0 and 1, and
gives the completeness as a function of position, \ie, the fraction of
all survey-selected galaxies for which survey quality redshifts are
actually obtained, while $\nbar(r)$ gives the radial selection
function.  As described in \sec{SystematicsSec}, 
this separability allows powerful tests for possible
systematic effects arising from extinction or calibration problems,
which would cause a purely angular modulation of density fluctuations,
or from a mis-estimate of the radial selection function, which would
cause a purely radial modulation of the density.  The SDSS
spectroscopic completeness is so good that we find no evidence for
weather-related effects breaking the separability as in the 2dFGRS
(Colless {\etal} 2001), and therefore do not need to perform corrections for this
effect as in THX02.  We will now describe our modeling of the two
factors $\nbar(\rh)$ and $\nbar(r)$ in turn.

\subsection{The angular selection function}
\label{MaskSec}

\subsubsection{Specification}
\label{MaskSpecificationSec}

The geometry of the survey is somewhat complex due to the fact that
the imaging and spectroscopy programs are carried out concurrently. To
supply targets for the spectroscopic program, periodically a ``target
chunk'' of imaging data is processed, calibrated, and has targets
selected. These target chunks never overlap, so that once a set of
targets is defined in a particular region of sky, it never changes in
that region. Thus, the task of determining the selection limits used
in any region reduces to tracking how the target chunks cover the sky.
This list of target chunks, their boundaries, and their selection
criteria is an important product of {\tt sample11}.

To drill spectroscopic plates which have fibers on these targets, we
define ``tiling chunks'' which in principle
can overlap more than one target chunk.
The tiling algorithm (Blanton {\etal} 2003) 
is then run in order to position tiles and assign fibers to them, after
which plates are designed (that is, any available fibers are assigned to
various classes of auxiliary targets as well as to sky and calibration stars)
and then drilled. In general, these tiling chunks {\it will} overlap
because we want the chance to assign fibers to targets which may have
been in adjacent, earlier tiling chunks but were not assigned a
fiber. For a target to be covered by a particular tile, it must be in
the same tiling chunk as that tile {\it and} be within the area of the
tile itself (because the edges of tiles can extend beyond the tiling
chunk boundaries, a particular region of sky can be within the area of a
tile but not ``covered'' by it as defined here). We then divide the
survey into a number of ``sectors'',
regions which are covered by a unique set of tiles and tiling chunks
(following the nomenclature of the 2dFGRS, Percival {\etal} 2001, the
``sectors'' are the same as the ``overlap regions'' defined in Blanton {\etal}
2001a).  For instance, two tiles overlapping each other but no other
tiles give rise to three sectors: the area covered only by the first
tile, the area covered only by the second tile, and the area covered
by both.

For \lsssample\, the sky area is covered by 669 circular tiles of
diameter $2.98^\circ$ and is split into 2489 disjoint sectors. This
decomposition is convenient since each sector has a unique sampling
rate.  The sampling rate of a sector is defined as the number of
redshifts of galaxy targets obtained in the sector (including the
galaxies assigned the redshift of a neighbor because of a fiber
collision) divided by the number of galaxy targets in the sector. The
sampling rate so calculated is about 95\% on average across the survey
area; about 95\% 
of the survey area has completeness greater than
90\%.
This is illustrated in \fig{aitoffFig}.

Two additional sets of geometric entities affect the angular selection
function $\nbar(\rhat)$: it vanishes inside each of 55 rectangular
holes (regions masked out due to bad data quality or tiling bugs from
early on in the survey) and outside the official survey region defined
by 83 rectangular bounding boxes (the boundaries of the target
chunks).  In summary, the angular selection function $\nbar(\rhat)$
equals the sampling fraction when inside the survey area, zero
otherwise.

An additional complication when evaluating $\nbar(\rhat)$ is that the
above-mentioned geometric entities are specified in three different
coordinate systems in various combinations: equatorial coordinates
(RA,Dec), SDSS survey coordinates $(\eta,\lambda)$, SDSS great circle
coordinates $(\mu,\nu)$.

\subsubsection{Spherical polygon representation}

Fortunately, we can convert the specification of $\nbar(\rhat)$ into
an equivalent but much simpler form in terms of spherical polygons 
which encodes all these complications. This
simplification is necessary since our
power spectrum estimation method involves the complex task of
expanding $\nbar(\rhat)$ in spherical harmonics.

All tile, hole and bounding box boundaries are simple arcs on the
Celestial Sphere, corresponding to the intersection of the sphere with
some appropriate plane.  This means that any convex spherical polygon
(a tile, hole, bounding box, sector, \etc) can be defined as the
intersection of a set of {\it caps}, where a cap is the set of
directions $\rh$ satisfying $\ahat\cdot\rh > b$ for some unit vector
$\ahat$ and some constant $b\in[-1{,}1]$.  Think of a cap as the area
defined by a plane slicing a sphere. For instance, a tile is a single
cap, and a rectangular hole is the intersection of four caps. The
angular selection function $\nbar(\rhat)$ (plotted in \fig{aitoffFig})
can be clearly be represented as a list of non-overlapping polygons
such that $\nbar(\rh)$ is constant in each one.  We construct the
polygon list using the {\it Mangle} software described in 
Hamilton \& Tegmark (2003) and
available at {\it
http://www/http://casa.colorado.edu/$\sim$ajsh/mangle/}, which
involves the following steps:
\begin{enumerate}

\item We generate a list of 807 polygons comprised of
669 tiles, 83 bounding boxes and 55 holes.

\item Whenever two gs intersect, we split them into
non-intersecting parts, thereby obtaining a longer list of 8484
non-overlapping polygons. Although slightly tricky in practice, such
an algorithm is easy to visualize: if you draw all boundary lines on a
sphere and give it to a child as a coloring exercise, using four
crayons and not allowing identically colored neighbors, you would soon
be looking at such a list of non-overlapping polygons.

\item We compute the completeness $n(\rh)$ for each of these new polygons,
using the 
scheme described in \sec{MaskSpecificationSec}.

\item We simplify the result by omitting polygons with zero weight
and merging adjacent polygons that have identical weight.

\end{enumerate}
The result is a list of 2914 polygons with a total (unweighted) area
of 2499 square degrees, and an effective (weighted) area
$\int\nbar(\rh)d\Omega$ of 2417 square degrees.  
These polygons are sometimes
sectors, sometimes parts of sectors.
This angular completeness map,
and the polygons into which it resolves,
are illustrated in Figure~\ref{aitoffFig}. 

\begin{figure} 
\vskip\smtopskip
\centerline{\epsfxsize=\figsize\epsffile{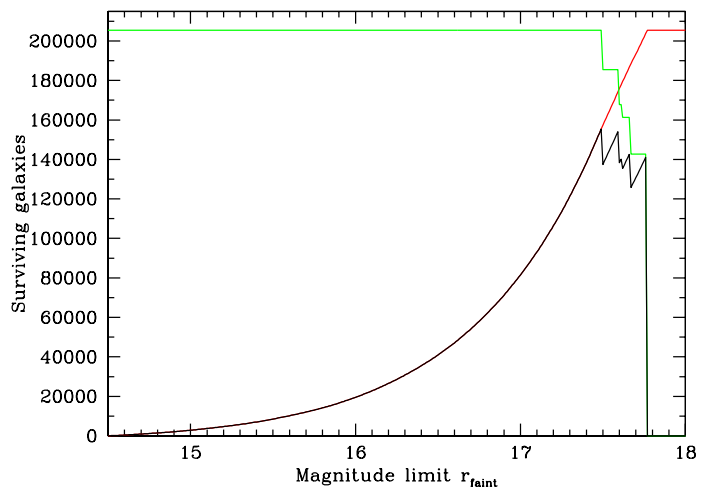}}
\vskip\smbotskip
\caption[1]{\label{survivorsFig}\footnotesize%
Jagged curve shows number of galaxies surviving as a function of
uniform magnitude cut, and is approximately shaped as the product of
the two other curves, which corresponds to our two cuts: the rising
curve shows the number of galaxies whose magnitude is brighter than
$\rdim$ and the falling staircase shaped curve shows the number of
galaxies in sectors whose magnitude limit is fainter than $\rdim$.}

\end{figure}

\subsection{Imposing a uniform magnitude limit}
\label{maglimSec}

As mentioned above, the faint magnitude limit varies in a known way
from target chunk to target chunk,
and is hence a known constant in each of our 2914 polygons.
We construct a uniform galaxy sample that is complete
down to a limiting magnitude $\rdim$ by applying the following two cuts:
\begin{enumerate}
\item Reject all galaxies
whose extinction-corrected magnitude $r$ is fainter than $\rdim$.
\item Reject all
sectors whose extinction-corrected magnitude limit is brighter than $\rdim$.
\end{enumerate}
\Fig{survivorsFig} shows the number of surviving galaxies as a function 
of $\rdim$. As we increase $\rdim$, the first cut eliminates 
fewer galaxies whereas the second cut eliminates more galaxies.
The result is seen to be a curve with peaks at
17.50, 17.60, 17.62, 17.67 and 17.77, corresponding to the
five magnitude limits used in spectroscopic target selection during the course
of the survey.
To maximize our sample size, we choose to cut at the highest peak
($\rdim=17.50$). 
This gives a sample of 157{,}389 galaxies, denoted {\tt safe0} in Table 1.
Since the optimal cut of 17.50 also happens to be the brightest magnitude limit used,
we need not reject any sectors (as per cut 2), so
the angular footprint of this uniform subsample has the same area as that of the full sample.
As the survey progresses further with its current magnitude limit of
17.77, this will eventually become the limit that yields the largest sample.

\subsection{The radial selection function}
\label{selfuncSec}

It is important to estimate the radial selection function $\nbar(r)$
as accurately as possible, since errors in it translate
into spurious large scale power.
Our estimate is plotted in \fig{zhistFig} and is computed as follows.

The radial window function of \lsssample\ is defined by the galaxy
luminosity function, the flux limits, and the absolute magnitude
limits of the sample in question.  As noted above, our sample is
limited at bright and faint apparent magnitudes: $14.5<r<17.5$.  Thus,
at any given redshift we can only observe galaxies in a given absolute
magnitude range. 
When making cuts based on absolute magnitude, we use the 
quantity $\band{0.1}{r}$ described in Blanton (2002), which refers to
the $r$-band magnitude $K$-corrected to its $z=0.1$
value. Thus evolution-corrected absolute magnitudes $M$ are calculated from
apparent magnitudes $m$ as follows:
\begin{equation}
M = m - \DM(z) - K_{0.1}(z) + Q(z),
\end{equation}
Here $K_{0.1}(z)$ is the galaxy $K$-correction, as calculated
using the code of Blanton {\etal} (2002), {\tt kcorrect v1\_11}.
$\DM(z)\equiv 5\log[r(1+z)] + 25$ is the distance modulus (see
Hogg 1999), where  $(1+z)r$ is the luminosity distance.
$Q(z)$ accounts for the average evolution in galaxy
luminosities in the recent past, and we use the fit $Q(z)=1.6(z-0.1)$ .
We will measure evolution in detail for different galaxy types in future
papers; however, for the present work, this simple fit for the evolution of all
galaxies is sufficient.

At any redshift, the fraction of objects in this absolute magnitude
range $\Mbri-\Mdim$ that are in the sample is
\beq{eqn:phi}
f(z) =
\frac{\int_{\Mbri(z)}^{\Mdim(z)} \Phi(M)dM}
{\int_{\Mbri(z)}^{\Mdim(z)} \Phi(M)dM},
\eeq
where $\Phi(M)$ is the luminosity function (number density of objects
per unit magnitude) and
\beqa{eqn:mminmax}
\Mbri(z) &=& {\rm max}[M_{\rm min}, 14.5-\DM(z)-K_{0.1}(z)+Q(z)], \cr
\Mdim(z) &=& {\rm min}[M_{\rm max}, 17.5-\DM(z)-K_{0.1}(z)+Q(z)].\nonumber
\eeqa
In this context, $K_{0.1}(z)$ is determined using the mean galaxy SED in the
sample. Equations~(\ref{eqn:phi}) and~(\ref{eqn:mminmax}) simply
express the fact that a galaxy must lie in our apparent magnitude
range and in our absolute magnitude range to be included in the
sample.  The luminosity function for our sample is determined in the
manner described by Blanton {\etal} (2001), using the step-wise maximum
likelihood method of Efstathiou {\etal} (1988), again using $Q(z) = 1.6(z-0.1)$.

We transform the galaxy positions
into the Local Group frame
assuming that the solar motion relative to the Local Group
is 306 km/s toward $l=99^\circ$, $b=-4^\circ$
(Courteau \& van den Bergh 1999).

\section{Power spectrum estimation details}
\label{powerdetails}

In this Appendix, we describe our power spectrum estimation procedure 
in sufficient detail for the reader interested in reproducing our analysis.

\subsection{Parameterizing our problem}

We parameterize the ratio of the three power spectra $\Pgg(k)$,
$\Pgv(k)$ and $\Pvv(k)$ to the prior 
as piecewise constant functions, each with
97 ``steps''.  Doing this rather than taking the power spectrum itself
to be constant avoids unnecessarily jagged spectra as discussed in
THX02.
The resulting
parameters $p_i$ are termed the {\em band powers}.  
As long as the prior agrees 
fairly well with the measured result, this has the advantage of giving 
better behaved window functions, as described in Hamilton \& Tegmark (2000).

We group these $3\times 97$ numbers into a
291-dimensional vector $\p$.
We choose our 97 $k$-bands to be 
centered on logarithmically equispaced $k$-values
$k_i = 10^{i-65\over 16}\hperMpc$, $i=1,...,97$,
\ie, ranging from $0.0001\hperMpc$ to $100\hperMpc$.   
This should provide fine enough 
$k$-resolution to  
resolve any spectral features
that may be present in the power spectrum.
For instance, baryon wiggles have a characteristic scale of order
$\Delta k\sim 0.1$, so we oversample the first one around $k\sim 0.1$ by
a factor $\Delta k/(k_{26}-k_{25}) \sim 16/\ln 10 \sim 7$.

This parameterization means that we can write
the pixel covariance matrix of \eq{xCovEq} as
\beq{CsumEq}
\C = \sum_{i=0}^{291} p_i \C,{_i},
\eeq
where the derivative matrix 
$\C,{_i}\equiv\partial\C/\partial p_i$
is the contribution to the covariance matrix from the $\ith$ band.
For notational convenience, we have included the noise term
in \eq{CsumEq} by defining $\C,_{0}\equiv\N$, corresponding
to an extra dummy parameter $p_{0}=1$ giving the shot noise normalization.

\subsection{Quadratic estimator basics}
\label{Qsec}

Quadratic estimators were originally derived for galaxy survey
applications (Hamilton 1997ab). They were accelerated and 
first applied to CMB analysis (Tegmark 1997b; Bond, Jaffe \& Knox 2000),
and have been a cornerstone of almost all recent CMB power spectrum analyses.

Our quadratic estimators $\ph_i$ defined by \eq{phDefEq} are designed
to measure the corresponding parameters $p_i$.
We choose the $\Q$-matrices to be of the form
\beq{QdefEq}
\Q_i = {1\over 2}\sum_{j=1}^m\M_{ij}\C^{-1}\C,_i\C^{-1}
\eeq
where 
$m=291$ is the number of bands (power spectrum parameters).
$\M$ is an $m\times m$ matrix that we will specify below.
In the approximation that the pixelized data has a 
Gaussian probability distribution 
(a good approximation in 
our case, because we are mostly in the linear regime),
the choice of \eq{QdefEq} has been shown to be lossless, distilling
all the power spectrum information
from the original data set into the quadratic estimator vector $\ph$ 
(Tegmark 1997b).
This is true for {\it any} choice of the matrix $\M$ as long as it is invertible: 
the result using a different matrix $\M'$ could always be computed
afterwards by multiplying the vector $\ph$ by $\M'\M^{-1}$.

The quadratic estimators $\ph_i$ have the additional 
advantage (as compared with, \eg, maximum-likelihood estimators)
that their statistical properties are easy to compute: their mean 
and covariance are given by equations\eqn{pMomentsEq1a}
and\eqn{pMomentsEq1b}, where the window
matrix $\W$ and the covariance matrix $\SS$ are 
\beqa{pMomentsEq2a}
\W_{ij}   &=&\tr\left[\Q_i\C,_j\right],\\
\SS_{ij}  &=& 2\,\tr[\Q_i\C\Q_j\C]\label{pMomentsEq2b}.
\eeqa
Substituting \eq{QdefEq}, this gives
\beqa{Weq}
\W	&=&\M\F,\\
\label{CovarEq}
\SS	&=&\M\F\M^t,
\eeqa
where $\F$ is the {\it Fisher information matrix} (Fisher 1935; Tegmark {\etal} 1997)
\beq{GaussFisherEq}
\F_{ij} = {1\over 2}\tr\left[\C^{-1}\C,_i\C^{-1}\C,_j\right].
\eeq
In conclusion, the quadratic estimator method takes
the vector $\x$ and its covariance matrix
$\C$ from \fig{xFig} and compresses it 
into the shorter vector $\ph$ in \fig{power_all3_binnedFig}
and its covariance matrix while retaining essentially 
all the cosmological information.

\subsection{Quadratic estimator variants: choosing the $\M$-matrix}

For the purpose of fitting models $\p$ to our measurements $\phat$, 
we are already done --- equations\eqn{pMomentsEq1a} and\eqn{pMomentsEq1b}
tell us how to compute $\chi^2$
for any given $\p$, and the result 
\beq{chi2Eq}
\chi^2 = (\phat-\expec{\phat})^t\SS^{-1}(\phat-\expec{\phat})^t
\eeq
is independent of the choice of $\M$.
However, since one of the key goals of our analysis is to provide
model-independent measurement of the three power spectra themselves, 
the choice of $\M$ is crucial. Ideally, we would like
both uncorrelated error bars (diagonal $\SS$) and 
well-behaved
(narrow, unimodal and non-negative) window functions $\W$ that
do not mix the three power spectra, $\W=\I$ being the ideal.

There are two separate issues of interest when choosing $\M$. 
The first involves the tradeoff between making $\SS$ well-behaved
and making $\W$ well-behaved.
The second involves the complication that we are measuring three power spectra 
rather than one, and that quadratic estimators tend to mix them, with 
estimators of one spectrum being contaminated by ``leakage'' from another.
The following two subsections discuss these two issues in turn.
For all choices below, we wish each window function (row of $\W$) to sum to unity
so that we can interpret $\phat_i$ as measuring a weighted average of the true power.
Because of \eq{Weq}, the rows of $\M$ are therefore normalized to satisfy 
\beq{MnormEq}
\sum_j (\M\F)_{ij} = 1
\eeq
for all $i$.

\subsubsection{Correlated, anticorrelated and uncorrelated band powers}
\label{Msec}

There are a number of interesting choices of $\M$ that each have their
pros and cons (Tegmark \& Hamilton 1998; Hamilton \& Tegmark 2000).
The simple choice where $\M$ is diagonal
gives the ``best guess'' estimates
in the sense of having minimum variance (Hamilton 1997a; Tegmark 1997a;
Bond, Jaffe \& Knox 2000), 
and also has the advantage of being independent of the number of 
bands used in the limit of high spectral resolution.
It was used for Figures~\ref{power_all3_correlatedFig} and~\ref{W0fig}.
Here the window functions are simply the rows of the Fisher matrix,
and are seen to be rather broad.
All entries of $\F$ are guaranteed to be positive as proven in PTH01,
which means not only that all windows are positive (which is good)
but also that all measurements are positively correlated (which is bad).

Another interesting choice is (Tegmark 1997b)
$\M=\F^{-1}$, which gives $\W=\I$.
In other words, all window functions are Kronecker
delta functions, and $\phat$ gives completely unbiased estimates 
of the band powers, with 
$\expec{\ph_i}=p_i$ regardless of what values the other band 
powers take. This gives an estimate $\phat$ similar to the maximum-likelihood
method (Tegmark {\etal} 1998), and the covariance matrix of
\eq{pMomentsEq1b} reduces to $\F^{-1}$.
A serious drawback of this choice is that that if we have
sampled the power spectrum on a scale finer than the inverse survey
size in an attempt to retain 
all information about wiggles {\etc}, this covariance matrix
tends to give substantially larger error bars 
($\Delta p_i\equiv\M_{ii}^{1/2}=[(\F^{-1})_{ii}]^{1/2}$)
than the first method, anti-correlated between neighboring bands.

The two above-mentioned choices for $\M$ both tend to 
produce correlations between the band power error bars.
The minimum-variance choice generally gives 
positive correlations, since the Fisher matrix cannot 
have negative elements, whereas 
the unbiased choice tends to give
anticorrelation between neighboring bands.
The choice (Tegmark \& Hamilton 1998; Hamilton \& Tegmark 2000) 
$\M=\F^{-1/2}$ with the rows renormalized
has the attractive property of making the errors uncorrelated,
with the covariance matrix of \eq{pMomentsEq1b} 
diagonal. The corresponding window functions $\W$ are
plotted in \fig{W1fig}, and are seen to be 
quite well-behaved, even narrower than those in
\fig{W0fig} while remaining positive in almost all cases.\footnote{
The reader interested in mathematical challenges will be interested to know
that it remains a mystery to the authors 
why this $\F^{-1/2}$ method works so
well. We have been unable to prove that the resulting window matrix $\F^{1/2}$ has no negative
elements (indeed, counterexamples can be constructed), yet the method
works like magic in practice in all LSS and CMB applications we
have tried.
}
This choice, which is the one we make in this paper, 
is a compromise between the two first ones:
it narrows the minimum variance window functions at the cost of
only a small noise increase, with uncorrelated noise as an extra bonus.
The minimum-variance band power estimators are essentially a smoothed version 
of the uncorrelated ones, and 
their lower variance was 
paid for by correlations which reduced the effective number of independent 
measurements.

\subsubsection{Disentangling the three power spectra}

The fact that we are measuring three power spectra rather than one
introduces an additional complication.
As illustrated by \fig{disentanglementFig}, an estimate 
of the power in one of the three spectra generally 
picks up unwanted contributions (``leakage'') from the other two,
making it complicated to interpret.
Although the above-mentioned $\F^{-1}$-method in principle eliminates
leakage completely (since it gives $\W=\I$), the cost in terms of increased error bars is found
to be prohibitive. We therefore follow HTP00 and THX02 in adopting the following 
procedure for disentangling the three power spectra:
For each of the 97 $k$-bands, we take linear combinations
of the $gg$, $gv$ and $vv$ measurements such that the unwanted parts
of the window functions average to zero. 
This procedure is mathematically identical to that
used in Tegmark \& de Oliveira-Costa (2001) for separating different types
of CMB polarization, so the interested reader is referred there for the
explicit equations. For the reader familiar with Bayesian statistics, our 
disentanglement procedure is tantamount to marginalizing over the amplitudes
of the other two power spectra, separately for each band. 

The procedure is 
illustrated in \fig{disentanglementFig}, and by construction has 
the property that leakage is completely eliminated if the 
true power has the same shape (not necessarily the same amplitude)
as the prior.
We find that this method works quite well 
(in the sense that the unwanted windows do not merely
average to zero) for the most accurately 
measured band powers, in particular the central parts of 
the gg-spectrum, with slightly poorer leakage elimination for 
bands with larger error bars.

The window functions plotted in \fig{Wfig} are the 
gg-windows after disentanglement.
Note that although our disentanglement introduces correlations
between the $gg$, $gv$ and $vv$ measurements for a given $k$-band, different
$k$-bands remain uncorrelated.

\begin{figure} 
\vskip\smtopskip
\centerline{\epsfxsize=\figsize\epsffile{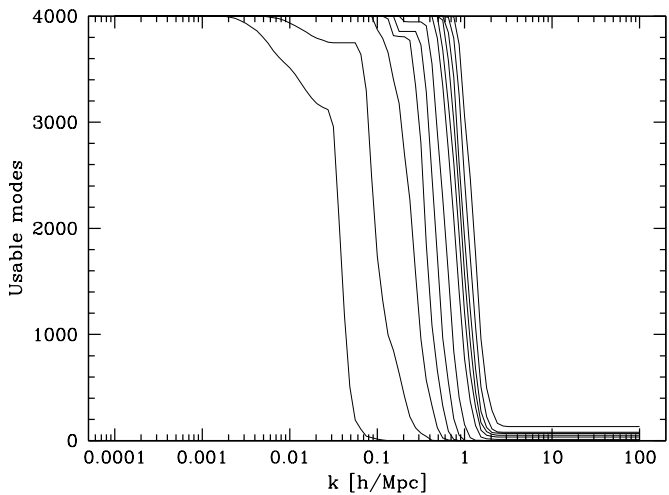}}
\vskip\smbotskip
\caption[1]{\label{usableFig}\footnotesize%
Numerical convergence. The figure shows for how many of our 4000 PKL modes
the numerical calculations are converged to accurately measure the power
up to a given wavenumber $k$.
From left to right, the 12 curves correspond to truncation at 
$\lcut=$20, 40, 60, 80, 100, 120, 140, 160, 180, 200, 220 and 240.
}
\end{figure}

\subsection{Numerical issues}

Our analysis pipeline has a few ``knobs'' that can be
set in more than one way. This section discusses the 
sensitivity to such settings.

\subsubsection{The prior}
\label{PriorSec}

The analysis method employed assumes a ``prior'' power
spectrum via \eq{CsumEq}, both to compute band power error bars
and to find the galaxy pair weighting that minimizes them.
An iterative approach was adopted starting
with a simple BBKS model for $\Pgg(k)$ (Bardeen {\etal} 1986), then 
shifting it vertically and horizontally to better fit the resulting measurements and 
recomputing the measurements a second time.
As priors for $\Pgv(k)$ and $\Pvv(k)$ we use 
equations\eqn{KaiserLimitEq1} and \eqn{KaiserLimitEq2} with 
$r=1$ and $\beta=0.5$, which provides a good fit to the measurements.

To what extent does this choice of prior affect the results?
On purely theoretical grounds
(\eg, Tegmark, Taylor \& Heavens 1997), one expects a grossly incorrect
prior to give unbiased results but with unnecessarily large
variance. If the prior is too high, the sample-variance 
contribution to error bars will be overestimated and vice versa.
This hypothesis has been extensively 
tested and confirmed in the context of power spectrum measurements
from both the Cosmic Microwave Background 
(\eg, Bunn 1995) and galaxy redshift surveys
(PTH01), confirming that the correct result is recovered on average even when using 
a grossly incorrect prior. 
In our case, the prior by construction agrees quite well with the actual
measurements (see \fig{power_all3_binnedFig}), so the quoted error 
bars should be reliable as well.

\subsubsection{Effect of changing the number of PKL modes}
\label{npixSec}

We have limited our analysis to the first $N=4000$ PKL modes
whose angular part is spanned by spherical harmonics with 
$\l\le 40$. This choice was a tradeoff between the desire to
capture as much information as possible about the galaxy survey
and the need to stay away from small scales where non-linear effects
invalidate the Kaiser approximation to redshift distortions. 
To quantify our sensitivity to these choices, we repeated the
entire analysis using the first 500, 1000, 2000 and 4000 modes.
Our power spectrum measurements on the very largest scales were
recovered even with merely 500 modes.
As we added more and more modes (more and more small-scale information), 
the power measurements converged to those in \fig{power_all3_binnedFig}
for larger and larger $k$.
\Fig{mode_keffFig} shows that our 4000 PKL modes
are all rather insensitive to cosmological information 
for $k\simgt 0.2$.

\subsubsection{Convergence issues}

A key step in our analysis pipeline is the computation of the 
matrices $\P_i\equiv\partial\SS/\partial p_i$ that give the contribution 
to the signal covariance matrix $\S$ from the $\ith$ band power.
This computation involves a summation over
multipoles $\l$ that should, strictly speaking, run from $\l=0$ to $\l=\infty$,
since the angular completeness map itself has sharp edges involving harmonics
to $\l = \infty$.
In practice, this summation must of course be truncated at some 
finite multipole $\lcut$. To quantify the effect of this truncation, we plot 
the diagonal elements of the $\P$-matrices as a function of $\lcut$
and study how they converge as $\lcut$ increases.
We define a given PKL-mode as having converged by
some multipole if subsequent $\l$-values contribute 
less than 1\% of its variance. \Fig{usableFig} plots the number
of usable PKL-modes as a function of wavenumber $k$,
defining a mode to be usable for our analysis only if it
is converged for all smaller wavenumbers $k'<k$ for 
all three power flavors ($\Pgg$, $\Pgv$ and $\Pvv$).
We use $\lcut=260$ in our final analysis, since this guarantees
that all 4000 modes are usable
for wavenumbers $k$ in the range $0-0.7\hperMpc$, \ie, 
comfortably beyond the large scales $0-0.3\hperMpc$ that are the
focus of this paper.  
With this cutoff, the computation of the $\P$-matrices (which scales as
$\lcut^2$ asymptotically),
took about a week on a 2 GHz linux workstation.
As a further test, we repeated our entire analysis with 
$\lcut=120$ and obtained almost indistinguishable power spectra.

\end{document}